\documentclass[a4paper,11pt]{article}
\pdfoutput=1 

\usepackage{jinstpub} 
\usepackage{multirow}

\def\aap{{A\&A}}               
\def\aj{{AJ}}                  
\def\apj{{ApJ}}                
\def\apjl{{ApJ}}               
\def\apjs{{ApJS}}              
\def\jcap{{JCAP}}              
\def\mnras{{MNRAS}}            
\def\prl{{PhysRevLett.}}            
\def\nat{{Nature}}            
\def\prd{{PhysRevD}}            
\def\nar{{New Astron. Rev.}}            

\newcommand{\crab}{{Tau\,A }}
\newcommand{\casa}{{Cas\,A }}
\newcommand{\cyga}{{Cyg\,A }}

\def\procspie{{SPIE Conference Series}}

\title{\boldmath LSPE-STRIP on-sky calibration strategy using bright celestial sources}

\author[a,b]{R.T. G\'enova-Santos,\note{Corresponding author.}}
\author[c,d]{M.~Bersanelli,}
\author[c,d]{C.~Franceschet,}
\author[e,f]{M. Gervasi,}
\author[a,b]{C.~L\'opez-Caraballo,}
\author[d]{L. Mandelli,}
\author[c,d]{S. Mandelli,}
\author[g]{M.~Maris,}
\author[c,d]{A.~Mennella,}
\author[a,b]{J.A. Rubi\~no-Mart\'{\i}n,}
\author[h]{F. Villa,}
\author[e,f]{M. Zannoni,}
\author[i]{C. Baccigalupi,}
\author[d]{B. Caccianiga,}
\author[c,d]{L. Colombo,}
\author[h]{F. Cuttaia,}
\author[i]{F. Farsian,}
\author[h]{G. Morgante,}
\author[c,d,l]{S. Paradiso,}
\author[j]{G. Polenta,}
\author[h]{S. Ricciardi,}
\author[h]{M. Sandri,}
\author[k]{A. Taylor,}
\author[h]{L. Terenzi,}
\author[c,d]{M. Tomasi}

\affiliation[a]{Instituto de Astrof\'{\i}sica de Canarias, E-38200 La Laguna, Tenerife, Canary Islands, Spain}
\affiliation[b]{Departamento de Astrof\'{\i}sica. Universidad de La Laguna (ULL), E-38206 La Laguna, Tenerife, Spain}
\affiliation[c]{Dipartamento de Fisica, Universit\`{a} degli Studi di Milano, via Celoria, 16, 20133 Milano, Italy}
\affiliation[d]{INFN--Sezione di Milano, Via Celoria 16, 20133 Milano, Italy}
\affiliation[e]{Dipartimento di Fisica, Universit\`{a} di Milano - Bicocca, Piazza della Scienza 3, 20126 Milano, Italy}
\affiliation[f]{INFN--Sezione di Milano Bicocca, Piazza della Scienza 3, 20126 Milano, Italy}
\affiliation[g]{INAF--Osservatorio Astronomico di Trieste, Via G.B. Tiepolo 11, Trieste, Italy}
\affiliation[h]{INAF--OAS Bologna, Via P. Gobetti 101, Bologna, Italy}
\affiliation[i]{SISSA, Astrophysics Sector, via Bonomea 265, 34136, Trieste, Italy}
\affiliation[j]{ASI, Via del Politecnico snc, 00133 Roma, Italy.}
\affiliation[k]{Sub-department of Astrophysics, University of Oxford, Denys Wilkinson Building, Keble Road, Oxford OX1 3RH, UK}
\affiliation[l]{Department of Physics and Astronomy, University of Waterloo,Waterloo, ON N2L 3G1, Canada}

\emailAdd{rgs@iac.es}

\abstract{In this paper we describe the global on-sky calibration strategy of the LSPE-Strip instrument. Strip is a microwave telescope operating in the Q-  and W-bands (central frequencies of 43 and 95\,GHz respectively) from the Observatorio del Teide in Tenerife, with the goal to observe and characterise the polarised Galactic foreground emission, and complement the observations of the polarisation of the cosmic microwave background to be performed by the LSPE-SWIPE instrument and other similar experiments operating at higher frequencies to target the detection of the B-mode signal from the inflationary epoch of the Universe. Starting from basic assumptions on some of the instrument parameters (NET, 1/f noise knee frequency, beam properties, observing efficiency) we perform realistic simulations to study the level of accuracy that can be achieved through observations of bright celestial calibrators in the Strip footprint (sky fraction of 30\%) on the determination and characterisation of the main instrument parameters: global and relative gain factors (in intensity and in polarisation), polarisation direction, polarisation efficiency, leakage from intensity to polarisation, beams, window functions and pointing model. 
}


\arxivnumber{} 

\proceeding{The LSPE/Strip instrument description and testing}

\begin{document}
\maketitle
\flushbottom

\section{Introduction}


The study of the anisotropies that were imprinted in the Cosmic Microwave Background at $z\approx 1100$ after the recombination of the Universe have been, over the last 25 years, one of the major sources of cosmological information. After the discovery of these temperature anisotropies by the COBE satellite \cite{smoot_1992}, a series of experiments were developed to measure their power spectrum from the ground with gradually increased angular resolution \cite{debernardis_00,lee_01,halverson_2002,dickinson_2004,readhead_2004,kuo_2004,tristram_2005}. These studies were continued from the space by the WMAP satellite \cite{bennett_2013}, and culminated with the measurements by the Planck satellite, that are cosmic-variance limited up to $\ell\approx 1600$ and produced the tightest constraints on cosmological parameters to-date, with sub-percent precision in most of them \cite{cpp2018-6}. The high fidelity of the Planck dataset also allowed, for the first time, to exploit the CMB polarization, in particular the curl-free E-mode component, to derive constraints on cosmological parameters with precision similar to that obtained from the temperature anisotropies. The detection of the other component of the CMB polarisation pattern, the curly B-modes, remains as of today one of the main challenges in cosmology. This B-mode signal should have been left imprinted in the CMB polarisation by gravitational waves created during the inflation of the Universe \cite{kamionkowski_1997,seljak_1997}. Detection of this faint signal would give access to physical processes occurring right after the Big Bang at energy scales of $\sim 10^{16}$~GeV, close to the Planck scale. Currently the best upper limit on this B-mode signal, parametrized through the ratio between tensor and scalar modes, is $r<0.036$ (95\% C.L.), and has been obtained from data from the BICEP/Keck experiment \cite{bicep2021-13}. Reaching or improving this limit requires not only an extremely good sensitivity (3 or 4 orders of magnitude below that required to measure the temperature anisotropies) and systematics control, but an extremely precise characterization in a wide frequency range of the spectral properties of Galactic foreground emission (mainly synchrotron and thermal dust emission) that completely overshadows the B-mode signal in any region of the sky and at any frequency \cite{cpp2015-10}. Several projects are being designed and developed to address these goals \cite{so_2019,litebird_2022}, and amongst them is the Large Scale Polarization Explorer (LSPE, \cite{lspe_2021}). 

Apart from the design and construction of large and extremely sensitive focal plane arrays, accessing the low expected amplitude of the B-mode signal requires development of a reliable and accurate calibration procedure to be implemented in the data processing pipeline. Inaccuracies in this process are inevitably transferred to the final processed CMB maps in the form of systematic errors, putting the fulfillment of the final scientific goals at risk. Therefore, improvements on detector sensitivity must go in parallel with improvements on the precision and reliability of the calibration strategy, which in some cases requires resorting to ideas or methods that have never been implemented before. One paradigmatic example, and perhaps one of the most challenging ones, is the calibration of the polarization direction, that should be determined with an accuracy of a few arcmin to reach $r\sim 10^{-3}$ \cite{vielva_2022}, something that is far from being achieved using \crab, the brightest polarised compact source on the sky \cite{aumont_2020}. 

In this paper, we address the on-sky calibration strategy of LSPE-Strip, one of the two instruments of SWIPE. It is part of a series of papers where we present a detailed description of different instrumental aspects of Strip. In section~\ref{sec:strip} we give a brief description of SWIPE, with emphasis on Strip. In section~\ref{sec:calibrators} we present a description of the main sources and regions that are visible in the Strip observed footprint and that we consider could be used for calibration purposes, and we discuss their calibration models. In section~\ref{sec:simulations} we have described the simulations of the Strip observations that are at the heart of the study of the expected accuracy achieved on different calibration parameters using on-sky data. Section~\ref{sec:strip_calibration} contains the main results of this study. Here we discuss the expected accuracy on the most important steps in the calibration process, namely: i) absolute grain calibration, that consists in converting the units of the signal measured by the instrument into a thermodynamic temperature, ii) relative gain calibration, that aims to correct internal gain variations that affect the output measured signal, iii) calibration of the global polarization direction, iv) calibration of the polarization efficiency, v) calibration and characterization of the leakage between intensity and polarization caused by non-idealities in the receivers, vi) calibration of the beam patterns and window functions, that determines the relation between the intrinsic spatial distribution of the sky brigthness and the observed antenna temperature distribution, and vii) calibration of the pointing model, that determines the relation between telescope coordinates and true spatial coordinates on the sky. Final remarks and conclusions are presented in section~\ref{sec:conclusions}.

\section{LSPE-STRIP}\label{sec:strip}
The Large Scale Polarization Explorer (LSPE) is a collaboration led by Italian institutes, with funding from the Italian Space Agency (ASI) and the Instituto Nazionale di Fisica Nucleare (INFN), and with the participation of various European and international partners \cite{lspe_2021}. The final goal of LSPE is to reach a sensitivity on the measurement of the tensor-to-scalar ratio of $r=0.03$ with 99.7\% confidence level (CL), and an upper limit of $r=0.015$ at 95\% CL. The instrumental setup to reach this goal consists of: 1) Strip, a HEMT-based polarimetric instrument at 43 and 95\,GHz that will observe from the Teide Observatory (Tenerife, Spain), and 2)  SWIPE, a bolometer-based instrument at 145, 210 and 240\,GHz from a stratospheric balloon. The idea behind this dual-instrument concept lies on the properties of the atmospheric spectrum. The Strip frequency bands can be covered from the ground, as the atmosphere transparency at those low frequencies is good enough, in particular from locations with a stable atmosphere with low precipitable water vapour like that of the Teide Observatory. On the contrary, the sky opacity raises steeply at high frequencies, and then the quality of the data obtained from a 15-day stratospheric flight surpasses what could be obtained from the ground even using a longer integration time. These two telescopes will be observing a similar sky patch, encompassing a large fraction of the Northern sky. Strip and SWIPE are fully complementary in the sense that combination of the wide frequency coverage provided by the two is crucial to achieve the targeted scientific goals. Strip low frequencies will be crucial to monitor the polarisation of the synchrotron emission produced by cosmic-ray electrons spiraling in the Galactic magnetic field, the high-frequency SWIPE channels will be sensitive to the polarised emission produced by thermal dust grains in the interstellar medium, while the 145\,GHz channel of SWIPE will provide the strongest constraining power on $r$ thanks to being more sensitive and closer to the foreground minimum. A thorough description of the project, of its instrumental setup, scientific goals and capabilities is presented in \cite{lspe_2021}. In the next subsection we will lay out a description of the main Strip parameters and scanning strategy that are relevant for this paper.

\subsection{The STRIP instrument}\label{sec:strip_instrument}

\begin{figure}
\centering
\includegraphics[width=14cm]{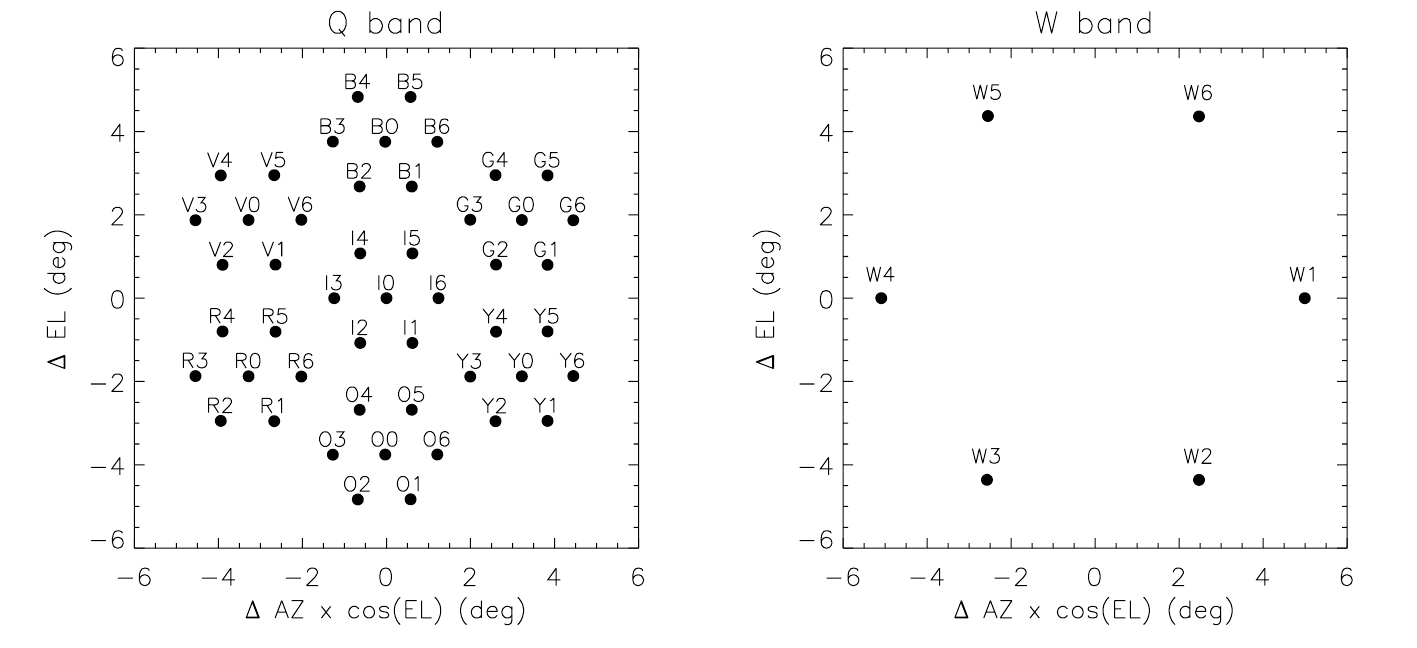}
\caption{Strip Q- (left) and W-band (right) focal plane configurations projected onto the sky. The angular separations, in AZ, EL coordinates, with respect to the pixel located on the centre of the focal plane (I0) are shown.}
\label{fig:strip_fp}
\end{figure}

The Strip instrument is composed of a total of 55 polarimeters that are able to directly measure Stokes Q and U parameters through a double-demodulation scheme that enables efficient $1/f$ noise rejection. 49 polarimeters operate in Q-band at a central frequency of 43\,GHz, while other 6 operate in W-band at 95\,GHz. The 6 W-band polarimeters, and 19 of the Q-band polarimeters were used in the QUIET instrument \cite{quiet_13}, while the rest of the Q-band modules have been manufactured following the same design. The Q-band receivers will be used to map the synchrotron polarisation, while the signal measured by the W-band receivers will be used to monitor and partially correct for atmospheric fluctuations. The layout of the 55 pixels on the Strip focal plane is depicted in Figure~\ref{fig:strip_fp}, where it can be seen that the size of the full field of view is around 10$^\circ$. The Strip focal plane is cryogenically cooled to a temperature of around 20\,K, and coupled to a fully steerable crossed-Dragone telescope
that was originally developed for the CLOVER experiment \cite{taylor_06}. This telescope has a parabolic primary mirror with an aperture of $1.5\,$m in diameter, providing angular resolutions of $\approx 20$~arcmin in Q-band and of $\approx 10$~arcmin in W-band. The telescope will be located at the Teide Observatory (16$^\circ$31' West, 28$^\circ$18' North, Tenerife, Spain) at an altitude of 2380~m a.s.l. This is a word-class astronomical observatory that benefits from a very stable atmosphere with median precipitable water vapour of 3.5\,mm \cite{castro_2016}. Its suitability for this kind of observations has been demonstrated by many other CMB experiments that have previously observed from this site, like the Very Small Array \cite{dickinson_2004}, or that are currently operating like QUIJOTE \cite{rubino_2023} or GroundBIRD \cite{honda_2020}. The basic parameters of the Strip instrument are listed in Table~\ref{tab:instrument_parameters}. Detailed information about the instrumental setup of Strip can be found in \cite{realini_2022,franceschet_2022,peverini_2022} and in the coming the papers of this series.

\begin{table*}
\caption{Basic parameters of the Strip instrument.}
\smallskip
\label{tab:instrument_parameters}
\centering
\begin{tabular}{@{}lcc}
\hline
 &	Q-band (43~GHz) & W-band (95~GHz)\\
 \hline
 Freq (GHz) & 43 & 95 \\
 Beam FWHM (arcmin) & 20 & 10 \\
 Number of polarimeters & 49 & 6 \\
 NET ($\mu$K$_{\rm CMB}\cdot $s$^{1/2}$) &  515 & 1139 \\
 Observing time & \multicolumn{2}{c}{2 years}\\
 Observing efficiency & \multicolumn{2}{c}{50\%}\\
 Sky coverage & \multicolumn{2}{c}{30\%}\\
 Geographical coordinates &  \multicolumn{2}{c}{$\lambda=+28.30^\circ$ , $\varphi=-16.1^\circ$} \\
 \hline
\end{tabular}
\end{table*}

\subsection{STRIP scanning strategy}\label{sec:scanning_strategy}

The baseline observing mode (henceforth ``nominal mode'') of the Strip telescope consists in continuous rotation of the telescope around the azimuth axis, with velocity $\omega=6$~deg/s at a constant elevation of $70^\circ$ (zenith angle $\beta=20^\circ$). This mode has many benefits: i) the telescope never undergoes accelerations, ii) constant elevation avoids variations in the signal baseline level due to different airmass, iii) highly uniform sky coverage, iv) fast scanning speeds minimises $1/f$ noise both from the instrument and from the atmosphere. With the latitude of the Teide Observatory being $28.3^\circ$ (see Table~\ref{tab:instrument_parameters}), this provides a sky coverage between declinations $8.3^\circ$ and $48.3^\circ$, while Earth rotation provides full coverage in right ascension after one (sidereal) day (see Figure~1 of \cite{lspe_2021}), allowing to achieve a coverage of $30.2\%$ of the sky and with a high overlap with the sky region that will be observed by SWIPE. 

Although the default is the ``nominal mode'', the Strip telescope will also allow observations in ``raster mode'' at elevations higher than 35$^\circ$, which is the lowest elevation that the telescope can reach (referred to the central detector, I0). The raster mode consists in moving the telescope back and forth between two different azimuths, either at a constant elevation or stepping the elevation by a small amount after each scan. This will allow reaching deeper integration times in specific sky regions, resulting in better sensitivities and then could be used to improve the signal-to-noise in specific calibrators, or to access interesting calibration sources like \casa that have elevations below $70^\circ$ and therefore are never picked up in the nominal-mode observations (see Figure~\ref{fig:planck_and_cal} and right panel of Figure~\ref{fig:sources_bp}).

\section{Bright celestial calibrators}\label{sec:calibrators}


To be used as calibrators, astronomical sources must fulfill various requirements like being: i) point-like, so that precise knowledge of the beam shape is not needed but just the beam solid angle to convert from flux density to specific intensity (antenna temperature) in the calibration process, ii) spatially isolated so that their measured flux densities, particularly in wide-beam telescopes, is not strongly biased by surrounding background structure, and iii) non-variable, so that a time-independent calibration scale can be defined. Sources meeting these criteria are not common. High-galactic latitude sources tend to be more isolated, but they are generally AGN/QSO that are prone to outbursts and variable. On the other hand, sources on the Galactic plane usually suffer from nearby background contamination (which is particularly harmful for low-resolution CMB experiments), and they tend to be either HII regions, which are usually extended, or SNR that tend to be variable. All these facts conspire so that the best achievable global calibration uncertainty based on point sources can hardly be better than 5\%. Orbital CMB missions resort to the (solar or orbital) CMB dipole to achieve calibration uncertainties well below 1\% \citep{jarosik_2011,cpp2013-5,cpp2013-8}.

\begin{figure}
\centering
\includegraphics[trim= 0mm 15mm 0mm 15mm, width=7.5cm]{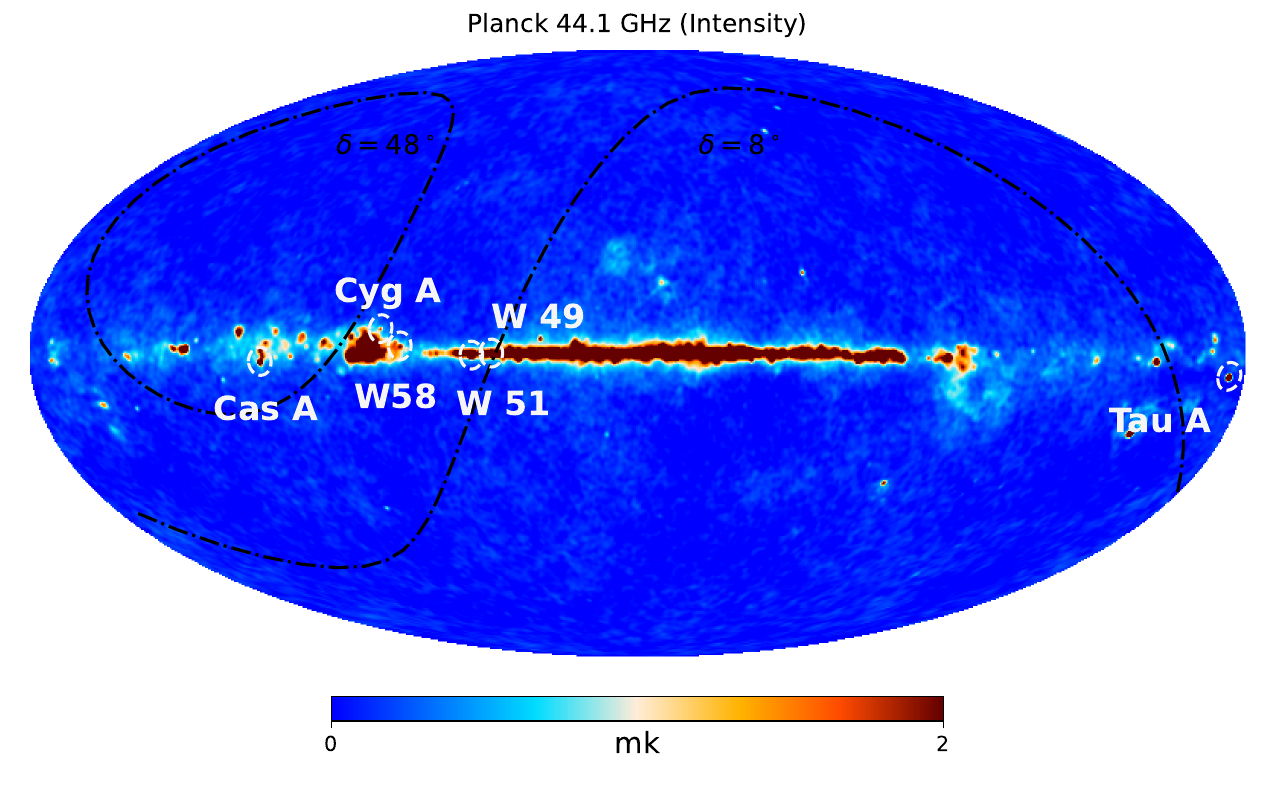}
\includegraphics[trim= 0mm 15mm 0mm 15mm, width=7.5cm]{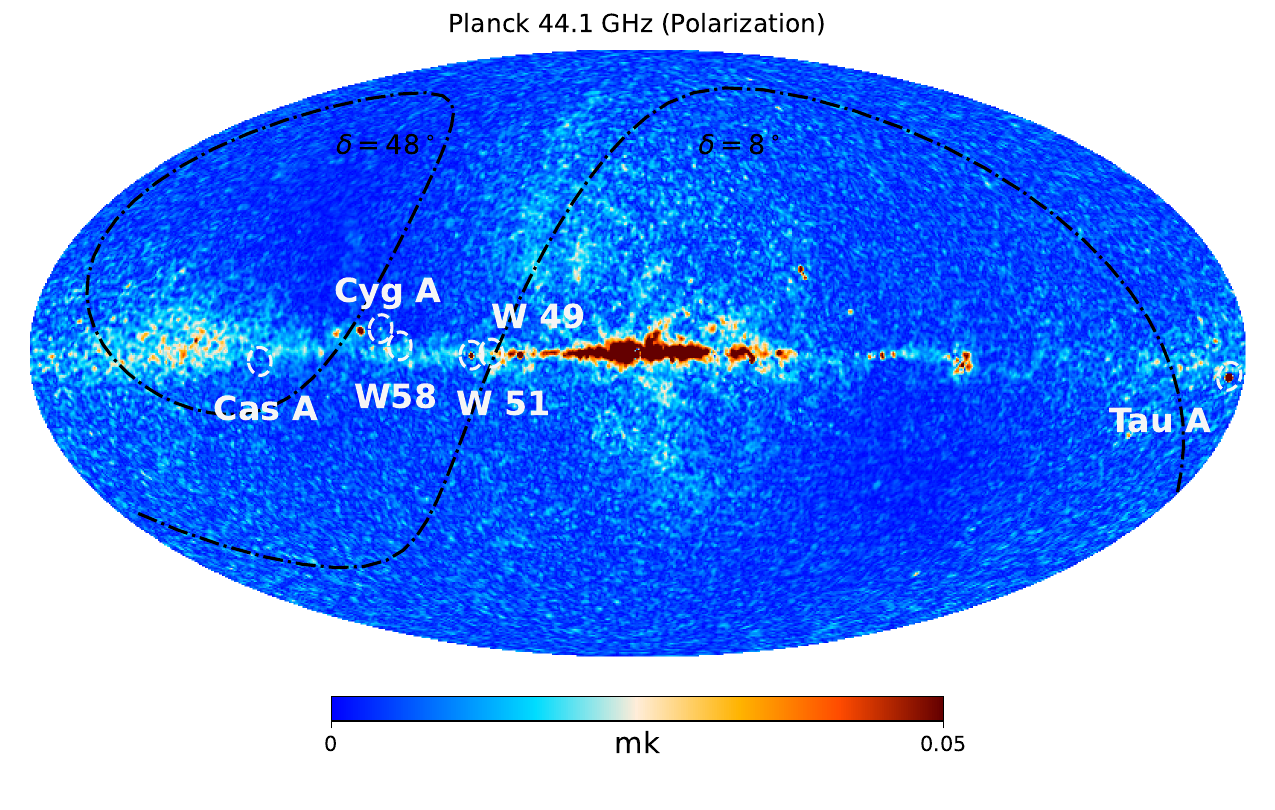}
\caption{Positions on the sky of the main calibrators considered in this work, overplotted on Planck $44.1\,$GHz intensity (left) and polarised intensity (right) maps in Galactic coordinates. Dashed lines delineate the two declination limits of the Strip survey.}
\label{fig:planck_and_cal}
\end{figure}

Given the nearness in frequency of the Planck LFI 44 and HFI 100\,GHz bands, we can use these surveys to identify the brightest sources on the sky lying on the Strip sky footprint (declinations between 8.3$^\circ$ and 48.3$^\circ$). In Table~\ref{tab:sources_pccs2} we have listed the brightest sources in this region of the sky, identified at these two frequencies in the second Planck catalogue of compact sources (PCCS2) \citep{cpp2015-26}. Several of these sources are embedded in the Cygnus X star forming complex, and are either extended or surrounded by strong background emission. Therefore we ignore these sources. W58 is not detected to high significance in the 44 GHz band, and is not too bright at 100 GHz. Therefore, of the sources of this table we will focus on \crab (also known as the ``Crab nebula''), Cyg\,A, W49 and W51. Although, due to its declination ($58^\circ$) it is not located in the Strip footprint, we will also consider Cas A, which is another bright radio source in the northern sky that could eventually be observed using targeted observations in raster-scan mode. In Figure~\ref{fig:planck_and_cal} we show the positions on the sky of these sources, on top of Planck $44.1\,$GHz total intensity and polarised intensity maps. We will also consider planets, in particular Venus and Jupiter, which are the brightest ones in radio. Finally, we will also consider the Sun and the Moon as beam calibrators. Apart from being extended, the Sun is so bright that cannot be observed through the main beam as it could cause damage to the detectors. Yet on-sky calibration of the beam structure away from the main beam (at levels of $\sim -40$\,dB or below) could be achieved with the Sun. The Moon could also be used for this purpose, in this case with the added benefit that thanks to being fainter it could observed directly through the main beam. In the next subsections we present a brief description of the basic properties, and spectral models, of these sources.

\begin{table*}
\caption{Brightest sources in the Second Planck Catalogue of compact sources, at 44 and 100 GHz, and observable within the STRIP-LSPE nominal footprint between declinations 8.3$^\circ$ and 48.3$^\circ$.}
\smallskip
\label{tab:sources_pccs2}
\centering
\begin{tabular}{@{}cccll}
\hline
R.A. (deg) & Dec. (deg) & Flux density (Jy) & Source name & Type \\
\hline
\multicolumn{5}{c}{44 GHz}\\
\hline
 83.63 &    22.03   &  292.7 $\pm$ 0.21	& Tau A  & Filled-centre SNR\\
290.89 &    14.47   &  105.3 $\pm$ 1.1	& W51	 & Giant molecular cloud (W51A,B)\\
&&&& and SNR (W51C)\\
 287.59 &     9.09  &    58.5 $\pm$ 0.8    &       W49	& Giant molecular cloud (W49A) and \\
 &&&&young SNR (W49B)\\
299.87 &    40.74   &	26.7 $\pm$ 0.51	& Cyg A  & Radio galaxy    \\
308.81 &    42.44   &	24.7 $\pm$ 2.9	& Cygnus & Extended\\			  
310.28 &    41.98   &	23.0 $\pm$ 2.8	& Cygnus & Extended\\
306.57 &    40.15   &	20.1 $\pm$ 3.1	& Cygnus & Extended\\
\hline
\multicolumn{5}{c}{100 GHz}\\
\hline
 83.62 &    22.02   &  215.2 $\pm$ 0.11	&  Tau A &\\
290.91 &    14.50   &	82.4 $\pm$ 0.9	&  W51A  &\\
287.57 &     9.11   &	48.4 $\pm$ 0.5	&  W49	 &\\
309.76 &    42.34   &	23.5 $\pm$ 0.4	&  Cygnus&\\
290.60 &    14.11   &	16.3 $\pm$ 0.1	&  W51	 &\\
300.45 &    33.55   &	15.4 $\pm$ 0.1	&  W58   &HII region\\
290.73 &    14.27   &	13.7 $\pm$ 0.1	&   W51B &\\
\hline
\end{tabular}
\end{table*}


\subsection{Tau\,A, \casa and \cyga}\label{subsec:sources1}

Amongst of the best-studied bright calibrators in the northern sky are Tau\,A, \casa and Cyg\,A. In Table~\ref{tab:cal_models} we present the basic properties of these sources. Both \crab and \casa are supernova remnants, while \cyga is an external galaxy that presents two jets extending out to about 3 arcmin. Note that, given their extent on the sky, \crab and \casa are partially resolved by Strip, particularly in W-band (beam FWHM=10$'$). Both \crab and \cyga are known to be polarised.
In particular, \crab is the strongest polarised compact source on the sky in the microwave range, and thence it will be our main polarisation calibration (polarised intensity and polarisation direction). The polarisation fraction of \cyga reaches $\approx 7\%$ at $\approx 10$\,GHz and decreases at higher frequencies as a consequence of the Laing-Garrington effect \cite{laing_1988} (see Figure~39 of \cite{rubino_2023}). \casa is in principle not polarised in the angular scales of the Strip beams. \citep{anderson_1995} detected polarised emission in the SNR ring at 5\,GHz, but with radial direction so then in scales beyond the ring size (5 arcmin) its polarised intensity should average to zero. \citep{weiland_2011} detected in WMAP data unexpected polarisation in \casa at the level of $\approx 0.3\%$, but with an irregular structure that sheds doubts on the reliability of this detection. In fact they mention the possibility of systematic beam effects that could be responsible for this detection and perform some tests to check this hypothesis. We will assume that this source is unpolarised, so then it will be used as one of our zero-polarisation calibrators.

\begin{figure}
\centering
\includegraphics[width=7.5cm]{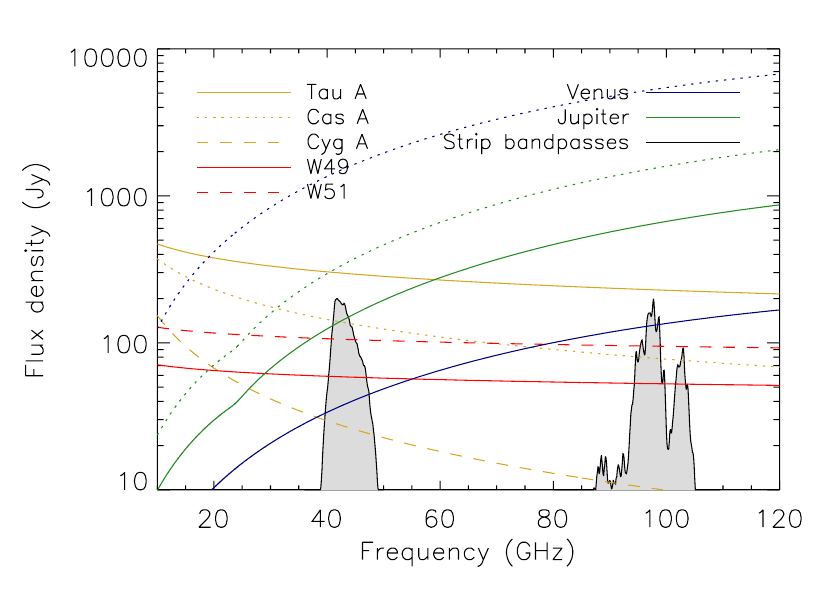}
\includegraphics[width=7.5cm]{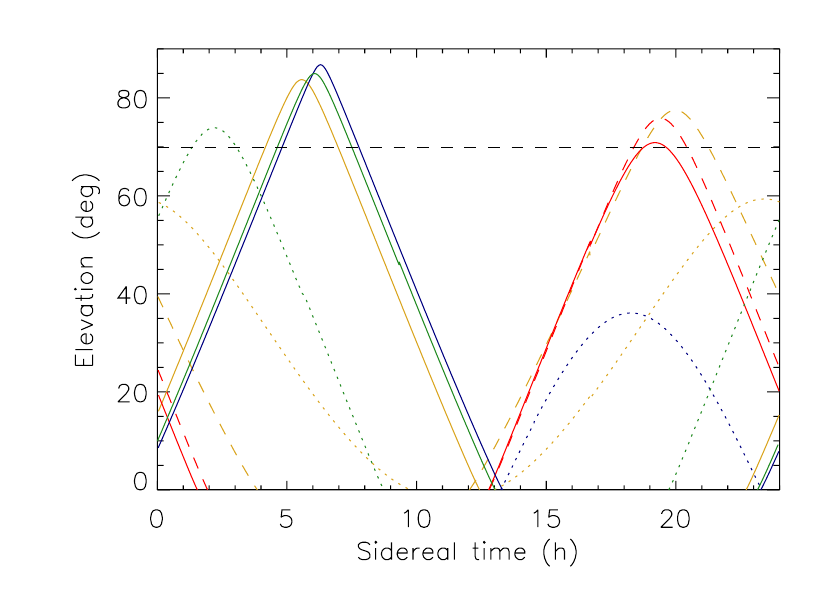}
\caption{Left: Flux-density spectra of bright calibration sources that are observable by LSPE-Strip: Tau\,A, Cas\,A, Cyga\,A, W49 and W51. We also show the minimum (solid lines) and maximum (dotted lines) flux densities of Venus and Jupiter, during the period January 2024 to December 2026. The Strip bandpasses for Q- and W-bands are shown in grey. Right: visibility curves of calibration sources, seen from the Teide Observatory. The colour coding is the same as in the left panel. For Venus and Jupiter we show the two extreme cases when they reach their minimum and maximum declinations during the period  January 2024 to December 2026. The dashed horizontal line marks the elevation of the telescope boresight during observations in nominal mode (70$^\circ$).}
\label{fig:sources_bp}
\end{figure}

In 1977 \citep{baars_1977} presented spectral models for these three sources, including a model for the frequency-dependence of the secular decrease of Cas\,A, which have been considered reference models for calibration ever since. Using WMAP data, which are calibrated off the CMB dipole, \citep{weiland_2011} updated theses models. More recently, we have re-analised all these data, in combination with recent flux densities derived from Planck data, and derived updated and more accurate models that have been used to calibrate the QUIJOTE-MFI experiment \cite{rubino_2023}. These models, which will be presented and discussed in detail in G\'enova-Santos \& Rubi\~no-Mart\'{\i}n et al. (in prep.), are shown in Table~\ref{tab:cal_models}. In the case of \crab and \casa these models are referred to year 2020, and we also show the fading rates. These rates are found to be valid at frequencies above 31.9\,GHz and 30.7\,GHz, respectively for \crab and Cas\,A, while at lower frequencies we found hints for dependency of these values with frequency. These fading rates are similar to those found by \citep{weiland_2011} or \citep{hafez_2008}. The differences between the flux densities predicted at 43 and 95\,GHz by these models and by those of \citep{weiland_2011} is around 1-2\%. This is within the global systematic uncertainty of these models that is $\approx 5\%$. As we will see, with Strip we should be able to determine flux densities of these objects to much better accuracy, and then the final global calibration uncertainty will be determined by this $5\%$ associated with the models. In the left panel of Figure~\ref{fig:sources_bp} we show the spectra of these three sources, while in the right panel we show their visibility curves. It can be seen that the maximum elevation of Cas A is too low as to be observed in the nominal mode. However this source could be observed in raster observations and then we still consider it as a possible calibrator in this study.

In Tau\,A, the weighted average of the polarisation fractions measured over the five frequency bands of WMAP \cite{weiland_2011} is $7.0\%$. We will use this value as a reference, from which we get the polarised flux densities shown in Table~\ref{tab:cal_flux_sen}. \cite{ritacco_2018} fitted the polarised spectrum of \crab using data between 20 and 350\,GHz and found $S_{\rm P, \nu}(\nu)=78.98\times (\nu/{}\rm 1 GHz)^{-0.35}$, referred to date 2018. Assuming the same secular decrease law in polarisation as in intensity, the expected polarisation fractions derived by the ratio of this model over our intensity model for \crab are $7.1\%$ and $6.9\%$ respectively at 43 and 95\,GHz (lower at high frequency due to the fact that the fitted spectral index is slightly steeper in polarisation). These values are consistent with our adopted value. Regarding the integrated \crab polarisation direction, the weighted average over the five values measured across the five WMAP bands \cite{weiland_2011} is $-87.8^\circ$ (in Galactic coordinates), which is exactly the same value obtained by \citep{ritacco_2018} through a weighted average of all available measurements between 20 and 150\,GHz. They assigned an error of $0.3^\circ$ to this measurement, which is the best ever achieved thanks to the combination of all these measurements. Here we will use as reference $-88.0^\circ$. The NIKA2 measurements of \cite{ritacco_2018} revealed significant spatial structure in the polarisation angle direction across the nebula (see also \cite{aumont_2010}). In principle this should be largely averaged out inside the Q-band beam (FWHM=20$'$), so it should be fine to just use the previous average polarisation direction as a reference, which has been derived from integrated Q and U flux densities and therefore corresponds to an average value across the nebula. However it should be studied the impact this may have in the higher angular-resolution W-band data.

\begin{table*}
\caption{Celestial calibration sources. The basic intensity properties of these sources and the reference models (referred for date 2020 for Tau A and Cas A) that will be used to calibrate LSPE-STRIP are shown.}
\label{tab:cal_models}
\smallskip
\smallskip
\centering
\begin{tabular}{@{}cccccc}
\hline
Name &	R.A. (J2000)&	Dec. (J2000)	&Size (arcmin) &	$S_{\nu}^{2020}(\nu)$ (Jy)   & $\frac{dS}{Sdt}$ (\%/year)\\

\hline
{\smallskip}
Tau A	    & 	05 34 31.95 &  +22 00 52.1 & 6	&	$297.3\left(\frac{\nu}{\rm 43 GHz}\right)^{-0.315}$  & $-0.218$\\
\smallskip
Cas A	    &	23 23 28.00 &  +58 49 03.0	& 5    &   $137.5\left(\frac{\nu}{\rm 43 GHz}\right)^{-0.682+0.005\,{\rm log}\left(\frac{\nu}{\rm 43 GHz}\right)}$  & $-0.524$ \\
\smallskip
Cyg A	    &	19 59 28.36 &  +40 44 02.1	& $2.8\times 0.8$ &   $27.1\left(\frac{\nu}{\rm 43 GHz}\right)^{-1.189}$     & --     	   \\
\hline
\end{tabular}
\end{table*}

\subsection{W49 and W51}\label{subsec:sources2}

W49 and W51 are two Galactic regions first discovered in a 22~cm survey by \cite{westerhout_1958}. \cite{tramonte_2023} have recently performed a detailed study of the microwave emission, in intensity and in polarisation, of these two regions using QUIJOTE-MFI information at 10-20\,GHz in combination with other datasets. A detailed discussion of their physical properties is presented in section~2 of that paper and references therein. W49 can be disentangled in two regions: W49A (G43.2+00.0), which is one of the sites with strongest star formation in our Galaxy, and W49B (G43.3-0.2), a young SNR. The emission in Planck maps at 44 and 100\,GHz is well centred in the coordinates of W49A (R.A.$=287.6^\circ$, Dec.$=9.1^\circ$), with no hints of emission on the location of W49B. Then we assume that the emission at the Strip frequencies is entirely dominated by W49A. The angular extent of this source has been measured to be 3.4~arcmin \cite{quireza_2006} and then it can be considered a point source both in Q- and W-bands. It being an HII region it's emission is expected to be dominated by free-free emission. \cite{tramonte_2023} have discussed the possibility of the presence of anomalous microwave emission (AME). However given the angular resolution of that analysis ($1^\circ$) the emission of the source gets diluted and it is difficult to tell whether the AME originates on the source or on the surrounding background. In any case, the level of AME detected in that study is low and does not cause a strong distortion of the free-free spectrum. Then, in our analysis we will take as reference the flux densities reported in the PCCS2 (see Table~\ref{tab:sources_pccs2}) at 44 and 100\,GHz, and scale them to the Strip central frequencies with a free-free spectrum with spectral index $\alpha=-0.13$ to get the reference flux densities shown in Table~\ref{tab:cal_flux_sen} \footnote{Note in any case that at the moment of calibrating the real data the source spectrum should be integrated on each polarimeter bandpass to get a reference amplitude that should be used as a reference to calibrate. For the purpose of this study, which is getting an idea of the level of precision we can get in the calibration, using central frequencies is enough.}. QUIJOTE-MFI, WMAP and Planck maps show no clear indication of polarised emission in this region (see \cite{tramonte_2023}), although the 100\,GHz Stokes Q map shows some emission with a cloverleaf spatial structure with a maximum polarisation fraction of $\sim 1\%$ that could be an indication of intensity to polarisation leakage. Based on these observational evidences, and on the fact that zero net polarisation is expected for free-free emission, we will assume that this source has no polarisation at Strip frequencies, and together with \casa we will use it as a control region for the intensity to polarisation leakage.

W51 can be separated into three different regions. W51A (G49.48-0.33) and W51B (G49.1-0.3) are embedded in a giant molecular cloud, with W51A harbouring the strongest star formation in the region. W51C (G49.2-0.7) has been identified as a SNR. W51A and W51B are clearly resolved out in the Planck 100 GHz map (FWHM$=9.7'$) while in the 44\,GHz map (FWHM$=27.9'$) they are partially blended showing an elongated structure. In fact the PCCS2 lists W51A (which is coined W51 in that catalogue) and W51B as two separate sources  (see Table~\ref{tab:sources_pccs2}). Even though this is a complex structure, with a flux-density spectrum difficult to model, we will study its potential as a calibrator. Similarly to W49, \cite{tramonte_2023} have discussed the possibility of the presence of AME in W51. But in this case AME is constrained to be even lower than in W49, and then the deviation from a pure free-free spectrum is even more marginal. As we did for W49, we extrapolate the PCCS2 44 and 100 GHz values using a free-free spectrum with $\alpha=-0.13$, to define the reference flux densities listed in Table~\ref{tab:cal_flux_sen}. QUIJOTE-MFI, WMAP and Planck polarisation maps show clear polarisation towards the position of the W51C SNR. Actually the intrinsic polarisation of this source seems rather strong although determining its exact polarisation fraction is complicated as the emission in intensity cannot be disentangled from that of W51A and W51B. In any case, its total flux density in polarisation at 40-90\,GHz is $\sim 2-3$~Jy \cite{tramonte_2023}, i.e. much fainter than \crab (just $\approx 15\%$ its polarised flux), in addition to it being difficult to model and therefore we do not consider it as a polarisation calibrator.


\subsection{Planets}\label{subsec:planets}

While the extra-galactic radio sources discussed in the previous sections are typically dominated by synchrotron or free-free emission, with spectral indices in the range $\sim -0.1$ to $-1.0$, planets are dominated by thermal emission whose specific intensity raises in frequency as $\nu^2$. Therefore they become brighter than most extra-galactic sources at around $\gtrsim 100$\,GHz. In addition they are compact (angular sizes below 1$'$), and as they move on the sky they are not affected by a constant background, something that can also be regarded as an asset. For these reasons they have also been traditionally used as calibrators. Planck HFI used Uranus and Neptune as photometric calibrators \cite{cpp2013-8} (Jupiter is too bright, in addition to having strong absorption features that makes its spectrum difficult to model as it is also the case of Saturn, while Mars is variable), and Jupiter and Saturn as calibrators of the beam response \cite{cpp2015-4}. In addition to the lower frequency range of Strip, beam dilution in their wider beams leads to smaller effective antenna temperature. However, as we will see below they provide very strong signal-to-noise, particularly in W-band, so can serve as gain and beam calibrators.

In Table~\ref{tab:planets} we show the main properties of Venus, Mars, Jupiter and Saturn. We do not consider Mercury, Uranus and Neptune as they are way too faint to be detected by Strip. We quote the minimum and maximum Earth-to-planet distance, angular diameter and solid angle subtended by each of these planets for the period between 1 January 2024 and 31 December 2025. To extract reference flux densities, for Venus we use the model that gives brightness temperature as a function of frequency that has been discussed in \cite{rubino_2023} (see Figure 38 of that paper), and which was obtained from a fit of the literature data to a power-law: $T_{\rm B}^{\rm Venus}(\nu)=1047.1\times(\nu/{\rm GHz})^{-0.25}$ K. This model gives Venus brightness temperature after subtraction of the CMB monopole, so then it can be directly compared with the data, which in practice also have the CMB monopole subtracted, as it is totally absorbed by the planet disk. This model gives slightly higher temperatures than physical models in the literature \cite{fahd_1992,bellotti_2015}. However, we prefer to stick to this phenomenological fit as it provides a better description of the observed data. Table~\ref{tab:planets} lists the (CMB monopole subtracted) brightness temperatures derived from this model at 43 and 95\,GHz.

In the case of Jupiter we chose to use as reference the ESA1 model, developed by Rapha\"el Moreno in the context of the calibration of the Herschel mission\footnote{See {\tt https://www.cosmos.esa.int/web/herschel/calibrator-models}}. Figure 38 of \cite{rubino_2023} shows a comparison of this model with that of \cite{karim_2018} and different observed data up to 40\,GHz. Table~\ref{tab:planets} lists Jupiter brightness temperatures predicted by this model at 43 and 95\,GHz after subtraction of the CMB monopole. As commented in \cite{pir52} a 5\% calibration error should be ascribed to this model, and judging from the scatter of the observed data a similar error should be attributed to the Venus model. Recent Jupiter measurements from Planck \cite{pir52,maris21} are fully consistent with the ESA1 model within a 5\%. Given that Venus and Jupiter flux densities may be recovered to better precision using total intensity data of Strip, global calibration errors in this case would be dominated by uncertainties in the calibration models, as it is also the case of the extra-galactic sources discussed in the previous versions. 

\begin{figure}
\centering
\includegraphics[width=15cm]{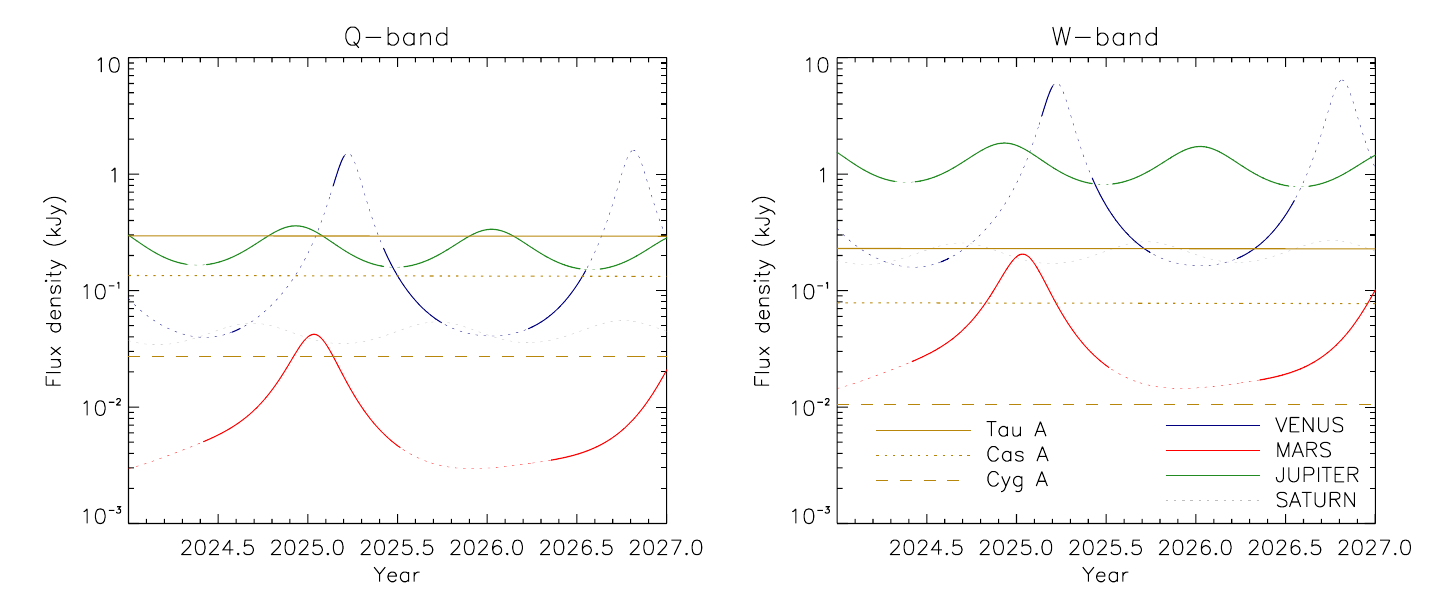}
\caption{Flux densities of planets as a function of time from 1 January 2024 to 1 January 2027 (gray), at 43 (left) and 95\,GHz (right). Solid curves show moments of time when i) the planet has a declination between 8.3$^\circ$ and 48.3$^\circ$ and therefore is observable inside the LSPE-Strip footprint, and ii) is at distance of at least $15^\circ$ from the Sun. For comparison we also show the flux densities of Tau\,A, Cas\,A and Cyg\,A.}
\label{fig:planets}
\end{figure}

We discard Mars as a calibrator because its temperature presents a strong variability that is difficult to model (see Figure~3 of \cite{weiland_2011}), and Saturn due to it being in principle too faint. We nevertheless show also the data for these planets in Table~\ref{tab:planets}. The quoted brightness temperatures are approximated values derived from Table~6 and from Figure~8 of \cite{weiland_2011}, respectively for Mars and Saturn. Based on the planets brightness temperatures at each frequency derived from the models, in Table~\ref{tab:planets} we list the minimum and maximum flux density of each planet in the period 2024-2025\footnote{We have considered that the Strip survey will start on 1 January 2024 and will extend for two years. A delay in the start of this survey will imply slightly different planets brightness and positions on the sky. These differences will be small in any case, and will affect outer planets more. For reference, a delay in one year will change the minimum and maximum flux densities of Jupiter and Saturn shown in Table~\ref{tab:planets} by $\approx 5\%$ and by $\approx 2.5\%$ respectively.}. Figure~\ref{fig:planets} shows the 43\,GHz and 95\,GHz flux densities of the planets as a function of time for this period, in comparison with those of Tau A, \casa and \cyga (constant horizontal line for Cya\,A, while for \crab and \casa we have applied the secular decrease rates shown in Table~\ref{tab:cal_models}). Solid lines show instants of time where i) the declination of the planet is such that it is observable in the nominal mode (elevation $70^\circ$, implying declinations between $8.3^\circ$ and $48.3^\circ$), and ii) the planet is at distance of at least $15^\circ$ from the Sun. Owing to it being an inner planet, in the case of Venus it gets close to the Sun precisely when its flux is maximum. Note also that its declination is outside the Strip range a significant fraction of the time. Due to the nearness of their orbits to that of the Earth, Venus and Mars are the ones exhibiting a stronger variation. Venus distance changes a factor $6.4$, resulting in a variation of its flux by a factor $42$. During its closest approach to Earth it is $5.4$ times brighter than \crab at 43\,GHz and 28 times brighter at 95\,GHz. Jupiter presents a milder variation of its flux, and it is observable most of the time. In Q-band its flux is comparable to that of Tau\,A, while in W-band it reaches a maximum flux that is 8 times that of \crab. Due to it being farther the variability of Saturn's flux is even less pronounced, its flux typically being comparable to the minimum flux of Venus. In any case, Saturn is never observable during this period (it will not reach declinations above $8.3^\circ$ until 2028), and is so faint that it would be hardly detected in a single raster-scan observation. Mars is much fainter, and its peak flux is comparable to the minimum of Venus.

\subsection{The Sun and the Moon}\label{sec:sun_moon}
Thanks to their nearness to the Earth, the Sun and the Moon are the two brightest radio sources on the sky, even if their peak brightnesses are located  in the optical and in the far infrared, respectively. As it was commented above, in the case of Strip the Sun is too intense as to be observed through the main beam. Despite being intense the response of Strip detectors to the emission of the Moon should in principle be inside the linear regime and then it could be used as a gain calibrator. However in this study the Moon and the Sun will be considered as near and far sidelobes calibrators. The effective brightness temperature of the Sun changes in the radio domain, increasing at low frequencies as the emission gradually becomes dominated by the outer regions of the Sun that are hotter. At frequencies above $\sim 100$\,GHz the emission starts to be dominated by the inner photosphere, the same situation as in the optical, and the brightness temperature gradually approaches 5778\,K, which is the thermodynamic temperature of the photosphere. The radio emission of the Moon is dominated by  thermal emission, with a brightness temperature of $\sim 215$\,K, similar to its thermodynamic temperature. The brightness temperature presents a mild decrease with frequency, as the emission becomes gradually dominated by the innermost regions (the emission is typically originated at depth of $\approx 10\lambda$ from the surface). More detailed information about the radio emission of the Sun and the Moon can be found in \cite{hafez_2014} and references and therein, or in any general Radio Astronomy textbook \cite{wilson_2013,condon_2016}. Here we will adopt the brightness temperature values quoted in \cite{hafez_2014}: 214\,K in Q-band and 210\,K in W-band for the Moon, and respectively 7100\,K and 6400\,K for the Sun. It must be noted that these are average values. The brightness of the Sun presents a variation associated with its 11.3-year sunspot cycle. On the other hand the emission from the Moon presents a variation associated with its phase. In the analyses presented in this paper we will work only with average values, although future analyses on real data should take these variations into account (especially that of the Moon in W-band, which is of $\sim 34$\%).

\begin{table*}
\caption{Planets parameters. We list radii and brightness temperatures at 43 and 95~GHz of Venus, Mars, Jupiter and Saturn, as well as minimum and maximum angular diameter, solid angle and flux densities at the two frequencies during 2024 and 2025.}
\smallskip
\label{tab:planets}
\centering
\begin{tabular}{@{}cccccc}
\hline
  && Venus & Mars  & Jupiter & Saturn  \\
  \hline
&Radius (km) & 6052 & 3386 & 69173 & 57316 \\   
&Distance (A.U.) & $0.27~/~1.74$ & $0.64~/~2.42$ & $4.09~/~6.30$ & $8.43~/~10.70$ \\ 
&Angular diameter (arcmin) & 0.16~/~1.02 & $0.06~/~0.24$ &  $0.50~/~0.78$ & $0.25~/~0.31$\\  
&Solid angle  (arcsec$^2$) & $73~/~2940$ & $12~/~166$ & $720~/~1711$ & $171~/~276$ \\   
\hline
\multirow{2}{*}{Q-band} & T$_{\rm B}$ (K) & 408.7 &  190.0 & 157.5 & 150.0 \\
& S$_\nu$ (Jy)  & $39.6~/~1604$ & $3.0~/~42.1$ & $151~/~360$ & $34.3~/~55.2$\\
\hline
\multirow{2}{*}{W-band} & T$_{\rm B}$ (K) & 335.2 &  190.0 &  165.8 & 150.0 \\
& S$_\nu$ (Jy)  & $159~/~6420$ & $14.4~/~205$ & $778~/~1848$ & $167~/~270$\\
\hline
\end{tabular}
\end{table*}

\begin{table*}
\caption{Expected final maps sensitivities per detector for Strip Q- and W-bands, and for i) a 2-year survey with 50\% observing efficiency, ii) one day observation in nominal mode and iii) one raster-scan observation. And expected flux densities of Tau A, Cas A and Cyg A (derived from the models of Table~\ref{tab:cal_models}, and referred to 2024 in the cases of Tau A and Cas A), and of W51 and W49 (obtained from the PCCS2 flux densities at 44 and 100~GHz and extrapolating with a free-free spectrum). In the case of W51, for W-band we show separately the flux densities associated with the W51A and W51B substructures.}
\smallskip
\label{tab:cal_flux_sen}
\centering
\begin{tabular}{@{}rrcc}
\hline
 &&	Q-band (43~GHz) & W-band (95~GHz)\\
\hline
\smallskip
\smallskip
\multirow{3}{*}{Sensitivity per detector (Jy/beam)} & 2-year survey & $0.06$ & $0.26$  \\ 
&  1 day & 1.20 & 4.90  \\
&  raster scan & 0.65 & 2.31 \\

\hline
{\smallskip}
\multirow{3}{*}{Tau A} & $S_{\nu,\rm I}$ (Jy) & $294.8$ & $229.6$ \\ 
&$S_{\nu,\rm P}$ (Jy) & $20.6$ & $16.1$ \\
& $\gamma$ ($^\circ$) & $-88.0$ &$ -88.0$ \\
\hline
{\smallskip}
Cas A & $S_{\nu,\rm I}$ (Jy) & $134.6$ & $78.5$ \\ 
Cyg A & $S_{\nu,\rm I}$ (Jy) & $27.1$ & $10.6$ \\ 
W51 & $S_{\nu,\rm I}$ (Jy) & 58.7 & 52.6 \\
W49 & $S_{\nu,\rm I}$ (Jy) & 105.6 & 83.0$~/~$13.8\\
\hline
\end{tabular}
\end{table*}

\begin{table*}
\caption{Expected sensitivities per detector of LSPE-Strip, in Q-band and W-band, on various calibration parameters, obtained on different point sources and from data in nominal mode observations (at elevation 70$^\circ$). We show sensitivities achievable during a one-day map (in the case of planets we consider their closest approach to Earth), and after 2-years of observations (with 50\% observing efficiency, i.e. 1 year of effective integration time). These values have been derived from simulations using white noise only (wn) and a combination of white and $1/f$ noise (wn+1/f).}
\smallskip
\label{tab:cal_int_pol_sen}
\centering
\begin{tabular}{@{}llccccccccccc}
\hline
&& \multicolumn{5}{c}{One day}&& \multicolumn{5}{c}{Full survey (2 years)}\\
\cline{3-7}\cline{9-13}
Source & Parameter & \multicolumn{2}{c}{Q-band} && \multicolumn{2}{c}{W-band} && \multicolumn{2}{c}{Q-band} && \multicolumn{2}{c}{W-band} \\
\cline{3-4}\cline{6-7}\cline{9-10}\cline{12-13}
            &                   & wn & wn+1/f && wn & wn+1/f && wn & wn+1/f && wn & wn+1/f\\
\hline
\noalign{\smallskip}
Tau A	& 	I/$\sigma_{\rm I}$	   &  68.1  &	12.2   &&  14.4   &   9.0  &&   2348  &    107  &&    393  &   48.0 \\
Tau A	& 	P/$\sigma_{\rm P}$	   &   7.4  &	 6.7   &&   2.2   &   2.1  &&    136  &   53.5  &&   27.2  &   27.9 \\
Tau A        &     $\sigma_{\rm \Pi}$ (\%)         &   0.9  &        1.0    &&     3.1   &   3.4    &&  0.05   &   0.13  &&    0.3  &    0.3  \\
Tau A	&      $\sigma_\gamma$ [deg]       &   3.3  &	 5.4   &&   19.9   &   21.0  &&	 0.3  &    0.7  &&    1.0  &	1.3 \\
W51     & 	I/$\sigma_{\rm I}$	   &  31.0  &	 5.5   &&   6.6   &   3.0  &&    636  &   22.8  &&    105  &   18.3 \\
W49	& 	I/$\sigma_{\rm I}$	   &  26.5  &	 8.0   &&   5.0   &   2.5  &&    323  &    9.0  &&   68.9  &   14.2 \\
W49	& 	$\sigma_{\rm \Pi}$ (\%)	   &   4.6  &	 2.9   &&  18.7   &  17.7  &&	 0.2  &    1.0  &&    1.3  &	1.1 \\
Cyg A	& 	I/$\sigma_{\rm I}$	   &  10.5  &	 1.7   &&   1.0   &   0.4  &&    287  &    9.5  &&   18.3  &	2.3 \\
\noalign{\smallskip}
\hline
\noalign{\smallskip}
Venus   &       I/$\sigma_{\rm I}$ 	   &   221  &	60.5   &&   458   &   209  &&   2926  &    266  &&   4230  &    734 \\
Venus	&       $\sigma_{\rm \Pi}$ (\%)	   &   0.4  &	 0.3   &&   0.1   &   0.2  &&	0.03  &   0.04  &&   0.01  &   0.02 \\
Jupiter &       I/$\sigma_{\rm I}$	   &  56.0  &	15.3   &&   149   &  68.1  &&    912  &   65.5  &&   2044  &    241 \\
Jupiter	&       $\sigma_{\rm \Pi}$ (\%)	   &   1.5  &	 1.4   &&   0.4   &   0.5  &&	 0.1  &    0.1  &&   0.03  &   0.05 \\
\noalign{\smallskip}
\hline
\end{tabular}
\end{table*}

\begin{table*}
\caption{Same as Table~\ref{tab:cal_int_pol_sen} but for a raster-mode observation, providing deeper integration time on the position of different sources.}
\smallskip
\label{tab:cal_int_pol_sen_1dr}
\centering
\begin{tabular}{@{}llccccc}
\hline
&& \multicolumn{5}{c}{One day (raster mode)}\\
\cline{3-7}
Source & Parameter & \multicolumn{2}{c}{Q-band} && \multicolumn{2}{c}{W-band}  \\
\cline{3-4}\cline{6-7}sss
            &                   & wn & wn+1/f && wn & wn+1/f\\
\hline
\noalign{\smallskip}
Tau A	& 	I/$\sigma_{\rm I}$	   &   173.0	& 13.8  &&   34.0  &   14.7   \\
Tau A	& 	P/$\sigma_{\rm P}$	   &	12.6	&  7.6  &&    7.7  &	4.9   \\
Tau A        &     $\sigma_{\rm \Pi}$ (\%) &     0.6  &  0.9 &&    0.9  &    1.4  \\
Tau A	&      $\sigma_\gamma$ [deg]       &	 2.8	&  5.1  &&    10.6  &	11.1   \\
W51     & 	I/$\sigma_{\rm I}$	   &	64.9	&  7.7  &&   15.5  &	3.9   \\
W49	& 	I/$\sigma_{\rm I}$	   &	27.2	&  4.9  &&    6.6  &	1.9   \\
W49	& 	$\sigma_{\rm \Pi}$ (\%)	   &	 1.5	&  4.7  &&    9.7  &   11.1   \\
Cyg A	& 	I/$\sigma_{\rm I}$	   &	21.6	&  3.4  &&    2.4  &	0.7   \\
Cas A	& 	I/$\sigma_{\rm I}$	   &	67.3	&  6.3  &&   16.6  &	2.6   \\
Cas A	& 	$\sigma_{\rm \Pi}$ (\%)    &	 1.1	&  1.7  &&    3.6  &	4.5   \\
\noalign{\smallskip}
\hline
\noalign{\smallskip}
Venus   &       I/$\sigma_{\rm I}$ 	   &    667    & 71.0  &&   1076  &    274   \\
Venus	&       $\sigma_{\rm \Pi}$ (\%)	   &    0.2    &  0.2  &&    0.1  &    0.1   \\
Jupiter &       I/$\sigma_{\rm I}$	   &    169    & 18.0  &&    350  &   89.4   \\
Jupiter	&       $\sigma_{\rm \Pi}$ (\%)	   &    0.6    &  0.9  &&    0.3  &    0.3   \\
\noalign{\smallskip}
\hline
\end{tabular}
\end{table*}

\section{Simulations}\label{sec:simulations}

The assessment of the precision achievable on the determination of the main instrument parameters using sky observations of bright astronomical sources requires the production of realistic simulations of sky signal and noise. The sky signal produced by specific sources can be simulated using their expected flux densities that are listed in Table~\ref{tab:cal_flux_sen}, derived from the models that were discussed in the previous section. The source amplitude should then be convolved with a given beam model. Producing noise simulations requires knowledge of the expected integration time per solid angle unit on the position of each source. This is achieved through simulations of the scanning strategy of the telescope. In combination with the NET values listed in Table~\ref{tab:instrument_parameters} this allows to estimate the variance of the thermal noise level in each position, from which a realisation of this noise can be generated. Most of the calibration steps presented in this paper are based on intensity data, where the $1/f$ noise component is important. Simulating $1/f$ noise requires some assumptions about its spectrum, which is usually parametrized by its knee frequency. Finally, the study of the ability to characterise the instrument beams requires some realistic beam simulations, which later could be tested through comparison with observations. The methods that we have followed to produce all these ingredients that are needed for our study are discussed in the next subsections.

\subsection{Sky simulations}

\subsubsection{Scanning strategy simulations}\label{sec:scan_strategy_sims}

\begin{figure}
\centering
\includegraphics[trim= 0mm 25mm 0mm 20mm, width=7.5cm]{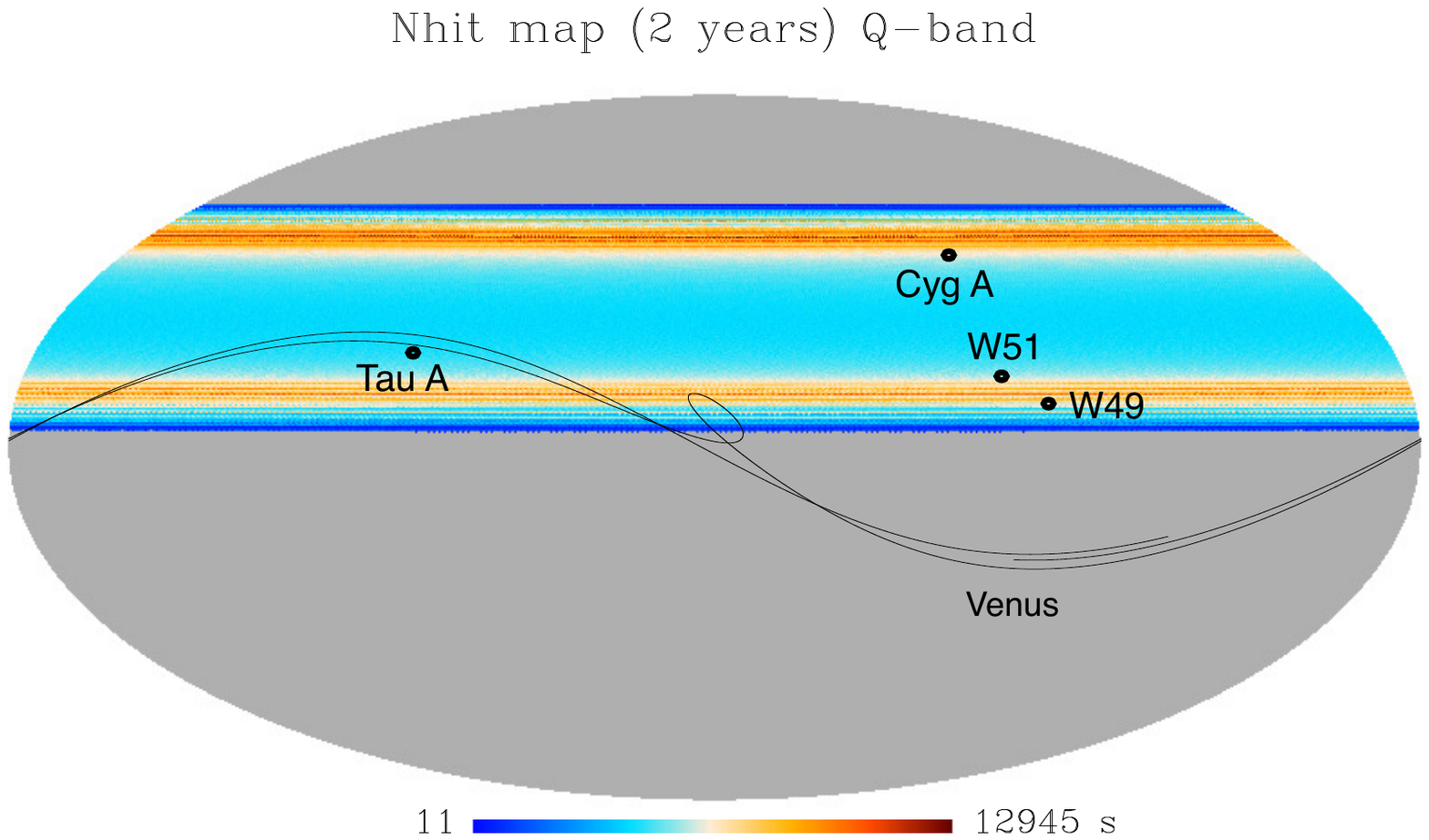}
\includegraphics[trim= 0mm 25mm 0mm 20mm, width=7.5cm]{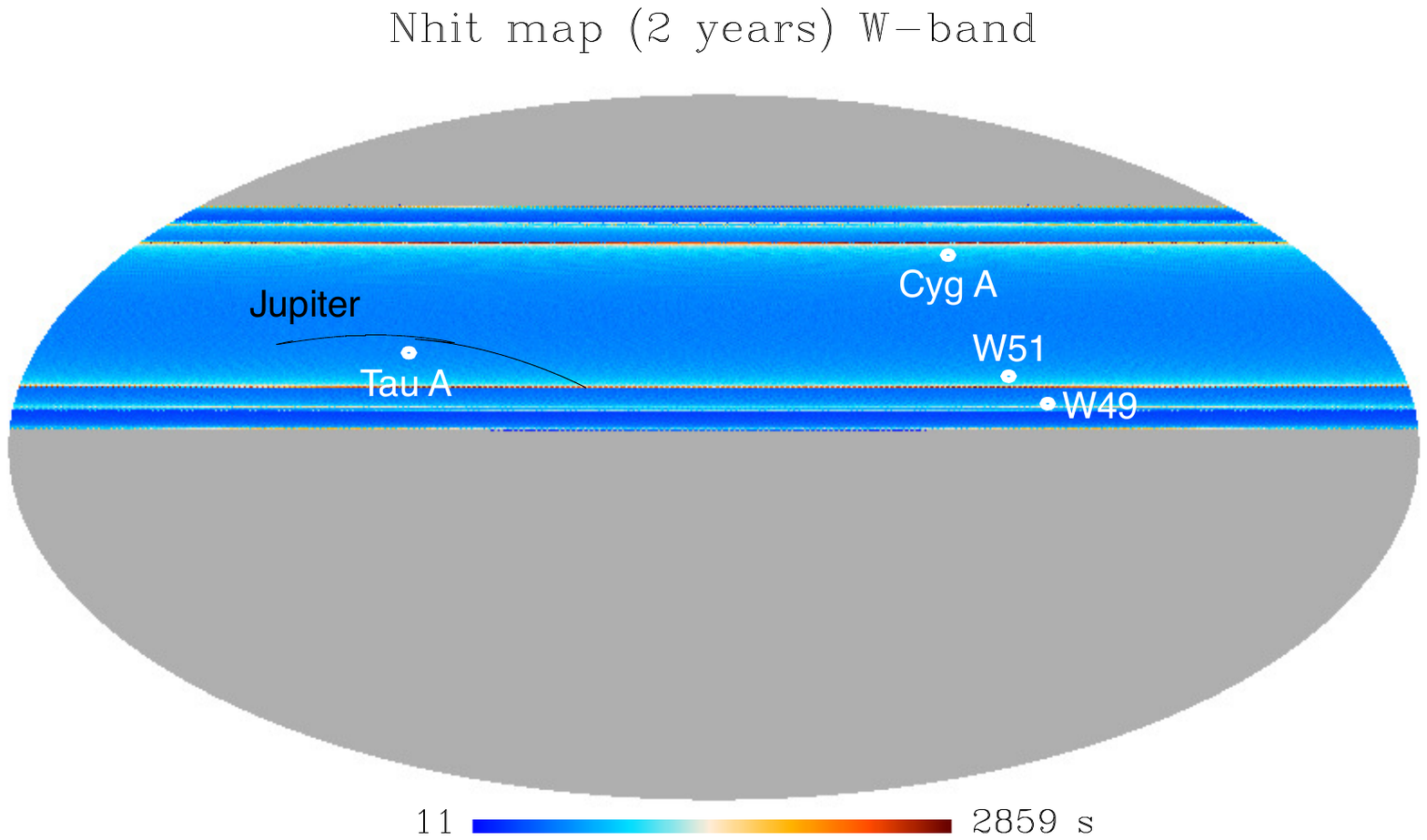}
\caption{Simulated hit maps, in seconds and projected in equatorial coordinates, for 2-year observations with 50\% observing efficiency (1 year effective integration time), corresponding to the full-array (all polarimeters) of Q-band (left) and W-band (right). We indicate the position on the sky of bright calibration sources, as well as the trajectories of Venus (left) and Jupiter (right).}
\label{fig:nhit_sky}
\end{figure}

As it was already discussed in section~\ref{sec:strip}, the default observing mode of Strip is the ``nominal mode'', in which the telescope spins at a fixed elevation with a constant velocity in azimuth. The azimuth of the telescope boresight at a given time $t$ is then given by AZ$=\omega\times t$, while EL$=70^\circ$. We then register the pointing of the telescope every one second, and using the geographical coordinates of the Teide Observatory we apply the standard coordinate transformations to derive equatorial coordinates (R.A. and Dec.) of the telescope. We project the hits using a full-sky Healpix pixelisation \cite{healpix} with $N{\rm side}=256$ (pixel size $13.7'$). We consider 2 years of observing time with a 50\% observing efficiency (1 year effective - see Table~\ref{tab:instrument_parameters}). We then ran a 2-year simulation, and multiplied the resulting hit map by 1/2 to account for the observing efficiency (which assumes that the time loss is constant with time). This would give the hit map for the pixel located on the centre of the focal plane (pixel I0 of Q-band). We repeated the same process for each pixel of the focal plane. To this aim, we calculate the pointing offsets for each pixel with respect to that on the centre of the focal plane by applying the necessary rotations using the azimuthal and polar angles listed in Table~3.4 of \cite{realini_2020}. We add the hit maps of all Q-band polarimeters, on the one hand, and of all W-band polarimeters on the other hand. The final hit maps are depicted in Figure~\ref{fig:nhit_sky}, where we can see the horizontal declination stripe that is covered by the observations. The fraction of the sky covered is 37.2\% for Q-band and 36.7\% for W-band. This is larger than the 30\% (see Table~\ref{tab:instrument_parameters}) covered by the radiometer on the centre of the focal plane, due to the different pointing directions of different radiometers on the focal plane. Considering the solid angle of one map pixel, one-year integration time, and that there are 49 radiometers in Q-band and 6 radiometers in W-band, the average integration time per pixel is 5278\,s in Q-band and 655\,s in W-band. These are more ore less the values of the pixels on the blueish regions around the centre of the filled strip, while as expected pixels towards the borders have larger integration times. By dividing these maps by 365 we can obtain hit maps corresponding to one day, which we will also use to perform a noise realisation corresponding to an observation of one day in nominal mode.

\begin{figure}
\centering
\includegraphics[trim= 0mm 25mm 0mm 20mm, width=7.5cm]{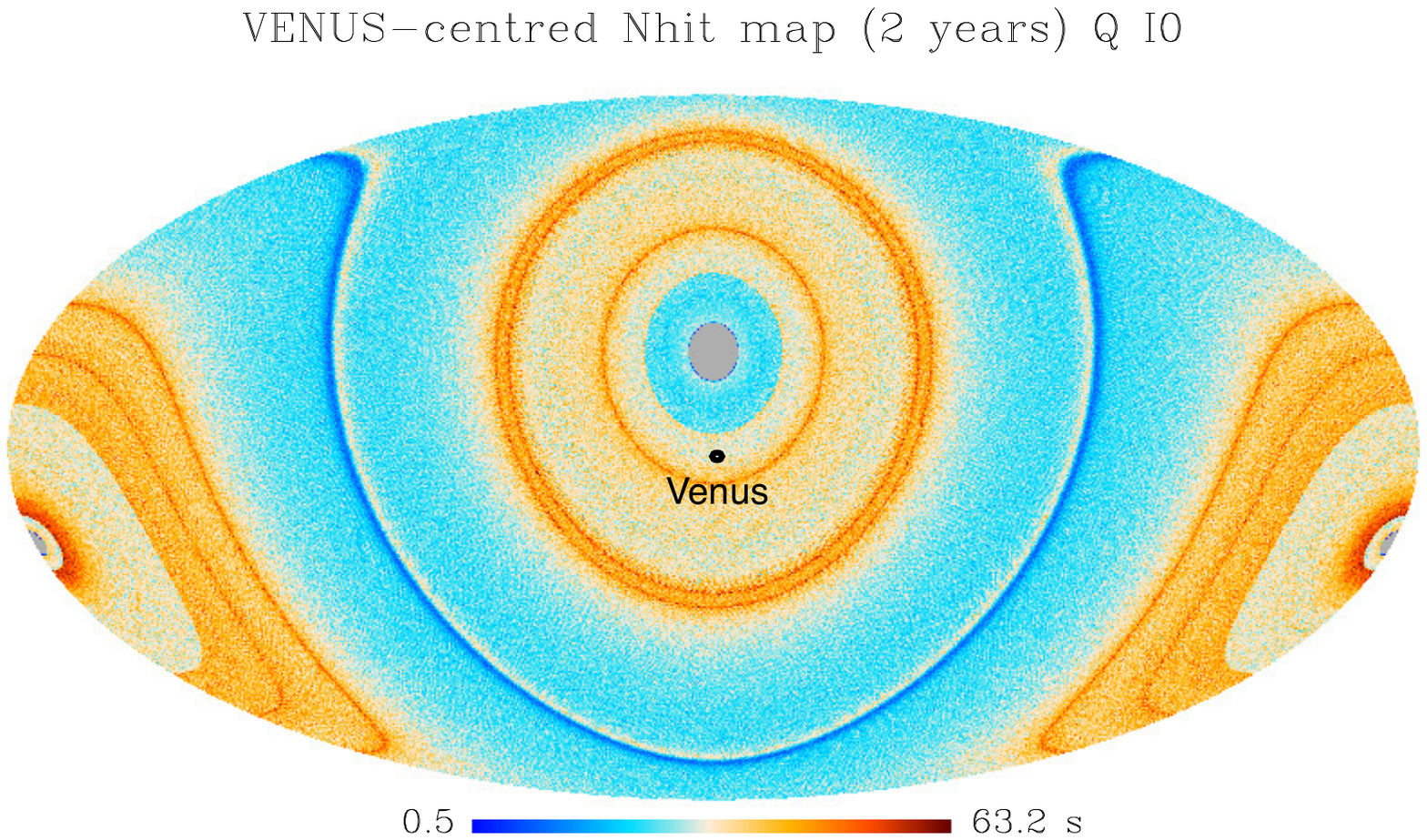}
\includegraphics[trim= 0mm 25mm 0mm 20mm, width=7.5cm]{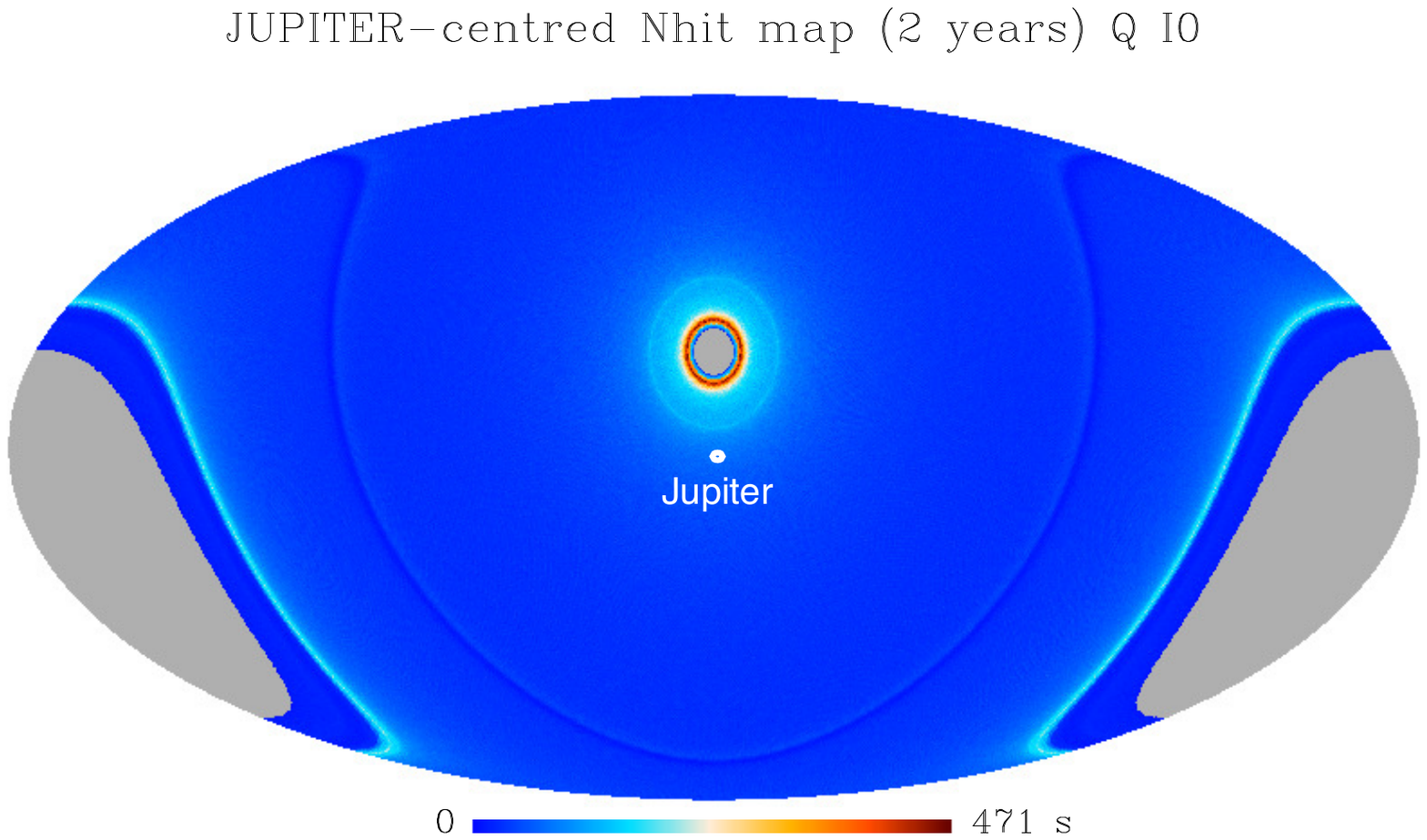}
\caption{Simulated hit maps centred on the positions of Venus (left) and Jupiter (right) for 2-year observations with 50\% observing efficiency (1 year effective integration time), corresponding to one single polarimeter located on the center of the focal plane (Q-band polarimeter I0). In both cases the position of the planet is (0,0) in this mollweide projection.}
\label{fig:nhit_planets}
\end{figure}

These maps can be used to estimate the integration time on the position of the different Galactic and extra-galactic sources that are indicated in Figure~\ref{fig:nhit_sky}. However, planets coordinates obviously change over the course of the observations. In fact, in that figure we have overplotted the trajectories of Venus and Jupiter during this period (2024-2025). To know the integration time around each planet we have to implement a different methodology in which we simulate hit maps not on sky coordinates but on planet-centred coordinates. If (AZ$_i^{\rm p}$,EL$_i^{\rm p}$) are the horizontal coordinates of the planet at a given time $t_i$, to which we associate a vector $\vec{v}_i$, and (AZ$_i$,EL$_i$) are the horizontal coordinates of the telescope at the same time $t_i$, we can find the position vector of the planet $\vec{v}_i^{\rm p}$ projected on the reference frame of the telescope through the following operation that involves the application of two rotation matrices:
\begin{equation}
\vec{v}_i^{\rm p}=[R]\vec{v_i}=
\left[\begin{matrix} {\rm cos(EL}_i)& 0 & {\rm sin(EL}_i) \\ 0 & 1 & 0 \\ -{\rm sin(EL}_i)& 0 & {\rm cos(EL}_i) \\   \end{matrix}\right]
\left[\begin{matrix} {\rm cos(AZ}_i)& {\rm sin(AZ}_i) & 0 \\ -{\rm sin(AZ}_i) & {\rm cos(AZ}_i) & 0 \\ 0 & 0 &1 \\   \end{matrix}\right] 
\left[\begin{matrix} {\rm sin(90^\circ-EL^p}_i){\rm cos(AZ^p}_i) \\ {\rm sin(90^\circ-EL^p}_i){\rm sin(AZ^p}_i)  \\ {\rm cos(90^\circ-EL^p}_i) \\   \end{matrix}\right] ~~,
\label{eq:rotation_planet}
\end{equation}
where the telescope boresight is along the $x$ axis, i.e. if the telescope is pointing to the planet we have $\vec{v}_i^{\rm p}=(1,0,0)$. We then calculate AZ$_i^{\rm p}$,EL$_i^{\rm p}$ of the planet at each time\footnote{Using the IDL routine {\sc planet\_coords}.}, apply the previous rotation and project the $\vec{v}_i^{\rm p}$ vector on a Healpix map with $N_{\rm nside}=256$. The final maps for Venus and Jupiter, corresponding to two years integration starting on 1 January 2024 and 50\% observing efficiency (we calculate the maps for 2 years and multiply them by $0.5$), and for Q-band pixel I0, are shown in Figure~\ref{fig:nhit_planets}.

The antenna temperature produced by the planet in pixel $j$ is calculated as
\begin{equation}
T_{{\rm A},j} = T_{\rm B}\frac{\Omega_{{\rm p},j}}{\Omega_{\rm A}}~~,
\label{eq:ta_planets1}
\end{equation}
where $T_{\rm B}$ is its brightness temperature (quoted in Table~\ref{tab:planets} for Q- and W-band), $\Omega_{{\rm p},j}$ is the solid angle subtended by the planet at time $t_i$ when pixel $j$ is observed, and $\Omega_{\rm A}$ is the solid angle of the antenna. As the distance to the planet, and its solid angle, do change over the course of the 2-year Strip observations, in addition to maps of the number of hits, maps of average distances are also needed. In each pixel $j$ the average ``effective distance'' to the planet is calculated as
\begin{equation}
d_{{\rm eff},j}=\left(\frac{\frac{1}{N}{\sum_i{\frac{1}{d_{\rm{p},i}^2}}}}{\frac{1}{N}{\sum_i{\frac{1}{d_{{\rm p},i}^4}}}}  \right)^{1/2}~~,
\label{eq:deff_opt}
\end{equation}
where the sums run over all $N$ samples lying in that pixel, and $d_{\rm{p},i}$ represents the distance to the planet for sample $i$ corresponding to time $t_i$.  This formula, whose derivation is explained in Appendix~\ref{ap:weighting}, results when the data are weighted using the signal-to-noise information, $w_i\propto 1/(\sigma_i\times d_{{\rm p},i})^2$, with the goal to improve the signal-to-noise of the final weighted map. In the case of Jupiter the variation of its distance is small (see Table~\ref{tab:planets} and Figure~\ref{fig:planets}) and then the improvement of the signal-to-noise is modest and barely noticeable. However, in the case of Venus, whose distance changes a factor $6.4$ (factor $42$ in flux density), thanks to using this optimal weighting, the signal-to-noise improves up to a factor $2.2$ (see Figure~\ref{fig:snr_planets} and related discussion in Appendix~\ref{ap:weighting}) with respect to the case of standard weighting ($w_i\propto 1/\sigma_i^2$).

\begin{figure}
\centering
\includegraphics[trim= 0mm 25mm 0mm 5mm, width=7.5cm]{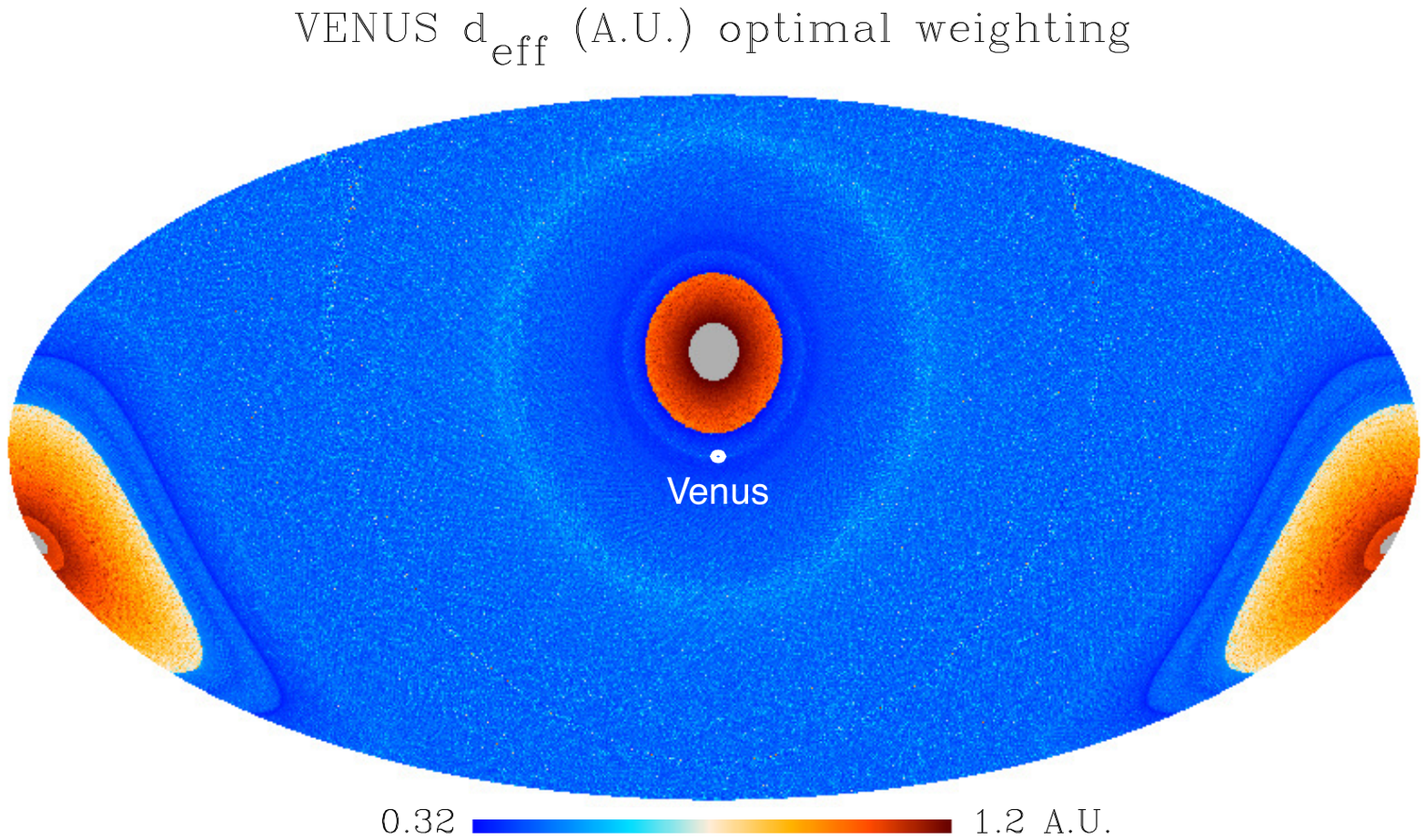}
\includegraphics[trim= 0mm 25mm 0mm 5mm, width=7.5cm]{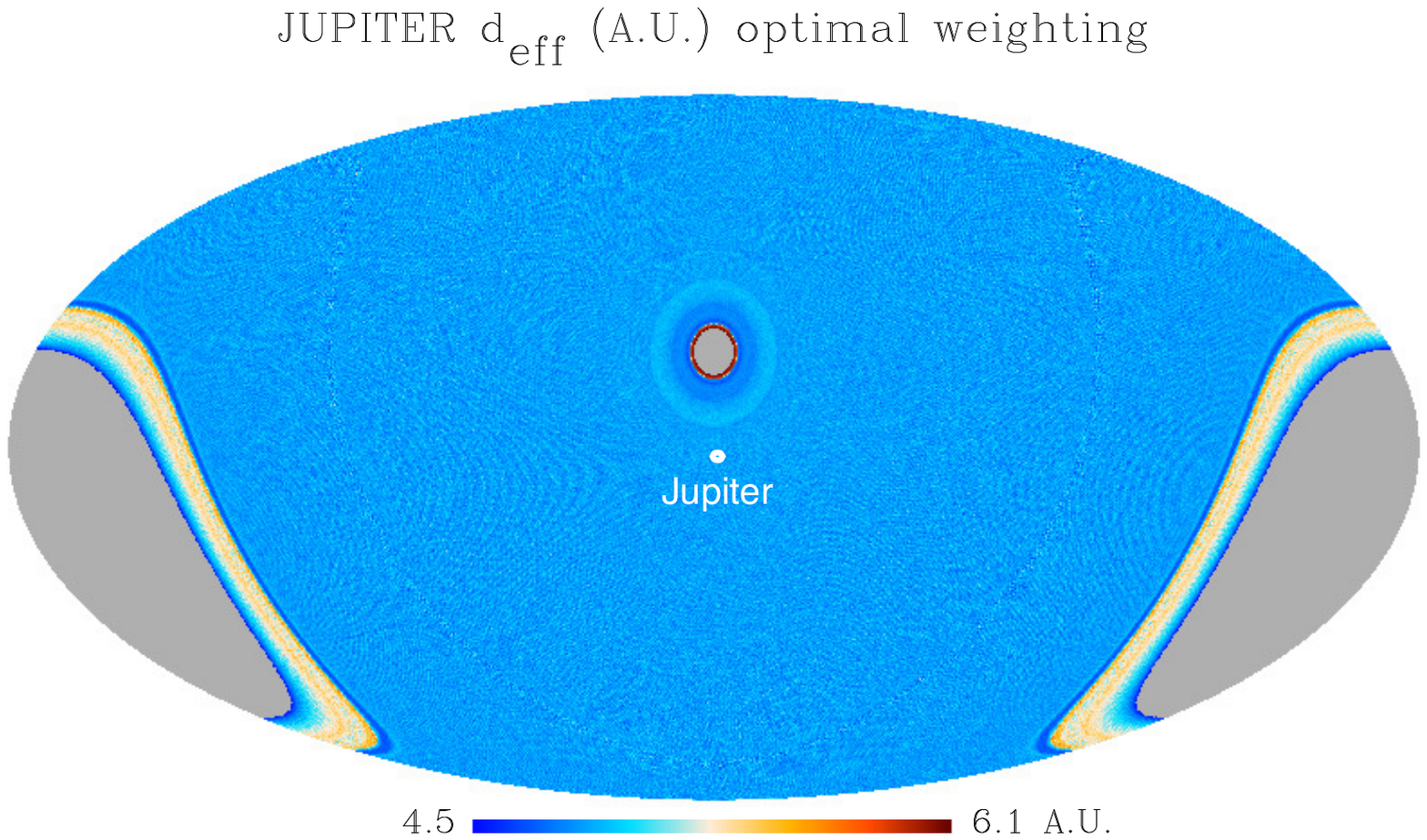}
\caption{Average effective distances (A.U.) for Venus (left) and Jupiter (right) for 2-year observations, starting on 1 January 2024. Maps correspond to one single polarimeter (Q-band polarimeter I0). In both cases the position of the planet is (0,0) in this mollweide projection.}
\label{fig:dist_opt_planets}
\end{figure}

The effective distance maps for Venus and Jupiter, calculated using equation~\ref{eq:deff_opt}, are shown in Figure~\ref{fig:dist_opt_planets}. The solid angle of the planet for each pixel $j$ is then calculated as $\Omega_{{\rm p},j}=\pi(r_{\rm p}/d_{{\rm eff},j})^2$, where $r_p$ is the radius of the planet (see Table~\ref{tab:planets}). These maps have the same projection as the hit maps shown in Figure~\ref{fig:nhit_planets}, i.e. centre of the map (coordinates (0,0)) show the average distance to the planet when the telescope boresight is pointing to it, whereas any other pixel of the map represents the average distance to the planet when it has coordinates ($\theta.\phi$) with respect to the telescope boresight. Of course for what concerns the derivation of global gain calibration factors having these maps in a close environment around the telescope boresight enclosing the main beam is enough. However having full-sky maps like these are useful to study how planets will allow to characterise the beam structure outside the main beam (see discussion in Section~\ref{sec:beams}).

\subsubsection{Maps simulations}\label{sec:maps_sims}

The results of the simulations of the scanning strategy explained in the previous section are maps containing integration time per pixel. The standard deviation of the noise in each pixel $j$ is then obtained as $\sigma_j=\sigma_0/\sqrt{t_j}$, where $\sigma_0$ is the standard deviation of the noise in one second (known as ``Noise Equivalent Temperature'', see values for Q- and W-bands in Table~\ref{tab:instrument_parameters}) and $t_j$ is the value of the hit map in pixel $j$, representing the total integration time in that pixel. In this study we consider three different hit maps, corresponding to three different types of observations:
\begin{itemize}
\item[1)] One-day observation in nominal mode. The hit ($t_j$) maps in this case represent integration time per polarimeter in one day. They are obtained by dividing the maps shown in Figure~\ref{fig:nhit_sky} by $(365\times N_{\rm pol})$, where $N_{\rm pol}$ is the number of polarimeters in Q- and W-band (see Table~\ref{tab:instrument_parameters}). Note that in this case we consider 100\% observing efficiency in that particular day. Note also that instead of scaling the full-array map, we should use the per-polarimeter hit maps, which were also generated. However, rather than studying properties of individual polarimeters we are interested in global averaged properties, so then we proceed in this way. We foresee small differences between polarimeters in any case.
\item[2)] Full survey in nominal mode. The hit maps in this case represent integration time per polarimeter in a 2-year survey with 50\% observing efficiency, so they are the maps shown in Figure~\ref{fig:nhit_sky} divided by $N_{\rm pol}$.
\item[3)] One-day in raster scan mode. Although the telescope will be observing most of the time in nominal mode, it will have the capability to perform raster-scan observations on specific regions (see section~\ref{sec:scanning_strategy}). This mode would also allow to visit sources like \casa that lie outside the nominal-mode sky footprint. The exact hit map in this mode depends strongly on the local coordinates of the observation.
For simplicity we consider reference cases of raster scans with lengths in azimuth of $8^\circ$ in Q-band and $6.5^\circ$ in W-band, with constant elevation, and duration of $25\,$min and $22\,$min respectively, which are approximately the times in which the sky moves in vertical direction a distance equal to the azimuth scan length. The result would then be a map of $8^\circ\times 8^\circ$ in Q-band and $6.5^\circ\times 6.5^\circ$. We chose these sizes as they are similar to the full focal-plane field of view at an elevation $\sim 60^\circ$ (see details of the focal plane configuration in \cite{realini_2020}). Assuming uniform sky coverage over the observed region, the integration time per pixel of $N_{\rm side}=256$ is $t_j=1.23\,$s and $t_j=1.64\,$s respectively in Q- and W-band. We use these values to generate white-noise realisations in this case. Note that, even if we have considered observations optimised for the two bands, in practice we observe in Q- and W-bands simultaneously so if we want all polarimeters to be covered we are limited by the observing parameters of Q-band.
\end{itemize}

The other important source of noise is the correlated $1/f$ noise, which has frequency spectrum 
\begin{equation}
P(f) = \sigma_0^2 \left[1+\left(\frac{f_{\rm k}}{f}\right)^\alpha \right]~~
\label{eq:one_over_f}
\end{equation}
where $f_{\rm k}$ is the knee frequency and $\alpha\sim 1$. This noise is due to fluctuations of the receiver gain, and also to time and spatial atmospheric fluctuations that cause the PWV to change spatially and in turn the integrated atmospheric load seen by the telescope as it scans the sky. As Strip has a differential receiver, its $1/f$ noise is largely suppressed in Stokes Q and U measurements. In addition atmospheric emission is known to be unpolarised, so the only $1/f$ component from the atmosphere may arise from intensity-to-polarisation leakage. Therefore, knee frequencies are expected to be small (less than few tens of mHz) in polarisation, but rather high in intensity (few Hz) \cite{lspe_2021}. This will have an important impact on the definition of the calibration strategy, as it will be discussed in section~\ref{sec:strip_calibration}.

The best procedure possible to assess the impact of $1/f$ noise at the map level involves end-to-end simulations starting with a simulation of sky signal plus the white and $1/f$ components at the time-ordered data level, using information of the instrument and atmospheric knee frequencies, and projection onto the sky using the same map-making algorithm that will be applied to the real data. Apart from being computationally demanding, realising the full potential of this process requires knowledge of the instrument and atmospheric noise properties to a precision that will not be achieved until real on-sky data will become available. For this reason, we chose to follow a more simple approach in which we simulate $1/f$ directly in harmonic space and then transform to the map space. To do this, we choose knee multipoles in harmonic space $\ell_{\rm k}=500$ in intensity and $\ell_{\rm k}=20$ in polarisation, and $\alpha=1.3$. These values are of the same order to what is found in real QUIJOTE-MFI data at 10-20\,GHz \cite{guidi_2021}. Introducing these parameters in equation~\ref{eq:one_over_f}, we compute the $C_\ell$ values that are used to simulate the $a_{\ell,m}$ from which we synthesise a real map\footnote{Using the Healpix \sc{idl} routine \sc{synfast}.}. Most of the calibration procedures discussed in this article rely on total intensity data, and thence a more realistic estimate of the instrumental $1/f$ is key to reassess the sensitivity estimates and conclusions that will be derived henceforth.

Apart from white and 1/f noise, in the case of planets it must be considered that, as the Earth rotates around the Sun, they change their position relative to the background emission. In principle this leads to a background confusion noise that in principle should be added on top of the previous two noise components. This effect will be more severe in Q-band, where Galactic background emission is stronger and planet emission is fainter than in W-band. In this band we have estimated that the background fluctuations will contribute to an RMS noise component that is $\sim 40\%$ of the combined white and 1/f noise components for a 2-year survey in nominal mode. However this contribution could be significantly alleviated by subtraction of a sky model (the Planck LFI 44\,GHz map could be used as a reference) before producing the planet-centred maps. Therefore, in our analysis we will neglect this noise contribution.

Finally, the signal of the source is simulated assuming Gaussian beams with widths $\theta_{\rm fwhm}$ given in Table~\ref{tab:instrument_parameters}:
\begin{equation}
m_j^{\rm s} = T_{\rm A, s} \,{\rm exp}\left(-\frac{8{\rm \ln}2}{2}\frac{\theta_j^2}{\theta_{\rm fwhm}^2}\right)~~,
\end{equation}
where $\theta_j$ represents the distance of pixel $j$ from the source position, and $T_{\rm A, s}$ is the antenna temperature of the source, in K$_{\rm CMB}$ units. For Tau\,A, Cas\,A, Cyg\,A, W49 and W51 this antenna temperature is derived from the flux-density models described in sections~\ref{subsec:sources1} and \ref{subsec:sources2}, evaluated respectively at 43\,GHz and 95\,GHz, as
\begin{equation}
T_{\rm A,s}=S_{\nu,s}(\nu)\frac{c^2}{2 k_{\rm B}\nu^2\,\eta_{\Delta T}(\nu)}\frac{1}{\Omega_{ \rm A}}~~,
\end{equation}
where $S_{\nu,s}(\nu)$ is the flux density predicted by the model at the reference frequency, $k_{\rm b}$ is the Boltzmann constant and $\eta_{\Delta T}(\nu)=x^2e^x/(e^x-1)^2$ represents the conversion factor between thermodynamic differential temperature (units K$_{\rm CMB}$) and Rayleigh-Jeans brightness temperature (units $K_{\rm RJ}$) with $x=h\nu/(k_{\rm b}T_{\rm cmb})$ the dimensionless frequency. In the case of planets, the antenna temperature is calculated as 
\begin{equation}
T_{{\rm A},j} = \frac{T_{\rm B}}{\eta_{\Delta T}(\nu)}\frac{\Omega_{{\rm p},j}}{\Omega_{\rm A}}~~,
\end{equation}
which is the same equation as equation~\ref{eq:ta_planets1} but now including the $\eta_{\Delta T}(\nu)$ factor to have the antenna temperature\footnote{Note that, while in the CMB jargon ``antenna temperature'' usually refers to temperature in RJ units (brightness temperature, $K_{\rm RJ}$), as opposed to thermodynamic temperature (K$_{\rm CMB}$), here we use this term to refer to the fact that it is temperature measured by the antenna (convolution between the intrinsic brightness temperature of the source and the beam pattern), regardless it is expressed in K$_{\rm CMB}$ as in this case or in $K_{\rm RJ}$. This is usually the definition of ``antenna temperature'' adopted in general Radio Astronomy textbooks.}  in units of K$_{\rm CMB}$. Note also that in this case we have included explicitly the dependence with the position on the sky, or with pixel $j$, owing to the variation of the effective solid angle of the planet in each pixel of the sky.

\begin{figure}
\centering
\includegraphics[width=7.5cm]{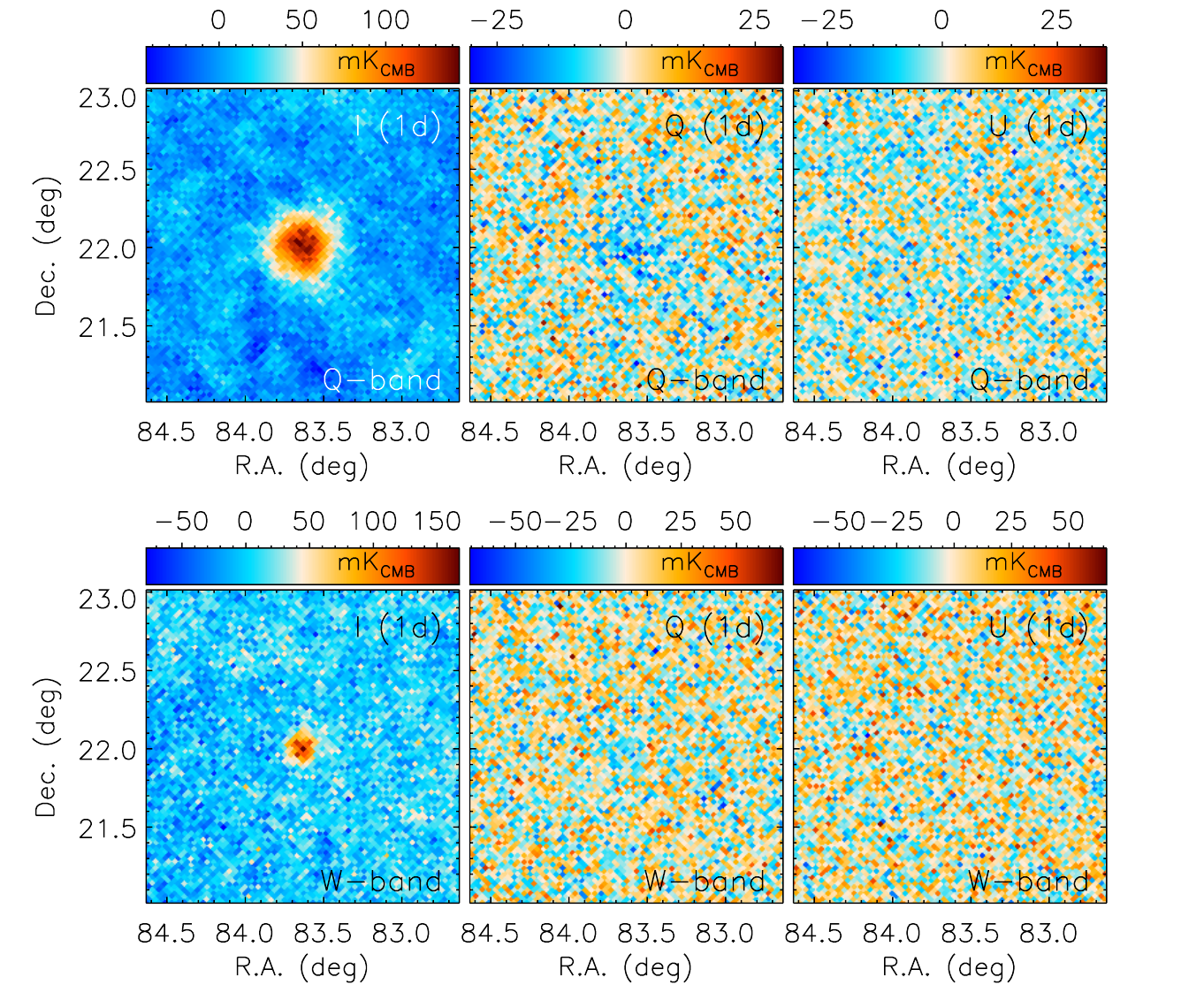}
\includegraphics[width=7.5cm]{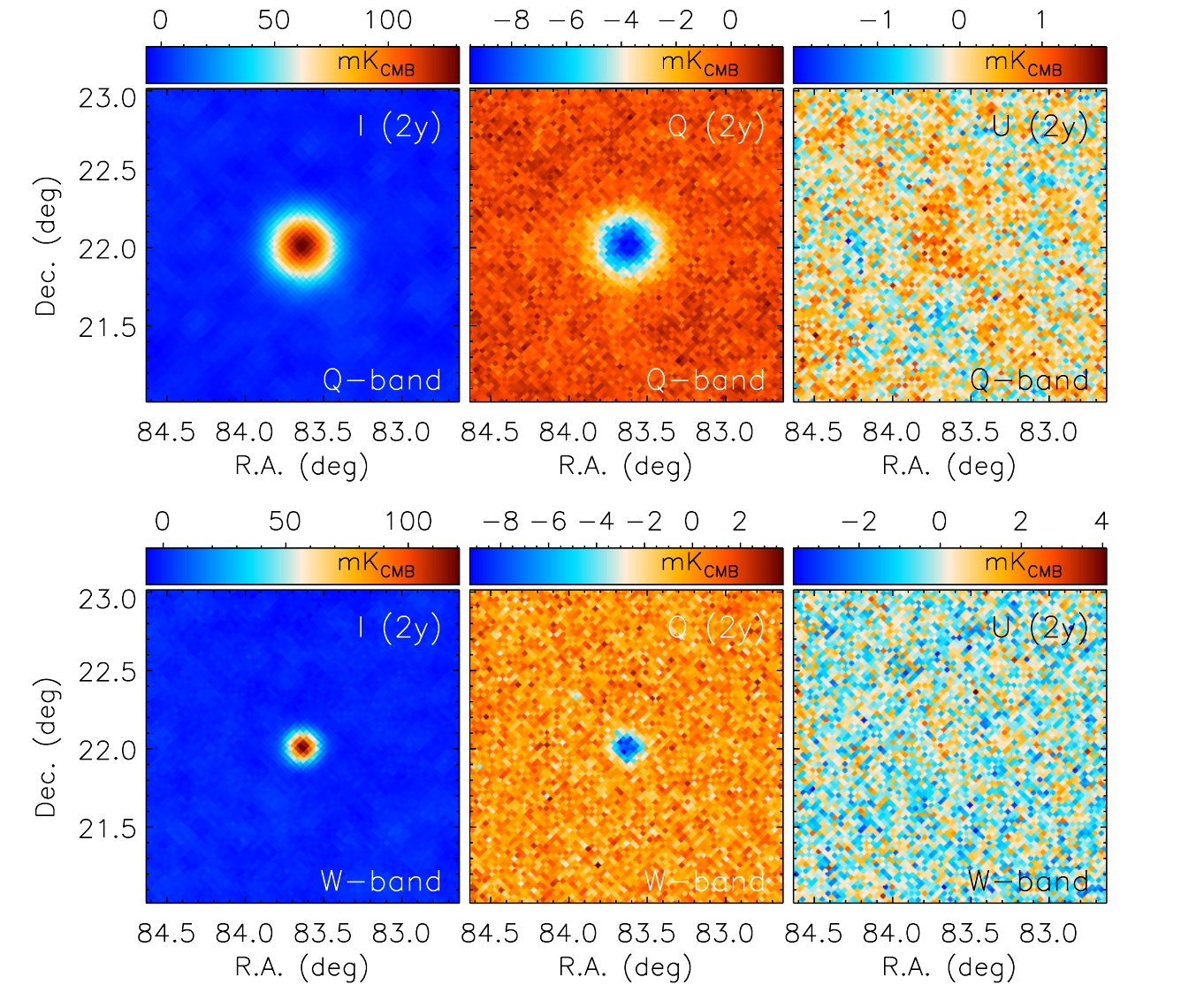}
\caption{Simulated \crab maps in intensity (Stokes I) and in polarisation (Stokes Q and U), as seen by one individual polarimeter of the LSPE-Strip instrument in Q (top) and W (bottom) bands in a one-day map (left-hand-side sets of plots) and full-survey (2 years) map (right-hand side set of plots).}
\label{fig:crab_simulation}
\end{figure}

\begin{figure}
\centering
\includegraphics[width=7.5cm]{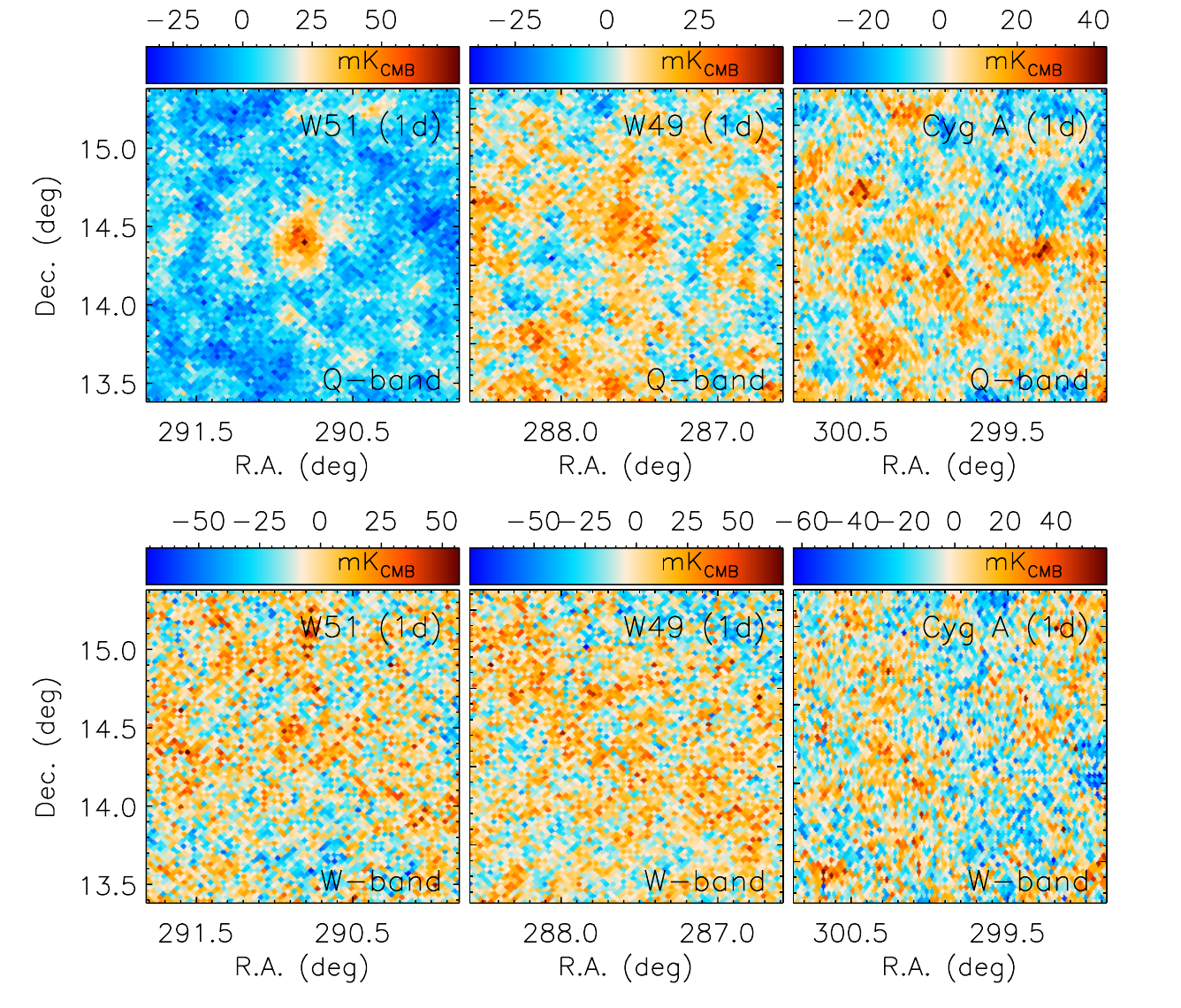}
\includegraphics[width=7.5cm]{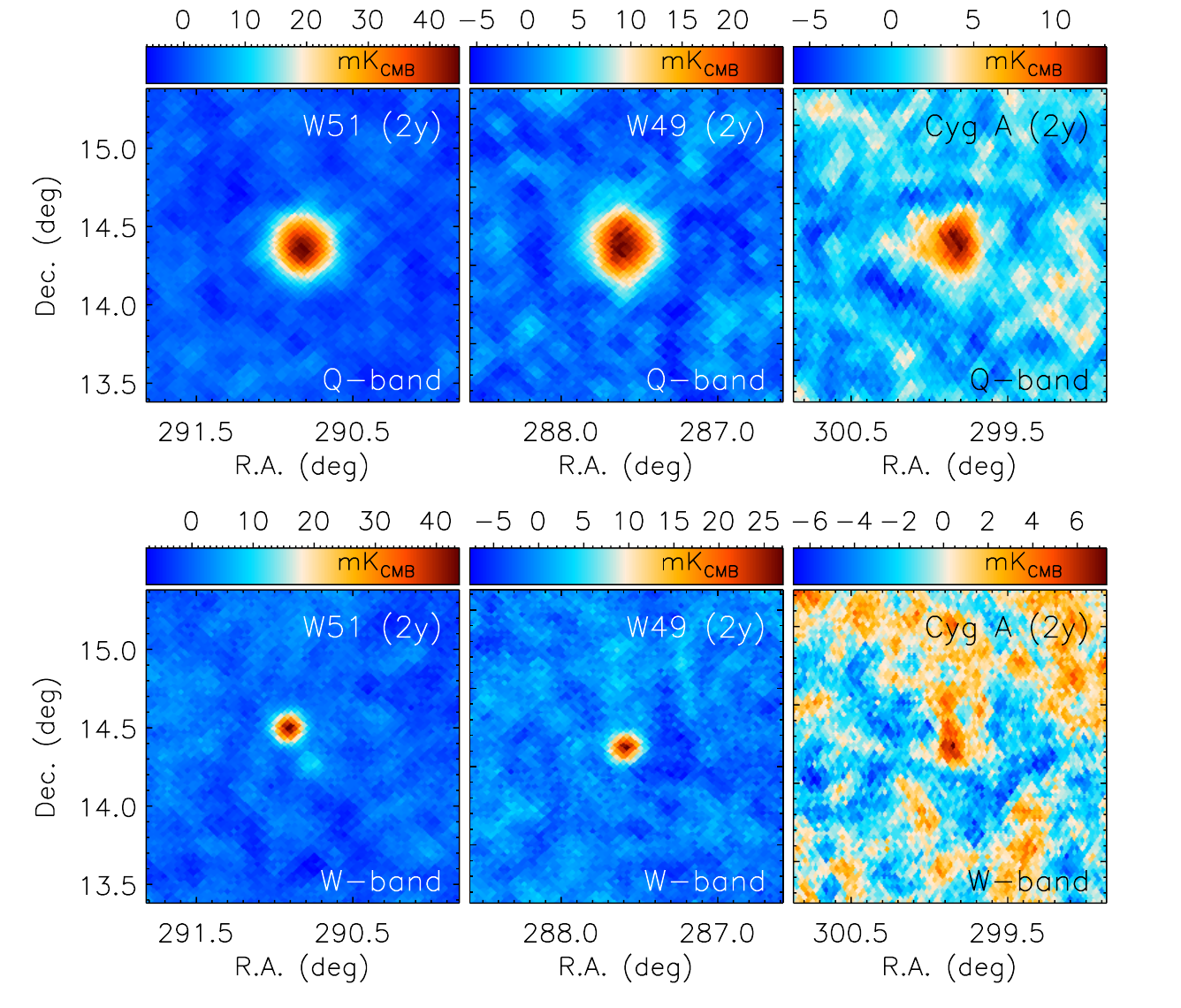}
\caption{Simulated maps of the sources W51, W49 and \cyga in intensity (Stokes I), as seen by one individual polarimeter of the LSPE-Strip instrument in Q (top) and W (bottom) bands in a one-day map (left-hand-side sets of plots) and full-survey (2 years) map (right-hand side set of plots). In these simulations we have considered a combination of white and 1/f noise. For W51 and W49 we have used the flux densities given in section~\ref{subsec:sources2}, while for \cyga we have considered the model described in section section~\ref{subsec:sources1}.}
\label{fig:sources_simulation}
\end{figure}

\begin{figure}
\centering
\includegraphics[width=7.5cm]{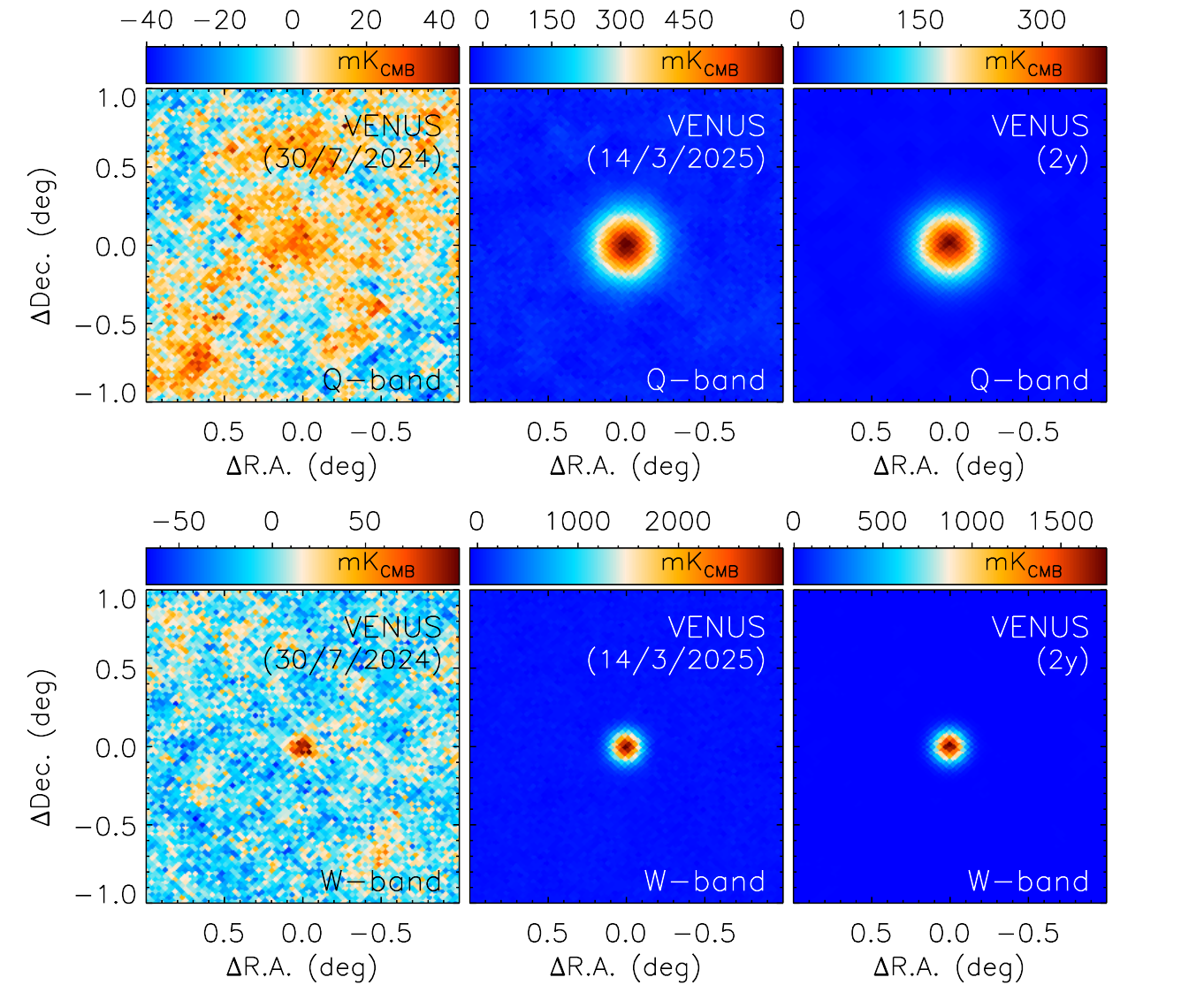}
\includegraphics[width=7.5cm]{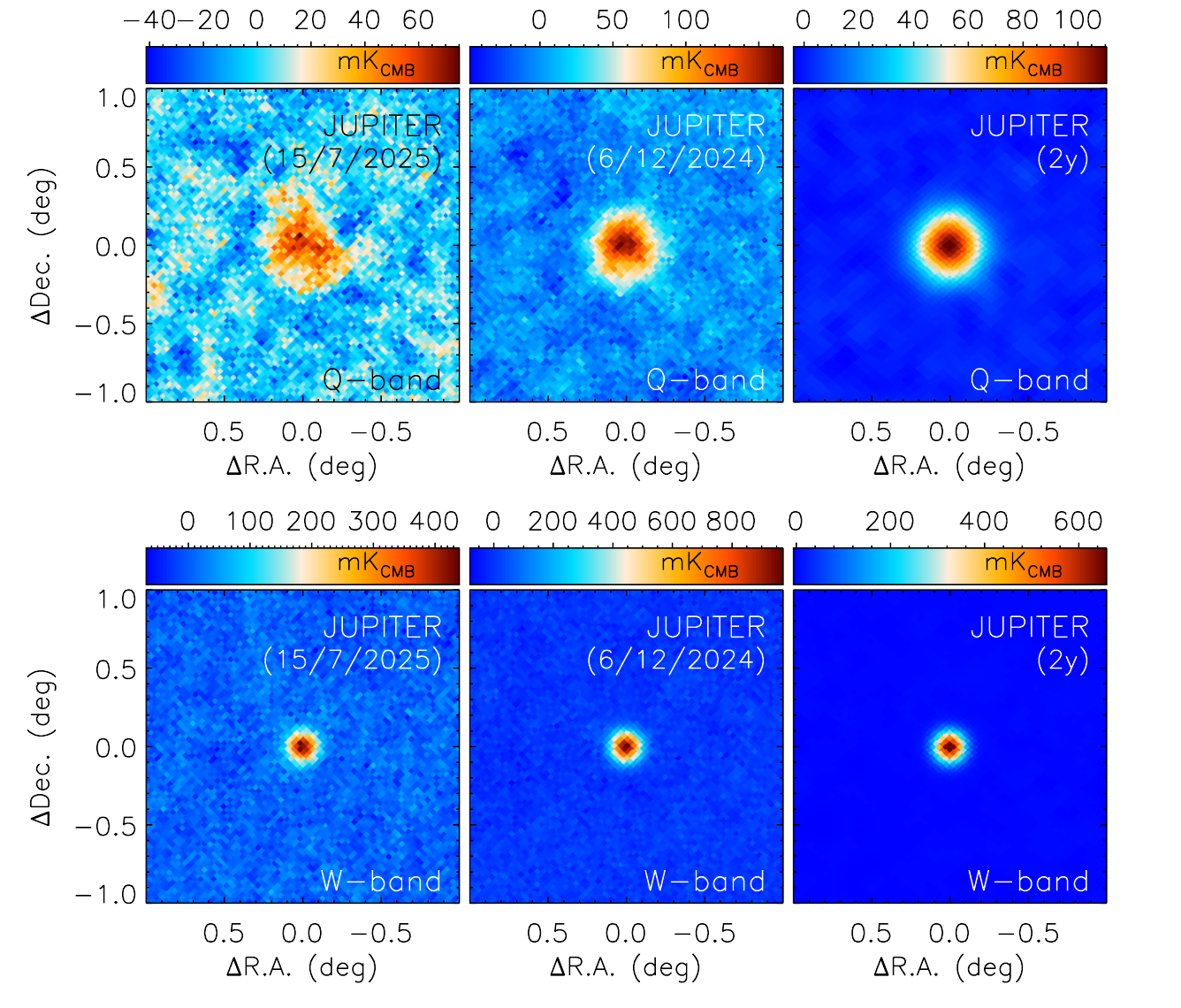}
\caption{Simulated maps of the Venus (left) and Jupiter (right) in intensity (Stokes I), as seen by one individual polarimeter of the LSPE-Strip instrument in Q (top) and W (bottom) bands. In each case we show three panels, corresponding to: 1) one day of data at the farthest separation between the planet and the Earth, 2) one day of data at the closest approximation between the planet and the Earth, 3) 2 year of observations (50\% observing efficiency) in planet-centred coordinates.}
\label{fig:planets_simulation}
\end{figure}

For \crab, which is our main polarisation calibrator, we simulate Q and U maps in the same way, using Q and U antenna temperatures calculated as $Q_{\rm A}=T_{\rm A}\frac{\Pi}{100}{\rm cos}(2\gamma)$ and $U_{\rm A}=-T_{\rm A}\frac{\Pi}{100}{\rm sin}(2\gamma)$\footnote{Note the minus sign in the equation for U. This is introduced when, as it is our case, Q and U are expressed following the COSMO convention (https://healpix.sourceforge.io/html/intro\_HEALPix\_conventions.htm), so that the derived polarisation direction is consistent with the IAU convention.}, where $\Pi=7\%$ and $\gamma=-88.0^\circ$ (see section~\ref{subsec:sources1}). Simulated maps containing signal and noise of Tau\,A, for intensity and for polarisation, for Q- and for W-band, and considering one day observation in the nominal mode and the full 2-year survey, are shown in Figure~\ref{fig:crab_simulation}. The source is detected with very high signal-to-noise in total intensity. In polarisation, given its polarisation direction, the bulk of the emission is contained in Q. This signal is detected marginally in one individual day, and with high signal to noise in the case of the full survey. The rest of the sources do have very small or no polarisation, so we focus on intensity simulations. Simulated maps for W51, W49 and \cyga are shown in Figure~\ref{fig:sources_simulation}. In one individual day only W51 is clearly detected, while in the full-survey maps all three sources are clearly detected. Simulated planets maps are shown in Figure~\ref{fig:planets_simulation}. Jupiter is always detected with high signal to noise, even in individual days. In the case of Venus, it becomes very faint at its farthest distance from Earth, and in this case its signal only remains above the noise in W-band. At its closest approach to Earth, and of course also in the full-survey combined data, it is detected at very high significance.

We calculate flux densities on the maps shown in Figures~\ref{fig:crab_simulation} to \ref{fig:planets_simulation} though the application of a standard aperture photometry technique. We integrate the signal inside an aperture with radius equal to the beam FWHM (see Table~\ref{tab:instrument_parameters}), $r_{\rm ap}=\theta_{\rm fwhm}$, and subtract a background level estimated through the mean of all pixels enclosed in an adjacent ring with internal radius $r_1=\theta_{\rm fwhm}$ and external radius $r_2=\sqrt{2}\,\theta_{\rm fwhm}$. We estimate errors through the dispersion of 10 flux-density estimates calculated using apertures with the same geometry but centred on 10 equally-spaced positions along a circle with radius $2^\circ$ around the source. We apply the same procedure in the I, Q and U maps, in order to derive also expected errors in the polarised intensity $P=\sqrt{Q^2+U^2}$, in the polarisation direction $\gamma=0.5\,{\rm atan}(-U/Q)$, and in the polarisation fraction $\Pi=P/I$. The final values for different sources are shown in tables~\ref{tab:cal_int_pol_sen} and \ref{tab:cal_int_pol_sen_1dr}, respectively for nominal mode maps and for a one-day observation in raster-scan mode. Note that to calculate the signal to noise shown in this table (I/$\sigma_{\rm I}$ and P/$\sigma_{\rm P}$) we have used as reference for the signal $I$ and $P$ the values from the models, and not the values derived from the simulations that are affected by noise.

\subsection{Beams simulations}\label{sec:beam_simulations}
In this work we use the Q- and W-band far-field main-beam radiation patterns simulations described in section 3.2 of \cite{realini_2022}. These simulations used the Physical Optics and Physical Theory of Diffraction methods implemented in the software GRASP (version 10.6.0), developed by TICRA. This is nowadays a standard design tool for reflector antennas, with extended use in other telescopes like Planck \cite{sandri_2010}. The far-field patterns were computed in ($u,v$) spherical grids, in which $u={\rm sin}\theta\,{\rm cos}\phi$ and $v={\rm sin}\theta\,{\rm sin}\phi$ ($\theta$ and $\phi$ are the spherical angles). The main beams for all 49 Q-band feedhorns were simulated within the range $-0.03<(u,v)<0.03$, corresponding to a maximum angular distance from the centre of the beam of $2.4^\circ$. Each grid was sampled using $301\times 301$ points, corresponding to a spatial resolution of about 29\,arcsec. In the case of the 6 W-band feedhorns a range of $-0.015<(u,v)<0.015$ was used, which gives a maximum distance from the centre of the beam of $1.2^\circ$. Also grids of $301\times 301$ points were used, corresponding in this case to a spatial resolution of about 15\,arcsec. In each point of the $uv$-grid the far field was computed in the co- and cross-polar basis using the Ludwig's third definition \cite{ludwig_1973}. To define the beam power amplitude in each $uv$ point we have assumed an ideal OMT with no cross-polarization, and we sum the co-polar and cross-polar amplitudes of each polarimeter output:
\[
B(u,v) = |E(u,v)_{\rm cp}|^2+|E(u,v)_{\rm xp}|^2 =  
\]
\begin{equation}
\left[Re(E(u,v)_{\rm cp})\right]^2 + \left[Im(E(u,v)_{\rm cp})\right]^2  + \left[Re(E(u,v)_{\rm xp})\right]^2 + \left[Im(E(u,v)_{\rm xp})\right]^2~~.
\end{equation}
More details about these simulations can be found in \cite{realini_2022}.

The GRASP outputs were later projected into Healpix by calculating the average of the co-polar amplitude of all $(u,v)$ points lying inside each Healpix pixel. We used resolutions of $N{\rm side}=2048$ (pixel size 1.7\,arcmin) for Q-band and of $N{\rm side}=4096$ (pixel size 0.9\,arcmin) for W-band. The main-beam maps for the three polarimeters that we will use in this analysis (I0, V4 and W1) can be seen in the top panels of Figure~\ref{fig:strip_beam_sim_venus} in logarithmic scale. The main beam and the near sidelobe structure are clearly seen. It becomes obvious the better beam quality, in terms of symmetricity, of the on-axis pixel (I0) as compared with the off-axis pixels (V4 and W1).

\section{STRIP calibration strategy}\label{sec:strip_calibration}

The calibration process can be separated in different steps that are applied to convert the time-ordered-data measured by the instrument into sky maps expressed in brightness units and with a well characterised beam transfer function. In the subsections below we describe each of these steps and, using the calibration models presented in section~\ref{sec:calibrators} and the simulations described in section~\ref{sec:simulations}, we study the accuracy with which each of these calibration steps can be executed using on-sky data.

\subsection{Absolute gain calibration}\label{sec:abs_cal}


The raw signal measured by a polarimeter $k$ can be expressed as
\begin{equation}
V_k = G_k\frac{\int g_k(\nu)\,T_A(\nu)~d\nu}{\int g_k(\nu)~d\nu}~~,
\label{eq:radiometer_output}
\end{equation}
where $G_k$ denotes the global gain of that polarimeter, $g_k(\nu)$ is the bandpass describing its relative response in frequency and is measured against a Rayleigh-Jeans emitter and $T_A(\nu)$ is the beam-averaged antenna-temperature spectrum of the observed sky, in units of Rayleigh-Jeans brightness temperature (K$_{\rm RJ}$). The absolute (or photometric) gain calibration consists in the determination of the global (average) gain factor $G_k$, which gives the conversion between the signal measured in arbitrary units (usually volts) and brightness. This is achieved through the measurement of the output signal $V_{k,{\rm c}}$ induced by a calibration source with a well-known spectrum $T_{\rm A,c}(\nu)$ from which we can estimate a reference calibrator band-averaged antenna temperature
\begin{equation}
T_{k,{\rm c}}=\frac{\int g_k(\nu)\,T_{A,c}(\nu)~d\nu}{\int g_k(\nu)~d\nu}~~,
\label{eq:cal_temp}
\end{equation}
so that the application of the calibration constant $K=T_{k,{\rm c}}/V_{k,{\rm c}}$ leads to fully calibrated data:
\begin{equation}
\tilde{T}_k=K\times V_k = T_{k,{\rm c}} \frac{\int g_k(\nu)\,T_A(\nu)~d\nu}{\int g_k(\nu)\,T_{\rm A,c}(\nu)~d\nu}~~.
\label{eq:calibrated_temp}
\end{equation}

The determination of $G_k$ relies on bright and compact celestial calibrators for which we know their flux density spectrum $S_{\nu,\rm c}(\nu)$ or, in the case of planets, their brightness temperature $T_{B,\rm c}(\nu)$ (see section~\ref{sec:calibrators}). Assuming point-like (unresolved) source the reference antenna temperature to be inserted in equation~\ref{eq:cal_temp} is calculated as
\begin{equation}
T_{A,c}(\nu) = \frac{c^2}{2 k_b\nu^2}\frac{S_{\nu,\rm c}(\nu)}{\Omega_A}~~,
\end{equation}
where $\Omega_A$ is the beam solid angle. In the case of planets, for which $T_{B,\rm c}(\nu)$ is known, equation~\ref{eq:ta_planets1} must be applied. Note that the determination of the calibrated temperature scale then requires knowledge of the beam solid angle. In the case of partially resolved sources, where the point-source approximation is no longer valid, the determination of $T_{A,c}(\nu)$ requires the application of a convolution between the source brightness and the beam profiles  (\crab and \casa are partially resolved by the W-band beams, see section~\ref{subsec:sources1}). Therefore in this case precise knowledge of the beam profile is also needed. Note that, in addition to the beam, a good characterisation of the bandpass response $g_k(\nu)$ is also needed to achieve an accurate calibration.

We can in principle envisage different ways of determining $G_k$. Daily data (in nominal or raster mode) or full-survey maps could be used. Even if there are relative gain fluctuations in short time scales, which are corrected or alleviated through relative gain calibration (see next subsection) the determination of the absolute gain calibration can be based on the evaluation of the full-survey maps on the position of bright calibrators. In fact, one could produce a map of the full-survey data in units of volts, and only apply the final absolute calibration at the end of the process. Therefore in assessing the precision in the calibration we will rely on the full-survey information contained in Table~\ref{tab:cal_int_pol_sen}. Given that Strip is a polarimeter with significant gain time variations in total intensity, using polarised sources like \crab could in principle seem the best option. However, being the brightest compact polarised source on the sky, the polarization fraction of \crab is $\approx 7\%$ (see section~\ref{subsec:sources1}), which means that as long as the final noise in intensity is less than $\approx 14$ times that in polarization the final signal-to-noise in total intensity would still be better. In fact, the values of Table~\ref{tab:cal_int_pol_sen} show that this is the case, and even in the case of considering white plus $1/f$ noise, the expected signal-to-noise in \crab is better in intensity (I/$\sigma_{\rm I}$) than in polarisation (P/$\sigma_{\rm P}$). Therefore, relying on the assumptions on noise properties used in section~\ref{sec:maps_sims}, we conclude that a calibration using \crab total intensity data provides a better precision. 

According to the values of I/$\sigma_{\rm I}$ shown in Table~\ref{tab:cal_int_pol_sen} we conclude that by using total-intensity data on \crab each polarimeter could be calibrated with a precision of $0.9\%$ in Q-band and of $2.1\%$ in W-band. Using Venus would allow to reach $0.4\%$ and $0.14\%$ respectively in Q- and W-band. Other sources like W51, W49, \cyga or Jupiter provide lower precisions, but could be used as cross-checks. Note also that it is important to identify as many bright calibrators as possible in order to have calibration information at different times during the day and then overcome or minimise systematic effects associated with daily variations (such as day/night temperature modulation). Polarisation data would be naturally calibrated with the same precision assuming the polarisation efficiency of the polarimeters to be of the order of 100\%. In order to avoid this assumption, data could be calibrated directly using \crab polarisation, and in this case the precision in the calibration of each polarimeter, from full-survey data, would be $1.9\%$ in Q- and $3.6\%$ in W-band. These precisions could be improved (by $\sqrt{49}$ in Q-band and by $\sqrt{6}$ in W-band) if, instead of applying an individual calibration of each polarimeter their are combined before calibration. Note anyway that the previous numbers are already better than the uncertainty in the calibration models, which is $\approx 5\%$ (see section~\ref{sec:calibrators}). We then conclude that the sensitivity of Strip data, both in intensity and in polarisation, suffices to achieve a precision on the absolute calibration better than the limiting factor of 5\% set by the precision of the calibrator models.

\subsubsection{Colour corrections}

As shown in equation~\ref{eq:calibrated_temp}, the final data are calibrated to a given reference spectrum, $T_{A,c}(\nu)$. As a result of the finite bandwidth of the filters, $g_k(\nu)$, the measured band-averaged antenna temperature is sensible to the spectrum of the observed source. To amend this, and assuming perfect knowledge of the instrument bandpass, a colour correction coefficient can be applied so that the corrected antenna temperature would represent the response of an ideal monochromatic detector observing at a reference frequency $\nu_0$. Let us assume that the final data are calibrated to thermodynamic brightness differential CMB temperature (K$_{\rm CMB}$), in the same way as WMAP \cite{hinshaw_2003}, Planck \cite{cpp2013-5,cpp2013-8} or ground-based experiments like QUIJOTE \cite{rubino_2023}, so that
\begin{equation}
\tilde{T}_k= \frac{\int g_k(\nu)\,T_A(\nu)~d\nu}{\int g_k(\nu)\,\eta_{\Delta T}(\nu)~d\nu}~~,
\label{eq:cmbcalibrated_temp}
\end{equation}
where $\eta_{\Delta T}(\nu)=x^2 e^x/(e^x-1)^2$ is the conversion factor from CMB-differential brightness temperature (K$_{\rm CMB}$) to Rayleigh-Jeans brightness temperature ($K_{\rm RJ}$), with $x=h\nu/k_bT_{\rm CMB}$ the dimensionless frequency (note that even if the data are calibrated using sources having a different spectrum, through multiplication by a constant factor they can be referred to a whatever other chosen spectrum, in this case the CMB).

Following the same convention and formalism as in \cite{cpp2013-5}, for a given reference frequency $\nu_0$ in this case the colour-correction coefficient can be defined as
\begin{equation}
C_k(\nu_0) = \frac{T_0}{\eta^0_{\Delta T}}\frac{\int g_k(\nu)\eta_{\Delta T}(\nu)~d\nu}{\int g_k(\nu)T_A(\nu)~d\nu}~~,
\label{eq:cc}
\end{equation}
so that the colour-corrected temperature $C_k(\nu_0)\tilde{T}_k=T_0/\eta_{\Delta T}(\nu_0)$ represents the temperature, expressed in K$_{\rm CMB}$ units, that would be measured by a monochromatic receiver operating at frequency $\nu_0$ when it observes a source with spectrum $T_A(\nu)$. In the case that the source is characterised by a power-law spectrum $T_A(\nu)=T_0(\nu/\nu_0)^{\alpha-2}$, equation~\ref{eq:cc} can be recast as
\begin{equation}
C_k(\alpha,\nu_0) = \frac{\int g_k(\nu)\eta_{\Delta T}(\nu)~d\nu}{\eta^0_{\Delta T}\int g_k(\nu)(\nu/\nu_0)^{\alpha-2}~d\nu}
\label{eq:cc_pl}
\end{equation}

\begin{figure}
\centering
\includegraphics[width=7.5cm]{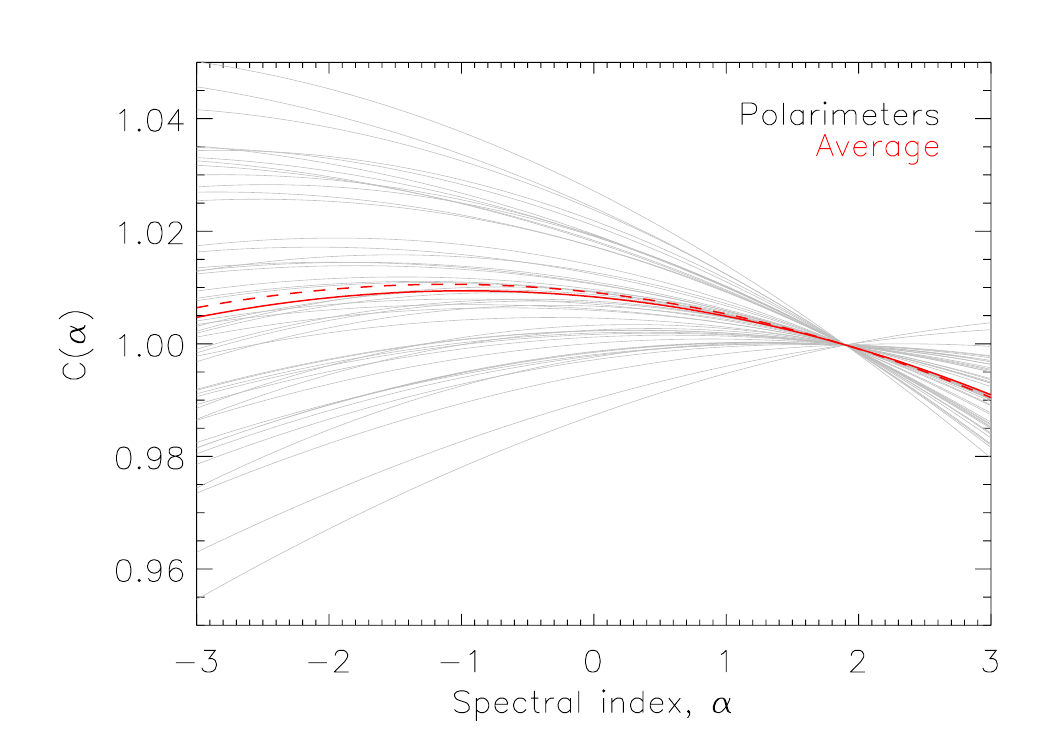}
\includegraphics[width=7.5cm]{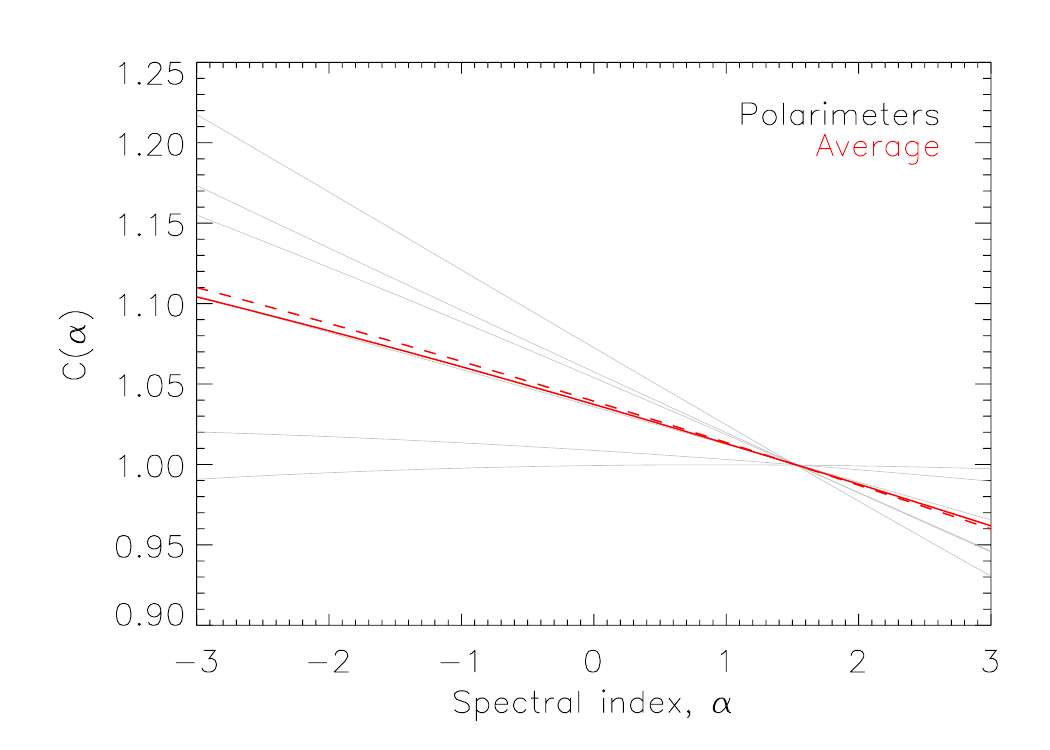}
\caption{Colour corrections coefficients for power-law spectra as a function of spectral index for Q- (left) and W-band (right). Grey lines correspond to individual polarimeters, while the red line shows the average colour-correction calculated through two different methods (see text for details).}
\label{fig:cc}
\end{figure}

Using the measured bandpasses for each Strip polarimeter, we have applied equation~\ref{eq:cc_pl} to compute the colour correction coefficients $C(\alpha,\nu_0)$ for a range of spectral indices $\alpha\in[-3,+3]$. The results are shown in Figure~\ref{fig:cc}, for Q- and W-bands, using respectively $\nu_0=43$ and 95\,GHz. Note that we have $C(\alpha_{\rm CMB},\nu_0)=1$ for $\alpha_{ \rm CMB}=1.91$ for Q-band and for $\alpha_{ \rm CMB}=1.55$ for W-band, which are the spectral indices of the CMB fluctuations for $\nu_0=43$ and 95\,GHz respectively ($\alpha_{ \rm CMB}={\rm dln}(\eta_{\rm \Delta T})/{\rm dln}\nu$+2). For $\alpha=-1$, a typical spectral index of synchrotron emission, the colour-correction coefficients are of order $\pm 2\%$. Red curves show the averaged colour corrections over polarimeters. The solid red curves correspond to the case in which an average bandpass, $g(\nu)=\sum_{k=1}^{N_{\rm p}}g_k(\nu)/{N_{\rm p}}$, is inserted in equation~\ref{eq:cc_pl}. The dashed red curves correspond to the case in which the individual colour-corrections (grey curves) are averaged as
\begin{equation}
C(\alpha,\nu_0) = \frac{N_{\rm p}}{\sum_{k=1}^{N_{\rm p}}1/C_k(\alpha,\nu_0)}~~.
\end{equation}
This is the equation that results for the case in which an averaged map is calculated from the maps of all individual polarimeters.

\subsection{Relative gain calibration}

It is well known that, as a result of $1/f$ fluctuations in the receiver, the instrument gain is not a constant function. In Strip knee frequencies are expected to be of the order of several Hz (see \cite{lspe_2021}). In practice this causes that the radiometer gain is a function of time, $G_k(t)$, implying that the radiometer output, $V_k(t)$, changes not only due to variations in sky emission but due to gain fluctuations. Given that Strip observes one map of the full sky in one day, gain variations in shorter time scales directly result in different parts of the sky map having different ``effective'' gains, and in fact different than on the position where the calibrators that are used to execute the absolute gain calibration (see previous subsection) are located. Therefore, absolute gain calibration must be combined with relative gain calibration in short time scales to alleviate the nasty effect of gain fluctuations. Mathematically this can be expressed as
\begin{equation}
G_k(t) = G_k (1+\delta(t))~~.
\end{equation}
Absolute gain calibration (previous subsection) consists in determining the constant factor $G_k$, while relative gain calibration consists in determining the $\delta(t)$ function (or ``gain model'') that accounts for relative gain variations. 

Strip implements an internal gain calibration system that consists in a stable signal injected on regular time intervals and generated by two thermally stabilised microwave generators (one in Q-band and another in W-band) installed in the optical assembly \cite{lspe_2021}. This signal is expected to allow a precision in the measurement of the $\delta(t)$ function better than $1\%$. Taking advantage that all calibration sources discussed in section~\ref{sec:calibrators} are observed daily, this ``internal'' calibration strategy can be combined with on-sky information. Table~\ref{tab:cal_int_pol_sen} shows that daily observations of \crab in nominal mode allow reaching sensitivities of $8.1\%$ and $11.1\%$ respectively in Q- and W-band. Venus and Jupiter allow reaching $\sim 5\%$, or even below $1\%$ at their closest approach to Earth. However planets have the drawback of not being always visible. Table~\ref{tab:cal_int_pol_sen_1dr} shows that deeper observations in raster mode can marginally improve these precisions. In addition to the low sensitivity achieved, it must be borne in mind that these calibration sources are visible (in nominal mode) during a given time interval every day, and then they are unable to provide continuous information about the gain. Other important issue is that there are gain fluctuations that are correlated with ambient conditions (temperature, time of the day), and given that any given calibration source is observed at virtually the same time in consecutive days, it cannot be used to correct for this type of fluctuations.

The atmosphere, apart from introducing $1/f$ noise in total intensity that can leak into polarisation via different effects, absorbs radiation. The relative fraction of absorption is given by $1-e^{-\tau/{\rm sin(EL)}}$, where $\tau$ is the atmosphere opacity at the zenith and EL is the elevation of the observation, EL=$70^\circ$ for nominal mode\footnote{The $1/{\rm sin(EL)}$ dependency in this equation results from the atmosphere plane-parallel approximation, which for elevation 70$^\circ$ is found to be accurate to better than $0.1\%$.}. If this opacity term was constant this effect would be naturally cancelled in the calibration process, as the calibrator would be observed under the same opacity conditions as the scientific data. Unfortunately this is not the case, and the opacity of the atmosphere changes mainly as a consequence of variations in the precipitable water vapour. This would then introduce a noise component similar to instrumental $1/f$ noise. PWV at the Teide Observatory has median value of 3.8\,mm with standard deviation of 3.2\,mm. This statistics has been computed using 2018 data recorded by a nearby GPS antenna, and is compatible with the data shown in \cite{castro_2016}. Taking as a reference the two extremes of this interval, PWV= 0.6 and 7.0\,mm, and using the Atmospheric Transmission at Microwaves (ATM) model \cite{pardo_2001} that gives $\tau$ at the zenith, we estimate that the opacity at the TO changes such that the transmitted fraction $e^{-\tau/{\rm sin(70^\circ)}}$ changes between 0.93 and 0.91 in Q-band and between 0.96 and 0.87 in W-band. This means that the atmospheric opacity could lead to maximum signal variations of the order of $\sim 2\%$ and of $\sim 9\%$ respectively in Q- and W-band. Given the expected sensitivities achieved in daily point source observations shown in tables~\ref{tab:cal_int_pol_sen} and \ref{tab:cal_int_pol_sen_1dr} we conclude that it will be not possible to correct these amplitude fluctuations using those calibration sources. Real-time PWV data, like those provided by the GPS antenna, should instead be used to try ameliorating this noise component. Note however that these PWV monitoring systems normally provide information about the integrated PWV along a specific line of sight, while there will be spatial variations of the PWV that will lead to signal fluctuations as the telescope scans the sky. These variations are expectedly smaller than PWV time variations, hopefully below the $2\%$ level,
and could be partially corrected using W-band data as a reference, that are conceived to serve as an atmospheric monitor.

\subsection{Polarisation direction}
The measured signal must be decomposed into the two linear polarisation Stokes parameters Q and U, in order to ultimately separate the polarisation pattern into the E- and B-modes that is the final goal of Strip observations. This requires accurate knowledge of the polarisation direction in a given reference system. The Strip polarimeters measure directly Stokes Q and U in a local reference system in horizontal coordinates (Strip has an alt-azimuth mount). The orientation of this system is dictated by the layout of the polarizer-OMT, and can be measured mechanically down to a precision of $0.5\%$ \cite{lspe_2021}. As shown in \cite{krachmalnicoff_2015} (see also Figure~19 of \cite{lspe_2021}) an uncertainty in the polarisation direction of $1^\circ$ leads to residual systematics well below the total expected foreground signal in Q-band, and below the primordial B-mode signal for $r=0.1$. The $0.5^\circ$ mechanical precision can then be deemed sufficient. However, the true polarisation direction could in principle change due to thermoelastic effects during cooldown. It is therefore important as a cross-check to execute a calibration of the polarisation direction on the sky.

To a first approximation the polarimeters polarisation direction could be assumed to be constant over time. This assumption would be derived from the fact that from a mechanical point of view the polarimeters positions and orientations on the focal plane are stable, and also that the properties of the phase switches do not change. In this case the calibration of the polarisation direction could rely on full-survey data. In table~\ref{tab:cal_int_pol_sen} we show that using the full 2-year survey data on \crab it is possible to calibrate the polarisation direction of each polarimeter with a precision of $0.7^\circ$ in Q-band and of $1.3^\circ$ in W-band. These values are comparable to $0.3^\circ$ that is, at best, the precision with which we know the intrinsic polarisation direction of \crab (see section~\ref{sec:calibrators}). However, daily observations in nominal mode lead to precisions on the angle of $5.4^\circ$ and $20^\circ$ in Q- and W-band respectively, while deeper observations in raster mode would allow to improve these numbers marginally (see table~\ref{tab:cal_int_pol_sen_1dr}). Therefore, testing the hypothesis of stability of the polarisation direction of the polarimeters would require the combination of various consecutive days of data: $\sim 15$ days might allow to reach sub-degree precisions in Q-band.

An alternative way of calibrating the angle on the sky, or at least to cross-check the \crab-based calibration, is the comparison with Q-band maps from WMAP (central frequency 40.6\,GHz) or from Planck LFI (central frequency 44.1\,GHz) on regions of bright diffuse Galactic emission. This method was applied in QUIJOTE (in that case, comparing with WMAP and Planck lowest-frequency channels) allowing to reach uncertainties of the order of $\sim 1^\circ$ (a bit larger on the 19\,GHz QUIJOTE channels due to higher noise on those QUIJOTE maps). A similar precision is expected in the case of Strip. 

\subsection{Polarisation efficiency}
As explained in section~\ref{sec:strip_instrument} each Strip polarimeter has four outputs measuring respectively I$\pm$Q, I$\mp$Q, I$\pm$U and I$\mp$U, and implements a fast switching strategy to continuously invert the sign of the measured linear polarisation Stokes parameter Q or U thence reducing the $1/f$ noise. Summing or subtracting two consecutive switch states then leads respectively to a measurement of I or of Q (U) for the first (second) two outputs. Therefore one can chose to i) take the sum of the data on a calibration source, calibrate using a given model, and then by pair differencing we would have calibrated polarisation data, or ii) take the difference of the data on a calibration source and calibrate directly in polarisation. We showed in section~\ref{sec:abs_cal} that in principle, given our assumptions on noise properties, intensity data on calibration sources like \crab provide better precision, and in this respect strategy i) would be the best option. However, this relies on the polarisation efficiency being 100\%. In practice, due to different instrumental non-idealities, 
the polarisation efficiency can be different from 100\% and would need to be calibrated. It can be calibrated by measuring the polarisation fraction of \crab and comparing with the model, which predicts it to be $7.0\%$ (see section~\ref{sec:calibrators}). If the polarisation efficiency is assumed to be constant in time, in principle we could resort to full-survey data. According to Table~\ref{tab:cal_int_pol_sen} using \crab this parameter could be measured with a precision of $0.13\%$ in Q-band and of $0.3\%$ in W-band. In daily observations in nominal mode we would have $1.0\%$ and $3.4\%$ respectively. 

\subsection{Leakage}

Cross leakage between I, Q and U (in all possible directions) can be sourced by different instrumental effects. Some of these effects are discussed and quantified in \cite{lspe_2021}. Cross-polarisation and amplitude imbalance in the polarizer-OMT system could lead to a leakage of $0.5\%$ between I and Q, while the leakage from I to U (in the local coordinates reference frame) is expected to be negligible. 
Asymmetries in the beam response are another important source of leakage. In Strip the cross-polarization discrimination factor (XPD), which is the ratio between the maximum directivity of the cross- and co-polar beams, is found to be $\approx -42$~dB for Q-band and $\approx -45$~dB for W-band \cite{realini_2022}. From this it is estimated that the sky spurious polarisation in Q-band should be of $\approx 0.01~\mu$K \cite{lspe_2021}. Asymmetries in the main beam are expected to be $\lesssim 3\%$ \cite{realini_2022} and this would lead to a cross-polar discrimination of $-40$~dB \cite{lspe_2021}.  

Although laboratory measurements and simulations are useful to have an idea of the magnitude of these effects (and eventually to implement a methodology to correct them in the data processing pipeline), it is also useful to have an on-sky characterisation of them. The polarisation angle of \crab in Galactic coordinates is $-87.8^\circ$ (see section~\ref{sec:calibrators}), which means that the bulk of its polarisation is projected into Q, while U is consistent with zero. Then the \crab map of Stokes U in Galactic coordinates can be used as a health test for the leakage from intensity to polarisation. It is important noting that, due to projection effects from local to sky coordinates, the signal measured in U on the sky will be a combination of the Q and U signals in local coordinates, and then the U map of \crab will give an idea of the leakage from I to both Q and U in the instrument reference frame. Using 2-year full survey data on the position of \crab this leakage could be measured with a precision of $0.13\%$ and $0.3\%$ respectively in Q- and W-band (see table~\ref{tab:cal_int_pol_sen}). Sources that are intrinsically unpolarised, like W49 or the planets (see section~\ref{sec:calibrators}), could also be used for this purpose. Using full-survey data W49 would allow to reach precisions of $1.0\%$ and $1.1\%$ respectively in Q- and W-bands. Venus, thanks to it being much brighter, would allow reaching $0.04\%$ and $0.02\%$ respectively in full-survey data, and $0.3\%$ and $0.2\%$ using one-day maps in nominal mode during its closest approach to Earth.

\subsection{Beams and window functions}
\label{sec:beams}

A precise characterisation of the beam patterns is one of the most important steps in the calibration process as they determine the sensitivity of the instrument to emission on different angular scales on the sky. In this paper we have considered absolute gain calibration based on point-sources (see section~\ref{sec:strip_calibration}). In this case, calculation of the reference antenna temperature of the source, that is used to calibrate, requires knowledge of the beam solid angle. Thence uncertainties on the value of the beam solid angle translates into a global systematic error on the global gain calibration scale. Furthermore, knowledge of the beam properties beyond the main beam, i.e. in the near and far-sidelobes, is crucial to assess the amount of stray-light the telescope is sensitive to, in particular RFI or ground pickup. Precise models of the beam patterns can be obtained through optical modelling implemented in packages like GRASP (see section~\ref{sec:beam_simulations}). However there are uncertainties on the input parameters required by these packages that could result in differences between the idealised simulated beam map and the real one. For this reason it is important to verify the optical models using on-sky observations on point sources (in the point-source approximation, and in the absence of any emission other than that from the source itself, the antenna temperature distribution is equal to the beam pattern). As we will see, while real on-sky data allow for a reasonably good characterisation of the main beam properties, measuring the beam structure outside the main beam is a major challenge.

In this section we study the accuracy with which the beam can be mapped using sky data, on the positions of bright calibrators. In particular we consider Venus, Jupiter and \crab that were also used as calibrators in previous sections. As it was commented already in section~\ref{sec:calibrators}, we do also consider the Moon and the Sun, which owing to being too bright and extended are not optimal gain calibrators but are yet useful to study the far sidelobe structure, precisely thanks to being so bright. To this aim we have used the same tools that were developed and described in previous sections, in particular the source-centred maps that were presented in section~\ref{sec:simulations}. These source-centred maps are important as they are able to preserve the beam 2D structure, as opposed to maps projected onto sky coordinates that lead to a 1D symmetrisation of the beam. One important issue to be borne in mind is that, given the Strip beams FWHMs (see Table~\ref{tab:instrument_parameters}), \crab (which is $\sim$5~arcmin across) is partially resolved. Therefore, in this case we have convolved the simulated GRASP beams, for both Q- and W-bands, with an idealised temperature distribution for \crab consisting of a uniform disc of 4.5~arcmin diameter. We have applied the same procedure for the Moon, but in this case using a uniform disc with a diameter of $0.52^\circ$. In the cases of Venus (its angular diameter changes between 0.16 and 0.99~arcmin, see Table~\ref{tab:planets}) and Jupiter (between 0.52 and 0.78~arcmin), their sizes remain sufficiently small compared with the beam and are considered point sources.

\subsubsection{Main beam and near sidelobes}\label{subsec:mb}

\begin{figure}
\centering
\includegraphics[width=14cm]{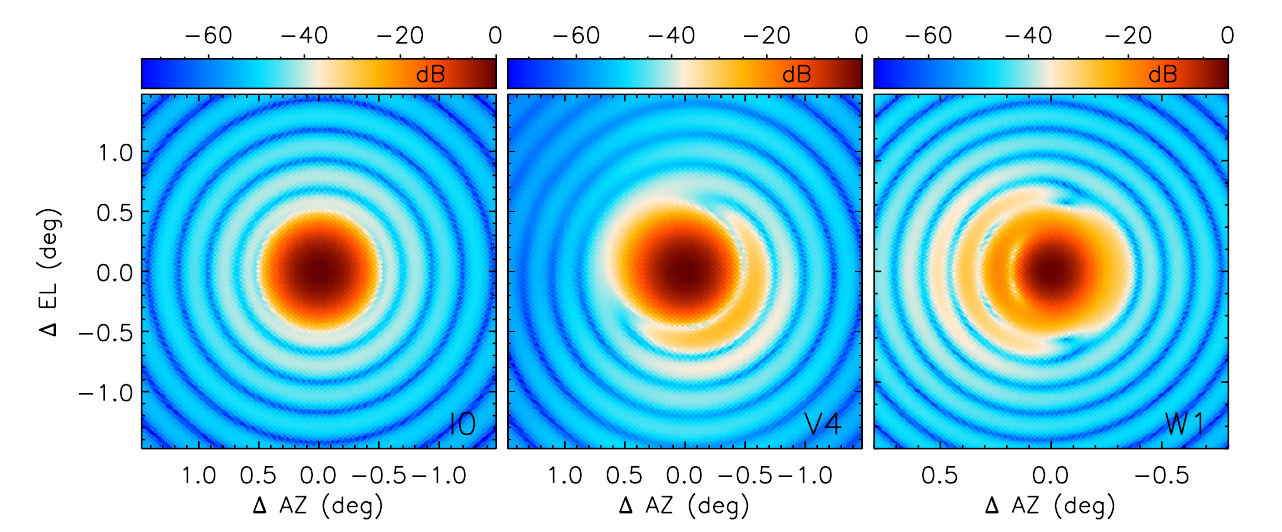}
\includegraphics[width=14cm]{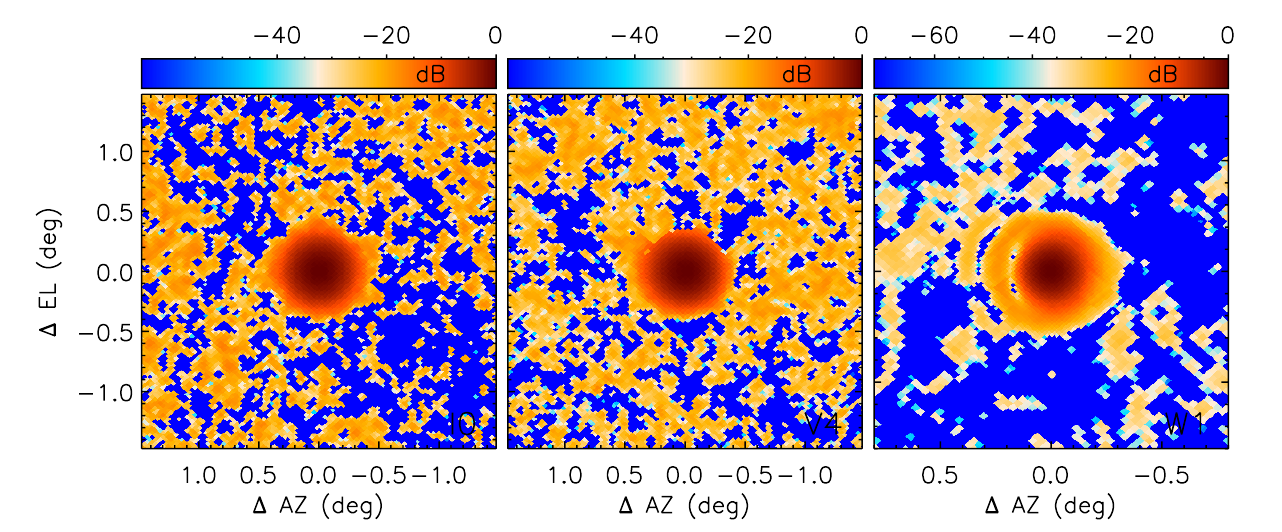}
\caption{2D maps of the main-beam GRASP simulations, for pixels I0, V4 and W1. The top row shows noiseless simulations, and the bottom row shows beam-convolved normalised VENUS observations including white and 1/f noise for a 2-year survey in nominal mode.}
\label{fig:strip_beam_sim_venus}
\end{figure}

\begin{figure}
\centering
\includegraphics[width=14cm]{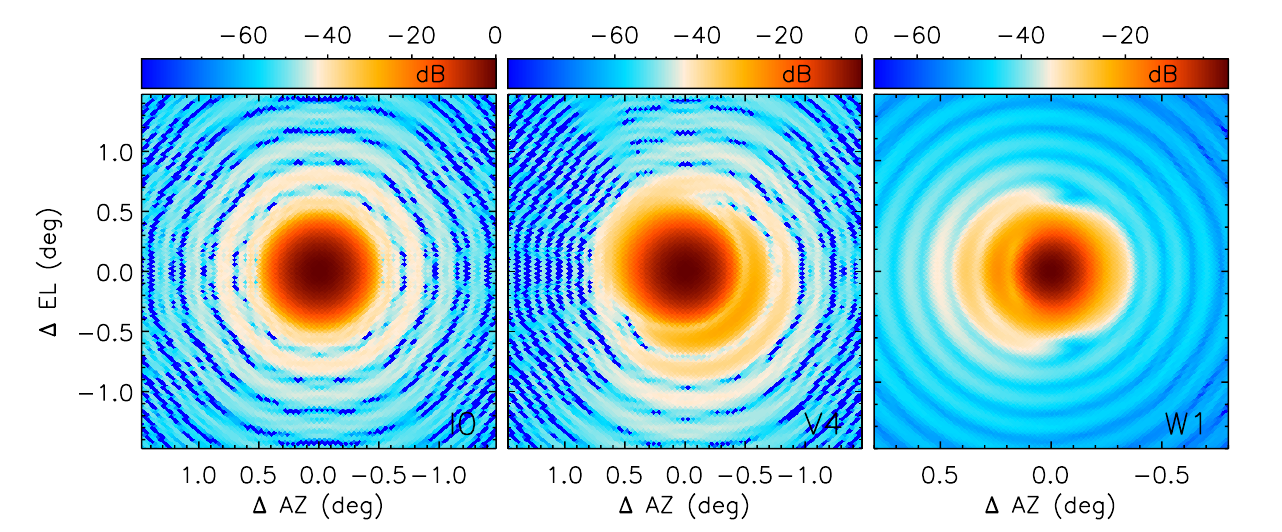}
\includegraphics[width=14cm]{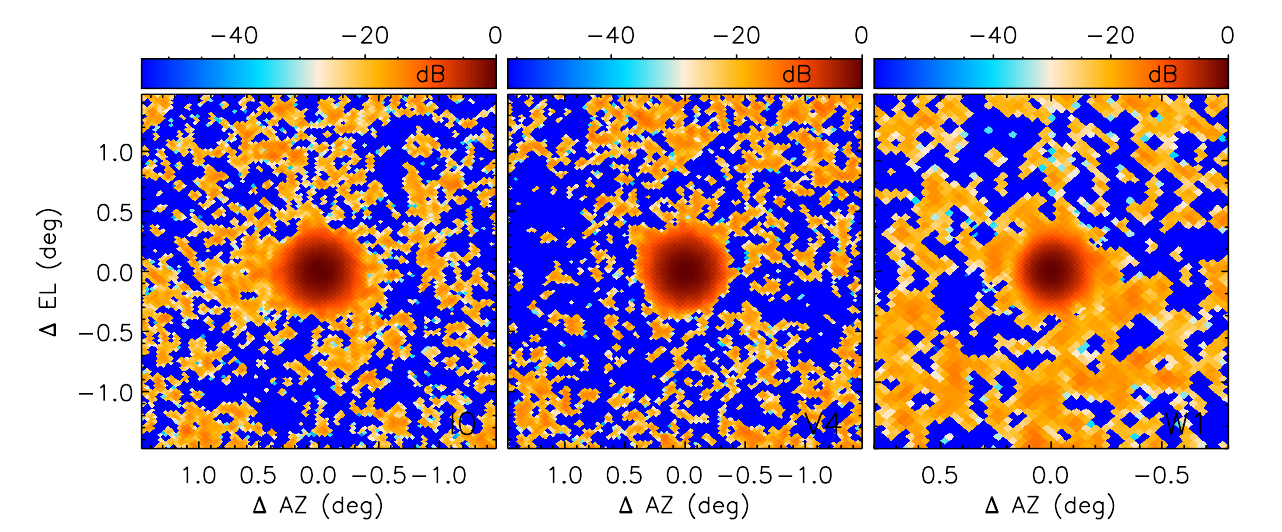}
\caption{Same as Figure~\ref{fig:strip_beam_sim_venus} but for Tau\,A. As in this case the point-source approximation is no longer valid the GRASP beams have been convolved with the spatial structure of \crab (approximated through a circle of radius $2.25$~arcmin).}
\label{fig:strip_beam_sim_crab}
\end{figure}

\begin{figure}
\centering
\includegraphics[width=14cm]{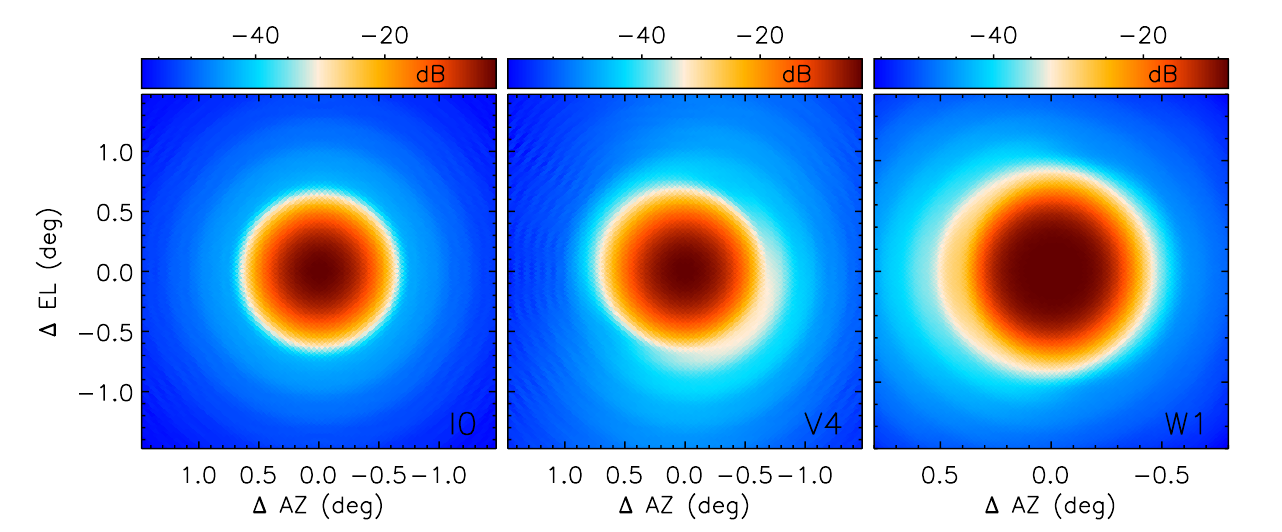}
\includegraphics[width=14cm]{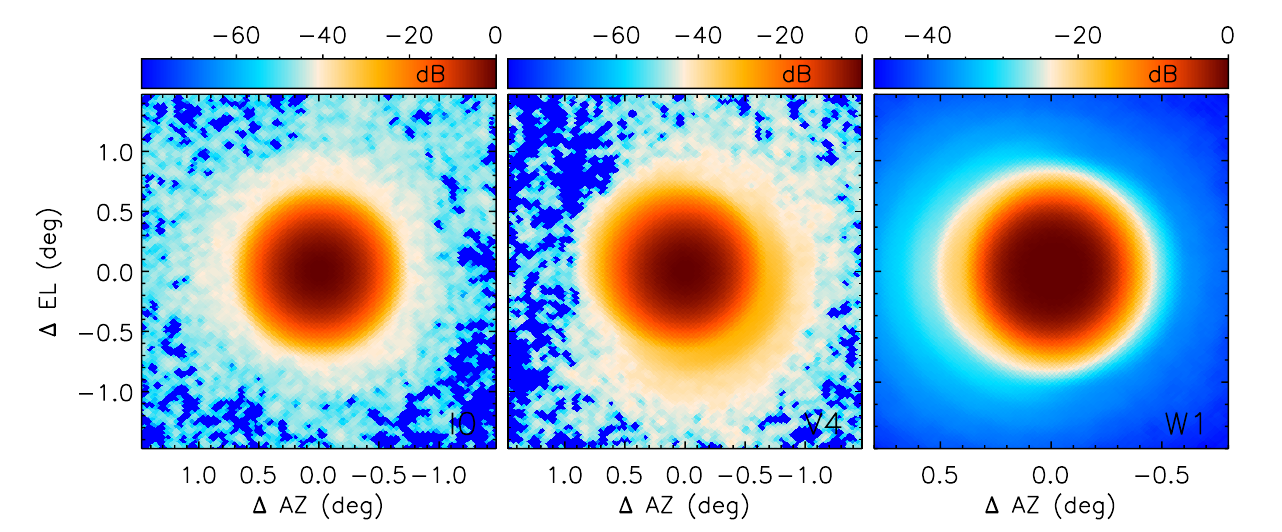}
\caption{Same as Figure~\ref{fig:strip_beam_sim_venus} but for the Moon. As in this case the point-source approximation is no longer valid the GRASP beams have been convolved with the spatial structure of the Moon (approximated through a circle of radius $0.26^\circ$).}
\label{fig:strip_beam_sim_moon}
\end{figure}

While in the analyses presented in previous sections we considered as reference pixels I0, for Q-band, and W1, for W-band, in this section we have also produced simulations for Q-band pixel V4, which is located on the border of the focal plane array (see Figure~\ref{fig:strip_fp}) and thence has beam properties notably different from those of I0 that is located right in the centre of the focal plane \cite{realini_2020,realini_2022}. In the three top panels of Figure~\ref{fig:strip_beam_sim_venus} we do represent the GRASP beam patterns\footnote{Venus is considered a point-source and then what is represented in these plots is directly the beam pattern.} respectively for pixels I0, V4 and W1, in logarithmic scale, where both the main beam and near sidelobe structure are clearly visible. While the beam pattern of pixel I0 shows a remarkable axial symmetry, those of V4 and W1 present asymmetric structure, with ellipticities respectively of 2\% and 3\%, due to being off-centred. The top three panels of Figure~\ref{fig:strip_beam_sim_crab} show the same maps but after convolution with the \crab temperature structure. Apart from a broadening of the main beam, this convolution leads to a partial smearing of the sidelobe structure. Obviously, due to the larger size of the source compared to the beam, these effects are more noticeable in W-band. In this case the FWHM of the beam broadens from 9.7 to 10.1~arcmin. Correspondingly, Figure~\ref{fig:strip_beam_sim_moon} shows the effect of the convolution with the Moon temperature pattern, which leads to an even stronger distortion of the beam. In this case the beam width is totally dominated by the size of the Moon, with an almost unnoticeable difference between Q- and W-bands.

In the bottom panels of Figures~\ref{fig:strip_beam_sim_venus}, \ref{fig:strip_beam_sim_crab} and \ref{fig:strip_beam_sim_moon} we show the simulated beam-convolved antenna temperature maps respectively of Venus, \crab and the Moon with realistic noise for a two-year survey in nominal mode (with 50\% observing efficiency). The noise simulation contains a combination of white noise and 1/f noise and has been produced following the recipe described in section~\ref{sec:maps_sims}. The maps have been normalised to the maximum antenna temperature in the centre of the beam and are represented in logarithmic scale (dB units).
These maps clearly show that, while all these sources allow to neatly map the main beam, only Venus in W-band has sufficient sensitivity to allow mapping the two first sidelobes. The Moon is much brighter but in this case the sidelobes are overshadowed not by the noise but by the convolution between the beam and the Moon brightness distribution.

\begin{figure}
\centering
\includegraphics[width=15cm]{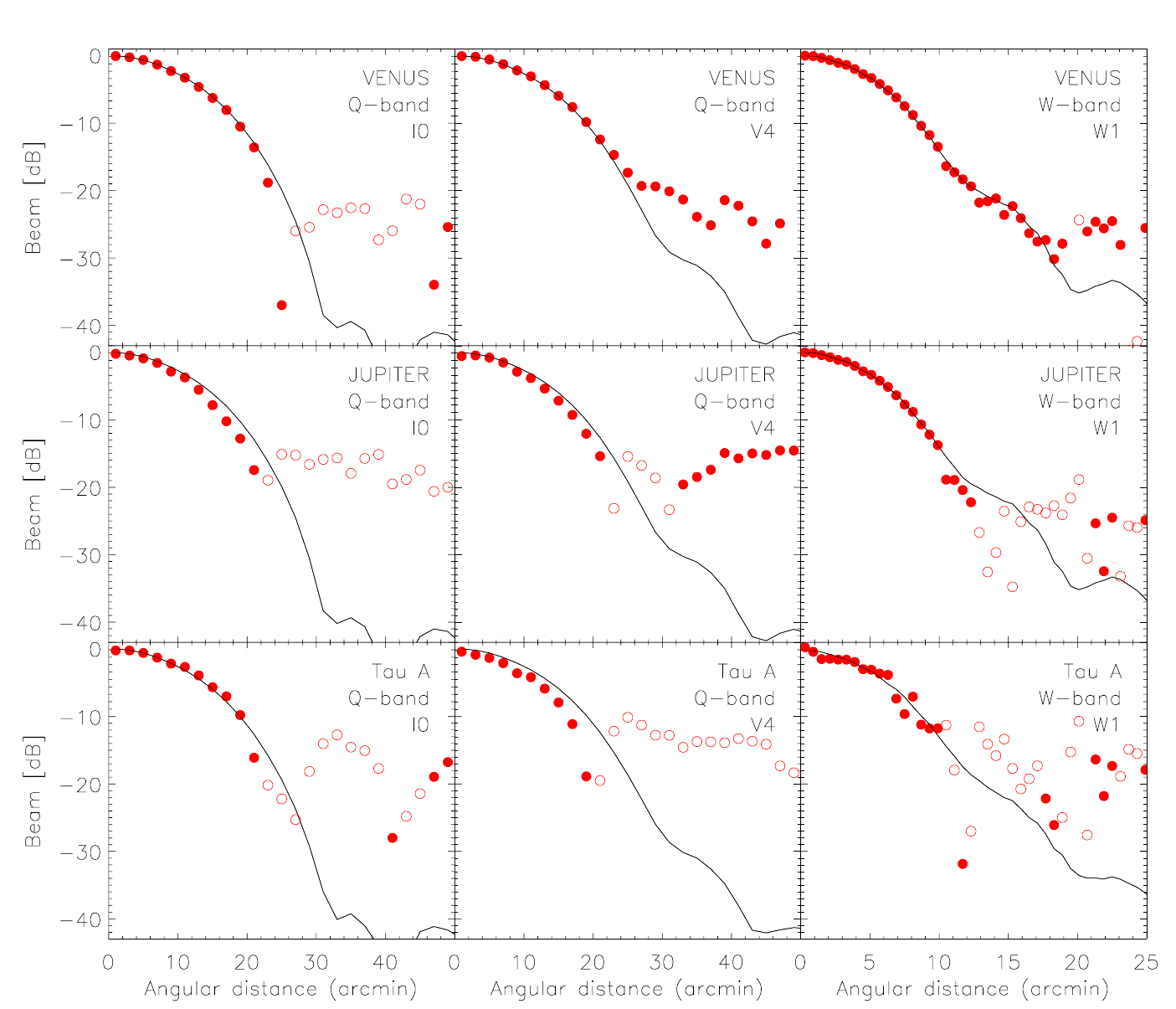}
\includegraphics[width=15cm]{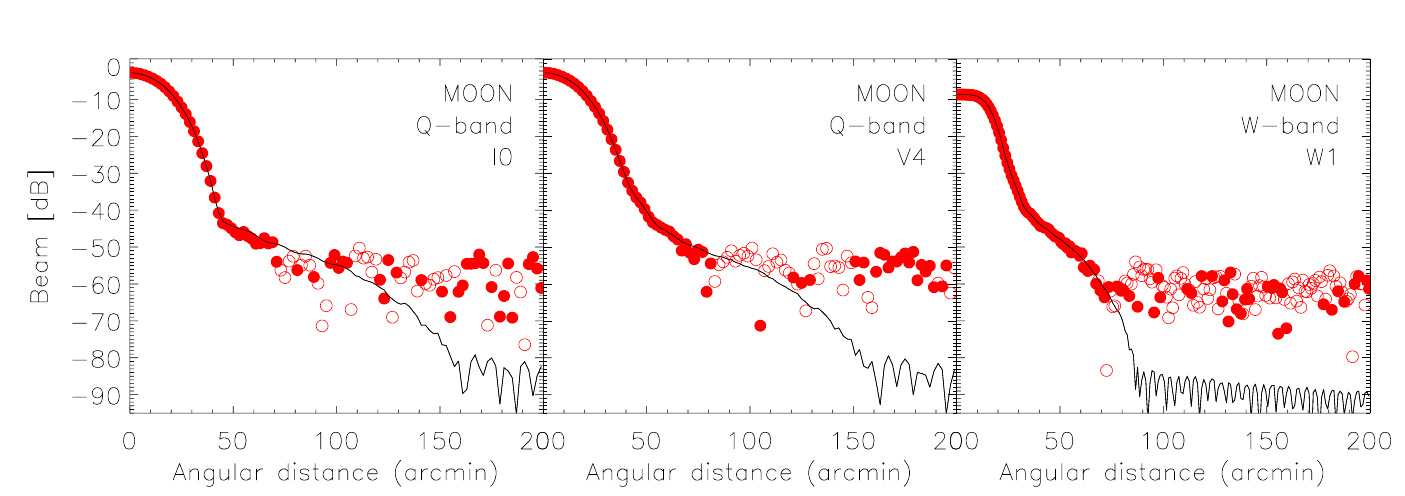}
\caption{Strip beam azimutally-averaged radial profiles derived from the GRASP optical simulation convolved with each source's profile (solid black curves) and from one simulated on-sky observation of one day in nominal mode of bright sources including realistic white and 1/f noise (red circles; filled and open correspond respectively to positive and negative values). Columns left to right show respectively pixels I0, V4 and W1. Rows from top to bottom correspond respectively to observations of Venus, Jupiter, \crab and the Moon.}
\label{fig:beams_sims_1d_rp}
\end{figure}

\begin{figure}
\centering
\includegraphics[width=15cm]{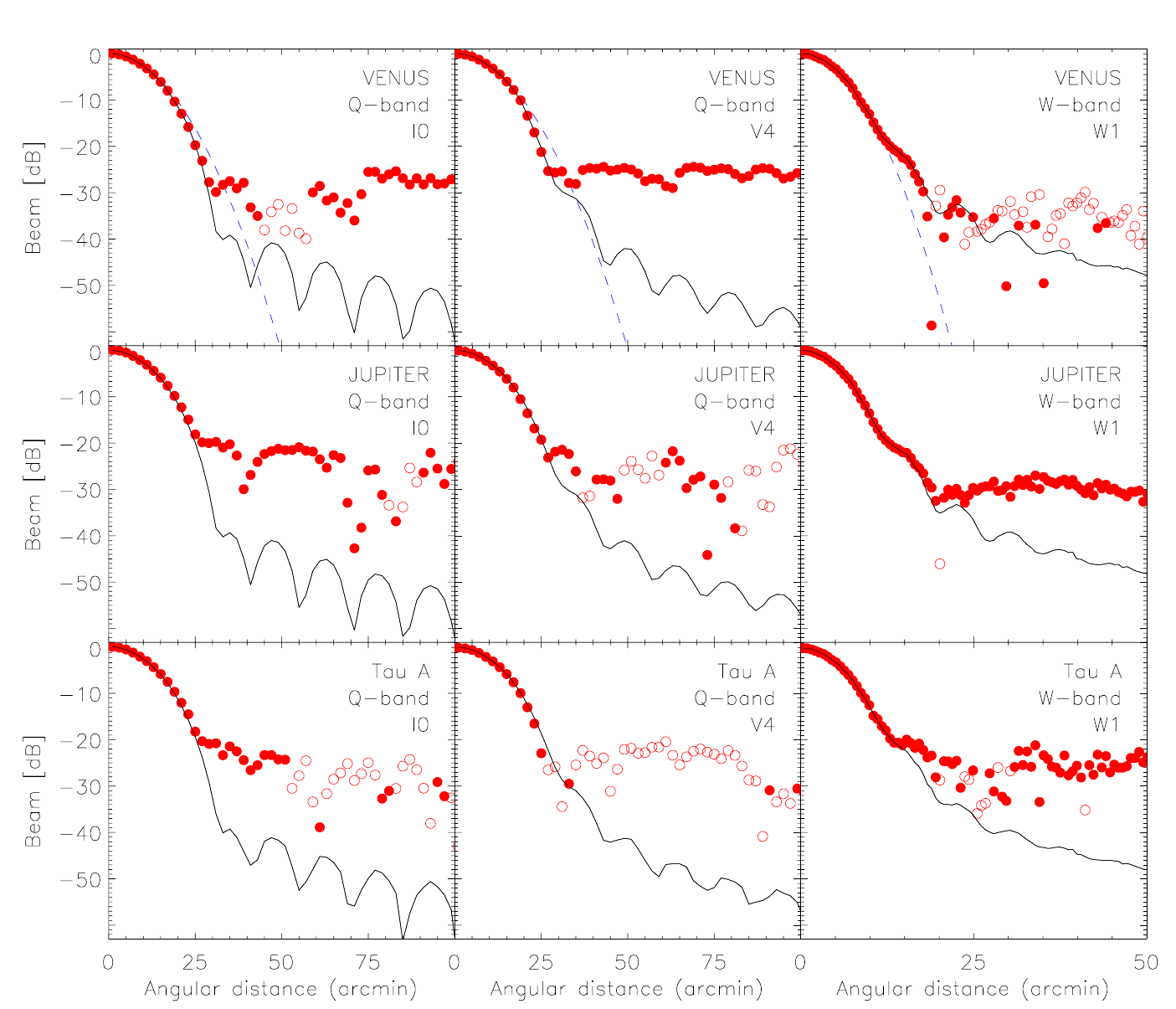}
\includegraphics[width=15cm]{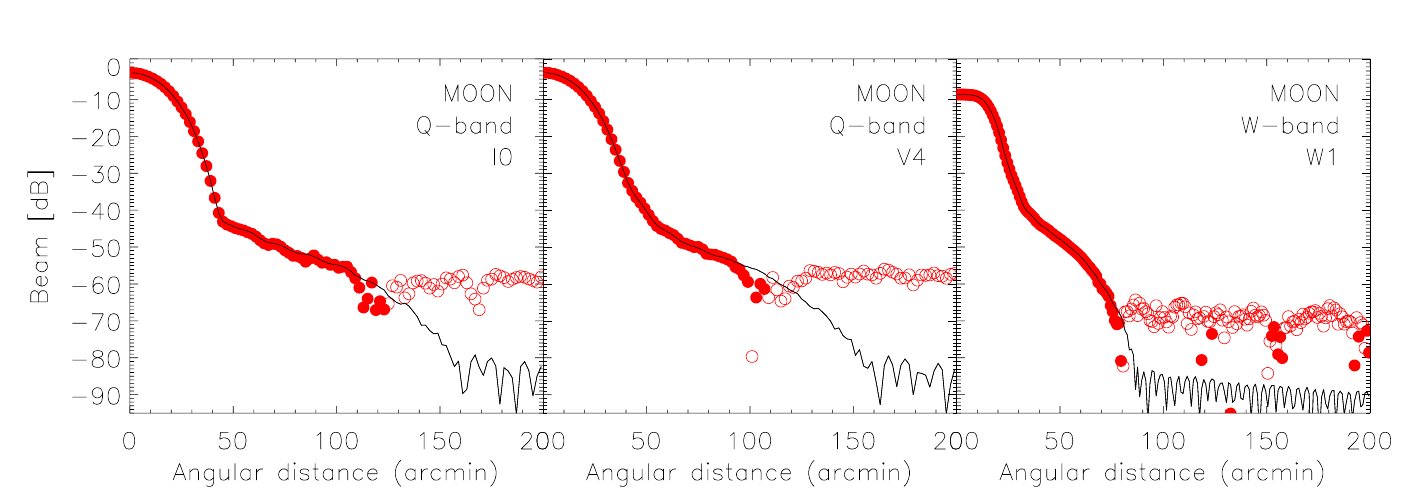}
\caption{Same as Figure~\ref{fig:beams_sims_1d_rp} but for the 2-year survey in nominal mode. For reference, we also plot (dashed blue lines in the top row) the radial profile of an ideal Gaussian beam with the same FWHM as that of the real beam.}
\label{fig:beams_sims_2y_rp}
\end{figure}

A more clear idea of the precision on mapping the beam allowed by these observations is provided by the azimuthally-averaged radial profiles of these maps. These radial profiles are represented in Figure~\ref{fig:beams_sims_1d_rp}, for a one-day observation in nominal mode, and in Figure~\ref{fig:beams_sims_2y_rp} for the stack of 2 years of data in nominal mode. The dark solid line represents the noiseless GRASP radial profile, while the red points represent the radial profiles that would be measured on the simulated noisy maps (for one day and for two years of data). These radial profiles have been normalised to the maximum of an the ideal antenna temperature beam-convolved map assuming a point source (i.e. assuming that all the source flux is introduced through the centre of the beam), which would give the maximum gain of the beam. For Venus and Jupiter, which are effectively point sources, the maximum of the normalised profile is 1 (0~dB). In \crab the effect of beam convolution is negligible in Q-band, while in W-band is barely visible and the maximum of the beam is at $-0.2$~dB. In the case of the Moon, which is totally resolved, the maximum of the radial profile is given by the ratio between the solid angle of the beam divided by the solid angle of the source, $\Omega_{\rm A}/\Omega_{\rm S}$, which is $-2.8$~dB in Q-band and $-8.7$~dB in W-band. 

In these plots, the departure of the measured data (red points) from the ideal noiseless beams (black curves) delimit the region where the noise starts to dominate. This shows that with a one-day observation sources like Venus, Jupiter or \crab allow to at best measure the beam down to $\approx -20$~dB in Q-band. This corresponds to around $2.5$ times the beam FWHM, so these observations are sufficient to provide a good characterisation of the main beam, but not of the first sidelobe that is located below $-30$~dB. In W-band, with Venus and Jupiter it is possible to reach around $-30$~dB or slightly below. The Moon, thanks to being brighter, allows reaching around $-50$~dB and $-60$~dB respectively in Q- and W-bands. Of course full-survey data (2 years with observing efficiency of 50\%) lead to improving these values although not dramatically due to the presence of considerable 1/f noise. Venus and Jupiter allow in this case reaching around $-35$~dB in W-band, and this allows measuring the first sidelobes, as it was seen in Figure~\ref{fig:strip_beam_sim_venus}. In this case the Moon allows reaching $\approx -60$~dB and $\approx -70$~dB in Q- and W-band respectively.

The bottom panels of Figures~\ref{fig:strip_beam_sim_venus}, \ref{fig:strip_beam_sim_crab} and \ref{fig:strip_beam_sim_moon} show one individual noise realisation that is added to the beam-convolved antenna temperature map. In order to assess the accuracy with which beam parameters can be recovered we have repeated 100 noise realisations, and for each one we calculate:  i) the beam FWHM, by measuring the position of the half power of the normalised beam 1D symmetrized profile, and ii) the solid angle inside the main beam, through integration on the normalised map out to radii 42.5' and 19.4' respectively for Q- and W-band, which corresponds approximately to two times the FWHM. Table~\ref{tab:beam_params} shows the scatter of the recovered FWHMs and main-beam solid angle, $\Omega_{\rm MB}$, when we add the white plus 1/f noise corresponding to a one day observation and to the stack of two years of data. We can see that both Venus and Jupiter allow measuring the beam FWHM with percent-accuracy or better. Due to being fainter, especially in W-band, \crab produces noisier measurements. We can also see that the measurement of the main-beam solid angle is intrinsically noisier as it entails integration out to the outer regions of the beam that are noise-dominated. We have not considered the Moon for this analysis because it is too extended to provide reliable estimates neither of the FWMH nor of $\Omega_{\rm MB}$.

\begin{table*}
\caption{Expected errors in the measurement of different beam parameters from a one-day observation (top) and from the two-year stacked maps (bottom) on the positions of Venus, Jupiter and Tau A, and for Q-band pixel I0 (left) and W-band pixel W1 (right). The numbers inside brackets show the relative (percent) error on the measurement of the beam FWHM and of the solid angle enclosed in the main beam.}
\smallskip
\label{tab:beam_params}
\centering
\begin{tabular}{@{}lccccccc}
\hline
\smallskip
 &	\multicolumn{3}{c}{Q-band (43~GHz)} &~& \multicolumn{3}{c}{W-band (95~GHz)}\\
\cline{2-4}\cline{6-8}
\smallskip
 & Venus & Jupiter & Tau A && Venus & Jupiter & Tau A \\
\hline
\smallskip
& \multicolumn{7}{c}{One day (1d)}\\
\cline{2-8}
FWHM [arcmin] 			& 0.29 (1.4)  &  1.22 (5.7)   &  1.56 (7.3) &&    0.07 (0.7) &	0.14 (1.4) & 2.78 (29) \\    
$\Omega_{\rm MB}$ [$\mu$sr]	& 2.93 (7.0)  &  11.1 (27)    &  13.8 (33)  &&    0.11 (1.2) &	0.26 (3.0) & 2.34 (27) \\    
$\Delta$AZ [arcmin]             & 0.13    	&  0.49   	  &  0.55	  &&    0.01  	     &	0.04	     & 1.99  \\
$\Delta$EL [arcmin]             & 0.12   	&  0.44   	  &  0.53	  &&    0.02         &	0.04	     & 3.29  \\
\hline
\smallskip
& \multicolumn{7}{c}{Two years (2y)}\\
\cline{2-8}
FWHM [arcmin] 			& 0.12 (0.6)  &  0.31 (1.5)   &  0.27 (1.3)   &&    0.01 (0.1) & 0.02 (0.2) & 0.13 (1.3) \\    
$\Omega_{\rm MB}$ [$\mu$sr]	& 1.08 (2.6)  &  2.97 (7.1)    &  2.43 (5.8)  &&    0.04 (0.4) & 0.08 (0.9) & 0.44 (5.0) \\    
$\Delta$AZ [arcmin]             & 0.04    	&  0.13   	  &  0.12     &&    0.003      & 0.007	     & 0.04  \\
$\Delta$EL [arcmin]             & 0.04   	&  0.13   	  &  0.11     &&    0.004      & 0.007	     & 0.04  \\
\hline

\end{tabular}
\end{table*}

\subsubsection{Far sidelobes}

\begin{figure}
\centering
\includegraphics[width=7.5cm]{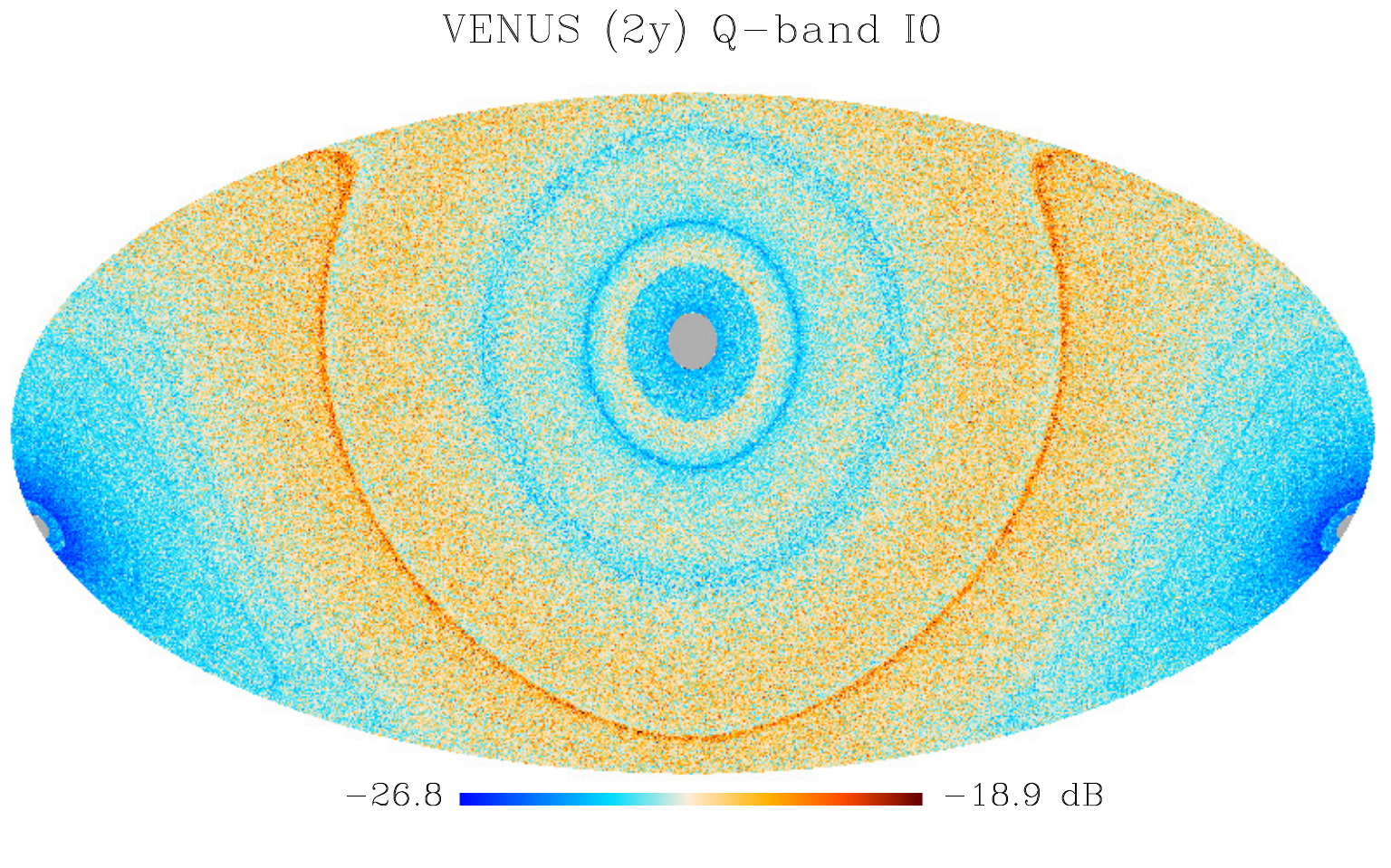}
\includegraphics[width=7.5cm]{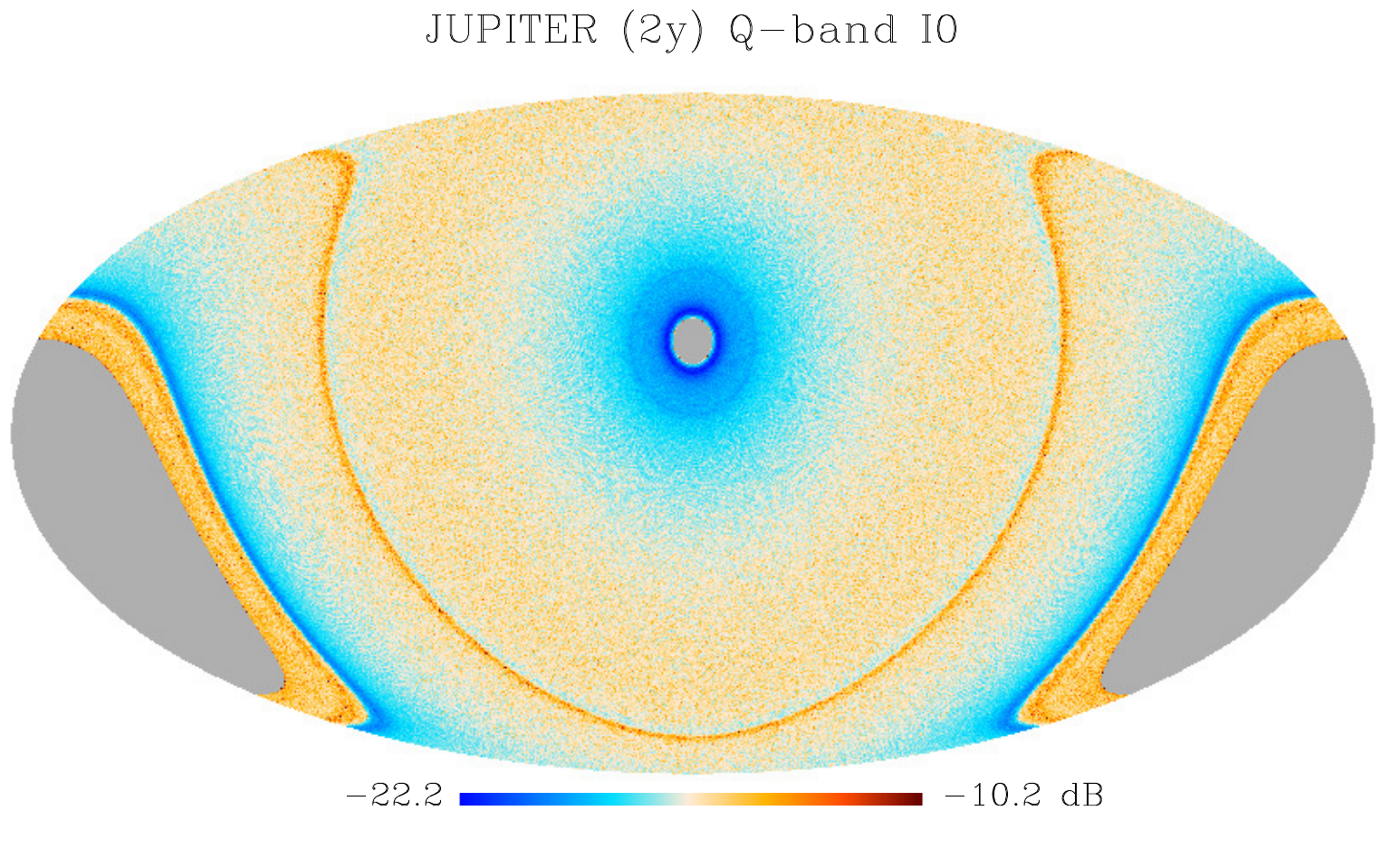}
\includegraphics[width=7.5cm]{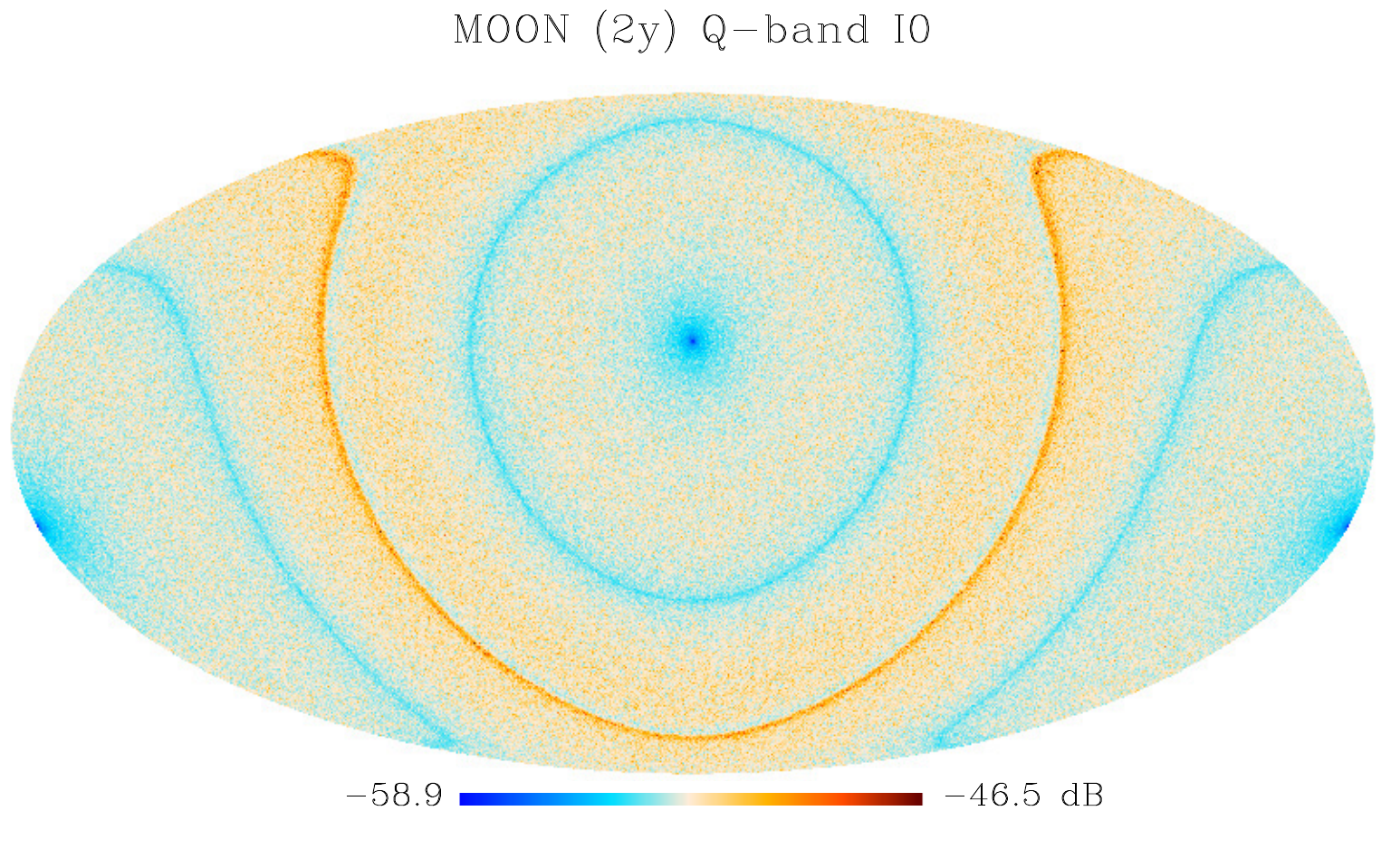}
\includegraphics[width=7.5cm]{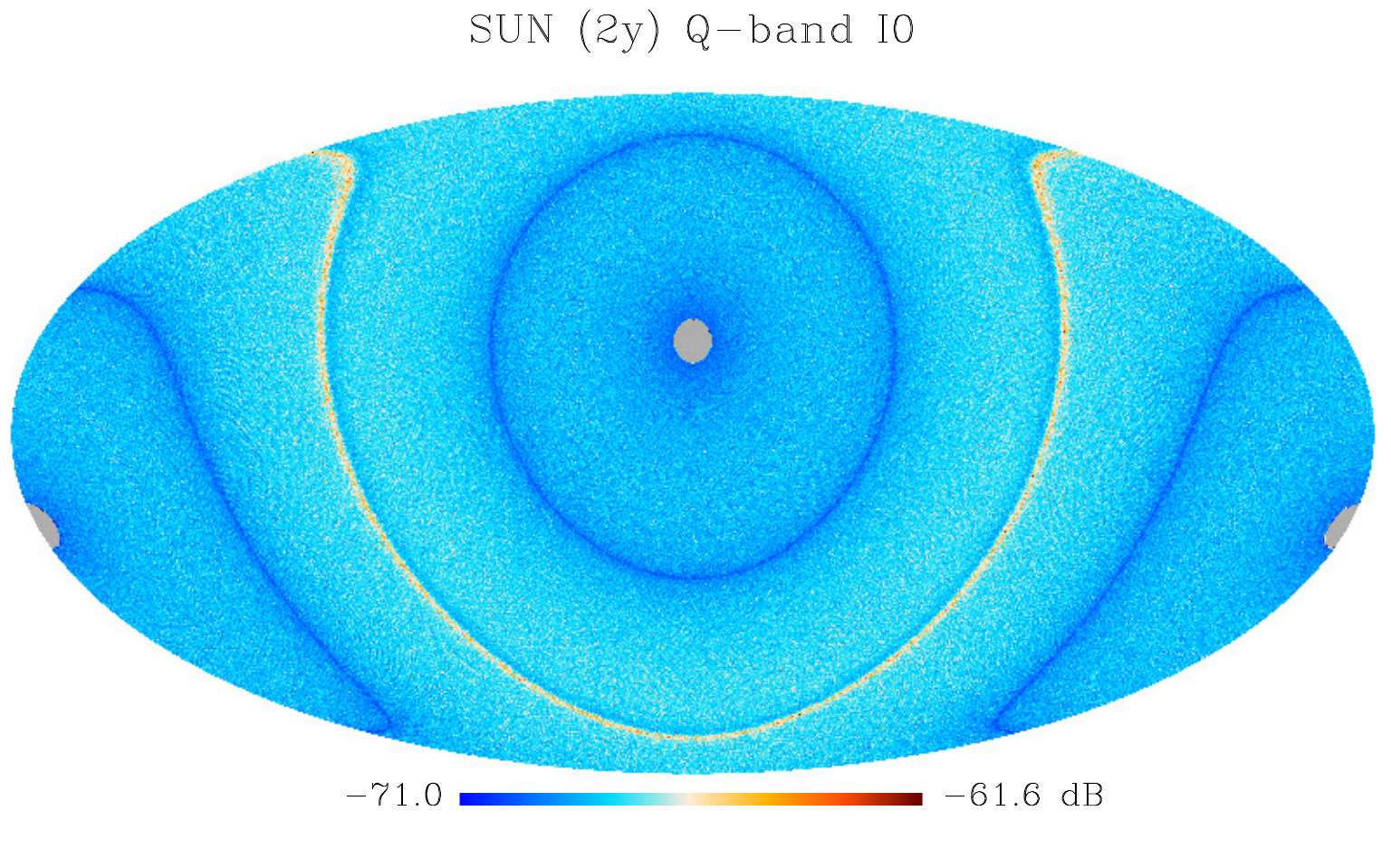}
\caption{Expected sensitivity (per pixel of size 6.9') in the measurement of the beam from the two-year Q-band stacked maps centred on the positions of Venus (top left), Jupiter (top right), the Moon (bottom left) and the Sun (bottom right), for Q-band pixel I0. The position of the main beam is (0,0), right on the centre of these maps.}
\label{fig:beam_sen}
\end{figure}

The GRASP beam simulations used in this work extend to just $1.7^\circ$ in Q-band and to $0.8^\circ$ in W-band (see section~\ref{sec:beam_simulations}), so they do not contain information about the far sidelobe structure. However, using the full-sky white plus 1/f noise simulations on the source-centred maps described in section~\ref{sec:maps_sims} we can estimate the sensitivity on the measurement of the beam in each pixel of the sky, $\sigma_{\rm dB}(p_i)$. Full-sky sensitivity maps in units of dB per pixel of  $N{\rm side}=512$ (pixel size $6.9'$), for the stack of two years of data, are shown in Figure~\ref{fig:beam_sen}. The value represented in each pixel $p_i$ is given by
\begin{equation}
\sigma_{\rm dB}(p_i) = 10\times {\rm log}\left(\frac{\sigma(p_i)}{{\rm max(M_{conv}})}\frac{{\rm max(M_{conv})}}{{\rm max(M)}}\right)~~,
\label{eq:sigma_db}
\end{equation}
where $\sigma(p_i)$ is the sensitivity in pixel $p_i$ of the map M$_{\rm conv}$ in units of mK$_{\rm CMB}$, M$_{\rm conv}$ represents the beam-convolved simulated noisy map, and M is the ideal beam-convolved map in which the source is assumed to be a point source, so that the maximum of the map represents the maximum gain of the beam. Equation~\ref{eq:sigma_db} then tells that the sensitivity map is normalised to this latter map. ${\rm max(M_{conv})}/{\rm max(M)}=1$ for a point source like Venus, and  ${\rm max(M_{conv})}/{\rm max(M)}=\Omega_{\rm A}/\Omega_{\rm S}$ for an extended source like the Moon. 

The maps of Figure~\ref{fig:beam_sen} show that using two-years Venus centred-maps the region around the main beam can be measured with sensitivities of $\sim -23$~dB (this is consistent with what is shown in Figure~\ref{fig:beams_sims_2y_rp}), while the far-sidelobes can be measured with sensitivities between -27~dB and -19~dB. Due to its averaged brightness being lower Jupiter centred-maps lead to a shallower mapping of the far sidelobes. Of course, due to being brighter, the Moon and the Sun allow to go deeper, in some regions down to -59~dB in the case of the Moon and to -71~dB in the case of the Sun. Table~\ref{tab:beam_sensitivities} shows a summary of these same values (main beam sensitivities and minimum and maximum far-sidelobe sensitivities) for all possible combinations of sources, bands and integration time. The 2$\pi$ beam simulations performed in \cite{realini_2022} show that the maximum power level of the beam far sidelobes is -59~dB for pixel I0 (see Figure~12 of that paper). This level can only be reached using the Moon or the Sun.

 \begin{table*}
\caption{Expected sensitivities on the measurement of the beam from a one-day observation (top) and from the two-year stacked maps (bottom) on the positions of Venus, Jupiter, Tau A, the Moon and the Sun, and for Q-band pixel I0 (left) and W-band pixel W1 (right). We list sensitivities on the positions of the main beam, and minimum and maximum sensitivities on the far-sidelobe region that these observations will allow to sample.}
\smallskip
\label{tab:beam_sensitivities}
\centering
\begin{tabular}{@{}lccccccccccc}
\hline
\smallskip
 &	\multicolumn{5}{c}{Q-band (43~GHz)} &~& \multicolumn{5}{c}{W-band (95~GHz)}\\
\cline{2-6}\cline{8-12}
\smallskip
 & \small{Venus} &  \small{Jupiter} & \small{Tau A} & \small{Moon} & \small{Sun} && \small{Venus} & \small{Jupiter} & \small{Tau A} & \small{Moon} & \small{Sun} \\
\hline
\smallskip
& \multicolumn{9}{c}{One day (1d)}\\
\cline{2-12}
Main beam    &  -19  &  -12  &  -13  &  -46  & -61 &&   -25  &  -21  &  -13  &  -54 & -69 \\
Minimum      &  -21  &  -16  &  -20  &  -52  & -64 &&   -28  &  -26  &  -20  &  -60 & -72 \\
Maximum      &  -13  &   -4.0  &   -4.6  &  -39  & -54 &&   -19  &  -13  &   -4.6  &  -47 & -62 \\
\hline
\smallskip
& \multicolumn{9}{c}{Two years (2y)}\\
\cline{2-12}
Main beam    & -23   &  -18  &  -19  &  -53 & -69 &&   -31  &  -28  &  -20  &  -63 & -79 \\
Minimum      & -27   &  -22  &  -26  &  -59 & -71 &&   -35 &  -33  &  -28  &  -69 & -82 \\
Maximum      & -19   &  -10  &  -11  &  -46 & -62 &&   -27  &  -20  &  -12  &  -56 & -72 \\
\hline
\end{tabular}
\end{table*}

\subsubsection{Beam transfer function and window function}

The beam transfer function (BTF), $B_\ell$, represents the transform of the beam in harmonic space, while the window function is the square of the BTF, $W_\ell=B_\ell^2$ and represents the ``filtering'' function by which the underlying power spectrum in harmonic space is multiplied when observed by a given telescope. The BTF can be calculated from the 1D symmetrised beam profiles, $b_{\rm S}(\theta)$, represented in Figures~\ref{fig:beams_sims_1d_rp} and \ref{fig:beams_sims_2y_rp}, through equation
\begin{equation}
B_\ell = \frac{2\pi}{\Omega_{\rm A}}\int b_{\rm S}(\theta) P_\ell({\rm cos}\theta) d({\rm cos}\theta)~~,
\label{eq:btf}
\end{equation}
where $P_\ell()$ is the Legendre polynomial of order $\ell$.

\begin{figure}
\centering
\includegraphics[width=15.5cm]{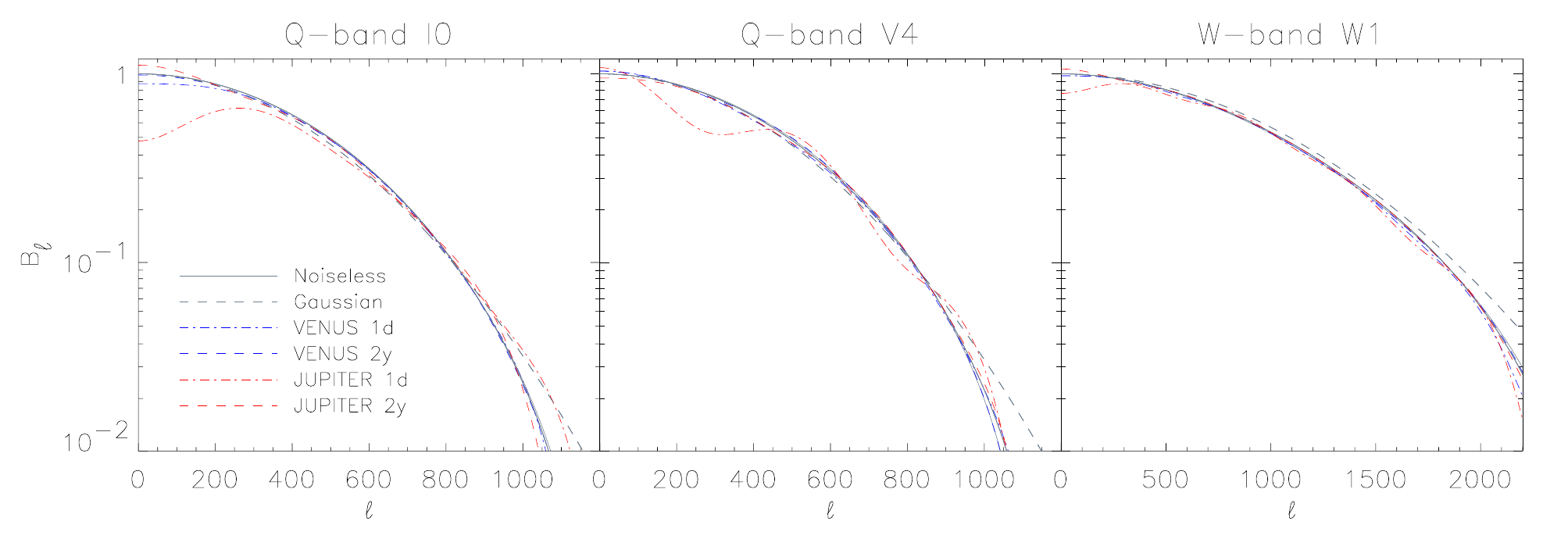}
\caption{Strip beam transfer functions for Q-band pixels I0 (left panel), V4 (centre) and for W-band pixel W1 (right). Grey solid curves show the BTFs derived from the theoretical GRASP beams (radial profiles showed in black in Figure~\ref{fig:beams_sims_1d_rp}), while the grey dashed curves show the analytical BTFs for an ideal 
Gaussian beam with the same FWHM as that of the real beam. Different coloured curves show the BTFs derived from simulated maps using one day and two years of data in nominal mode on the positions of Venus and Jupiter.}
\label{fig:beams_sims_wf}
\end{figure}

Figure~\ref{fig:beams_sims_wf} shows the BTFs for pixels I0, V4 and W1 computed, using equation~\ref{eq:btf}, from the noiseless GRASP beams in comparison with those calculated from the noisy one-day and two-year maps of Venus and Jupiter. These plots show that even one day of data allow for a very precise measurement of the BTF out to multipoles of $\ell\sim 1000$ in Q-band and of $\sim 2000$ in W-band. 

\subsection{Pointing model}

In any real telescope the transformation between azimuth and elevation coordinates read by the telescope encoders and true azimuth and elevation coordinates on the sky is affected by non-idealities and torsions in the telescope structure. This transformation, that depends on a set of $\sim 10$ parameters that account for those non-idealities, must be precisely and properly calibrated so to minimise the pointing errors in the final maps.  A star tracker is specifically being developed for Strip to serve this purpose (Maris et al. in prep.). In parallel, a system based on an emitter mounted on a drone (Virone et al. in prep.) is also being developed for beam characterisation and pointing-model reconstruction. However, the same Strip observations of bright calibrators, ideally covering a wide range of elevations, can also be used to aid calibration of the telescope pointing model. This would require the determination of the observed (AZ,EL) coordinates of these sources in real data, that will latter be used either as input to fit the parameters of the pointing model, or to characterise the accuracy with which the pointing model has been fitted with other methods (eg. star tracker or drone-based observations) or by using other calibration sources. It is then necessary to assess with what precision the (AZ,EL) positions of those sources can be measured using real data. We have then used the same noise simulations that were used in section~\ref{subsec:mb} to study the precision on the recovery of FWHM and $\Omega_{\rm A}$ to study the precision on the measurement of the source central coordinates. To this aim, in each simulation we have fitted a Gaussian beam on the source position and measured the differences $\Delta$AZ and $\Delta$EL with respect to the input coordinates. Table~\ref{tab:beam_params} shows that the scatter of these two quantities in 100 simulations is well below the 1-arcmin level even when the noise of a one-day observation in nominal mode is considered. In particular, Venus allows reaching $\sim 0.1'$ in Q-band and $\sim 0.02'$ in W-band. These values are below 0.5\% of the beam FWHM and can be regarded as sufficiently precise. Using full-survey data allows reaching $\sim$ arcsecond precision in Q-band and sub-arsecond precision in W-band. 

\section{Conclusions}\label{sec:conclusions}
This paper has presented a forecast study of the prospects for on-sky calibration of the LSPE-Strip instrument. Calibration of the different instrument parameters is a crucial step, if not the most crucial step, in the data processing pipeline. This study is based on noise simulations that consider the presence of both white and $1/f$ noise components, which are characterised by the noise equivalent temperature of each detector and by its knee frequency. While the former is relatively well known, we have used approximated values for the latter that must be confirmed using real-data analyses. We also realistically simulate the scanning strategy of the telescope, assuming a nominal observing mode consisting of constant-elevation scanning of the sky with elevation $70^\circ$ from the horizontal. We have also considered the possibility of deeper raster-scan observations on specific regions of the sky. As sky calibrators we have used the brightest compact sources accessible by Strip, namely Tau\,A, Cas\,A, Cyg\,A, W49, W51, Venus and Jupiter. Although not compact, we have also considered the Moon and the Sun as calibrators for the far-sidelobe structure of the beam patterns. We have also used as inputs GRASP simulations of the beam radiation pattern around the main beam. 

We find that using full-survey data (2 years of observations with efficiency of 50\%) the absolute gain of each detector can be calibrated with precision of 0.9\% in Q-band and of 2.1\% in W-band using Tau A, while Venus would allow to reach 0.4\% and 0.14\% respectively. These values are well below the typical uncertainty of the calibration models, which is of the order of 5\% and in fact sets this value as the minimum global calibration uncertainty achievable from ground-based observations that, owing to the presence of large angular scale atmospheric fluctuations, can not measure the CMB dipole that is the default calibrator of space CMB missions. Improving the global calibration uncertainty below 5\% is then difficult through on-sky calibration, and alternative artificial calibration methods should be envisaged to overcome this limitation. Short time-scale gain fluctuations produced by $1/f$ noise are important to be corrected. In this case, Tau A daily observations allow reaching precisions of 8.1\% and 11.1\% respectively in Q- and W-bands, while Venus and Jupiter typically lead to $\sim 5\%$ and below the 1\% level at their closest approach to Earth. Given these precisions, and the fact that sky sources are visible only during a given time interval every day, we conclude that relative gain calibration should better rely on the internal gain calibration system that is implemented in Strip. Being a polarimeter that targets the measurements of the E- and B-modes from the Galactic foregrounds and from the CMB, measurement of the polarisation direction is equally important. Using full survey data we find that the polarisation direction can be calibrated on Tau A with precision of $0.7^\circ$ in Q-band and of $1.3^\circ$ in W-band. These values are close to the best uncertainty of the model that is $0.3^\circ$. On the other hand daily observations lead to much worse sensitivities so it will be difficult to test the hypothesis of the non-variability of the position angle of the polarimeters unless periods of $\sim 15$ days of data are averaged. Being the brightest polarised compact source in the microwave regime Tau A is also considered as the main calibrator for the polarisation efficiency. We find that this can be calibrated to a good precision of $0.1\%$ and of $0.3\%$ using full-survey data in Q- and W-bands respectively. The intensity-to-polarisation leakage should rely on unpolarised sources. This could be measured with precision of around 1\% using W49, and well below the percent level using Venus.

Beam calibration is one of the most important steps in the calibration process, as the beam patterns determine how the calibration scale is transferred between different angular scales. Also, characterisation of the sidelobe structure is of paramount importance to determine how stray-light will affect the final maps, or to correct for it. We find that on-sky observations of bright calibrations are able to provide a good characterisation of the main beam.  However, measuring the sidelobes, which are typically below $-40$\,dB, is more challenging. Main beam FWHMs can be measured using one-day observations in nominal mode with accuracy of  $\sim 1$\,arcmin in Q-band and, using Jupiter and Venus, with accuracy of $\sim 0.1$\,arcmin in W-band. Using full-survey data would would lead to $\sim 0.1$\,arcmin and $\sim 0.01$\,arcmin in Q- and W-bands respectively. While daily observations would allow to map the beam only down to around $-20$\,dB, full-survey data would allow to reach around $-30$\,dB and, in the case of W-band, Venus-centred maps would allow reaching the $-40$\,dB and then measuring the first sidelobe. The Moon and the Sun have the obvious drawback of being resolved, leading to a distortion of the beam structure. They are useful however to study the far sidelobes. We find that full-survey Moon-centred or Sun-centred maps would allow to measure the far sidelobes with precisions typically between $-60$ and $-80$\,dB depending on the region of the sky and on the considered band. Finally, we have also considered the reliability of the information that can be extracted from these observations to calibrate the pointing model of the telescope. We find that using daily observations the position of Venus can be measured with accuracy of $\sim 0.1'$ in Q-band and of $\sim 0.02'$ in W-band. Full-survey data on the other hand would allow reaching $\sim$ arcsecond precision in Q-band and sub-arcsecond precision in W-band.

\acknowledgments
The LSPE-Strip instrument has been developed thanks to the support of ASI contract I/022/11/1 and Agreement 2018-21-HH.0 and to funding from INFN (Italy). We acknowledge the usage of the HEALPix \cite{healpix}. RTGS, CLC and JARM acknowledge financial support from the Spanish Ministry of Science and Innovation (MICINN) under the project PID2020-120514GB-I00, and from the ACIISI, Consejer\'{\i}a de Econom\'{\i}a, Conocimiento y Empleo del Gobierno de Canarias and the European Regional Development Fund (ERDF) under grant with reference ProID2020010108.

\newpage
\appendix

\section{Optimal weighting of planet maps}\label{ap:weighting}

The temperature of each pixel $j$ of the map is calculated through the weighted mean of all $i$ measurements with the telescope having coordinates lying on that pixel:
\begin{equation}
m_j = \frac{\sum_i w_i T_{{\rm A},i}}{\sum_i w_i}= \frac{\sum_i w_i T_{\rm B}\Omega_{{\rm p},i}/\Omega_{\rm A}}{\sum_i w_i}
\end{equation}
where $T_{{\rm A},i}$ is the antenna temperature corresponding to sample $i$, and $w_i$ is a weight associated with that measurement. The error of $m_j$ is then
\begin{equation}
\sigma(m_j) = \frac{1}{(\sum_i w_i)^{1/2}}
\end{equation}
As it is well known, $w_i'=1/\sigma_i^2$ with $\sigma_i$ being the error of sample $i$, leads to minimum variance of the final map. Henceforth we call this ``standard weighting''. Given that we consider a constant time sampling rate, we can assume all samples do have the same error so then $\sigma_i=\sigma=\sigma_0/\sqrt{t_s}$ where $\sigma_0$ is the NET (see Table~\ref{tab:instrument_parameters}) and $t_s$ is the sampling rate of the telescope (we consider $t_s=1\,$s). In this case, using $\Omega_{{\rm p},i}=\pi(r_{\rm p}/d_{{\rm p},i})^2$, we can write
\begin{equation}
m_j' = \frac{T_{\rm B}\pi r_{\rm p}^2}{\Omega_{\rm A}N}\sum_i\frac{1}{d_{{\rm p},i}^2}=\frac{T_{\rm B}\pi}{\Omega_{\rm A}}\left(\frac{r_{\rm p}}{d_{\rm eff}'}\right)^2~~,
\end{equation}
where $N$ is the number of samples lying in that pixel and we have defined an ``effective distance'' as 
\begin{equation}
d_{\rm eff}'=\left( \frac{N}{\sum_i\frac{1}{d_{{\rm p},i}^2}}\right)^{1/2}~~.
\label{eq:deff_std}
\end{equation}
Combining the previous equations we can write the signal-to-noise of pixel $j$ as:
\begin{equation}
{\rm SNR_j'}= \frac{T_{\rm B}\pi r_{\rm p}^2}{\sigma\sqrt{N}\Omega_{\rm A}}\sum_i \frac{1}{d_{\rm{p},i}^2} = \frac{T_{\rm B}\pi r_{\rm p}^2\sqrt{N}}{\sigma\Omega_{\rm A} (d_{\rm eff}')^2}~~.
\label{eq:snr_std}
\end{equation}

\begin{figure}
\centering
\includegraphics[trim= 0mm 25mm 0mm 5mm, width=7.5cm]{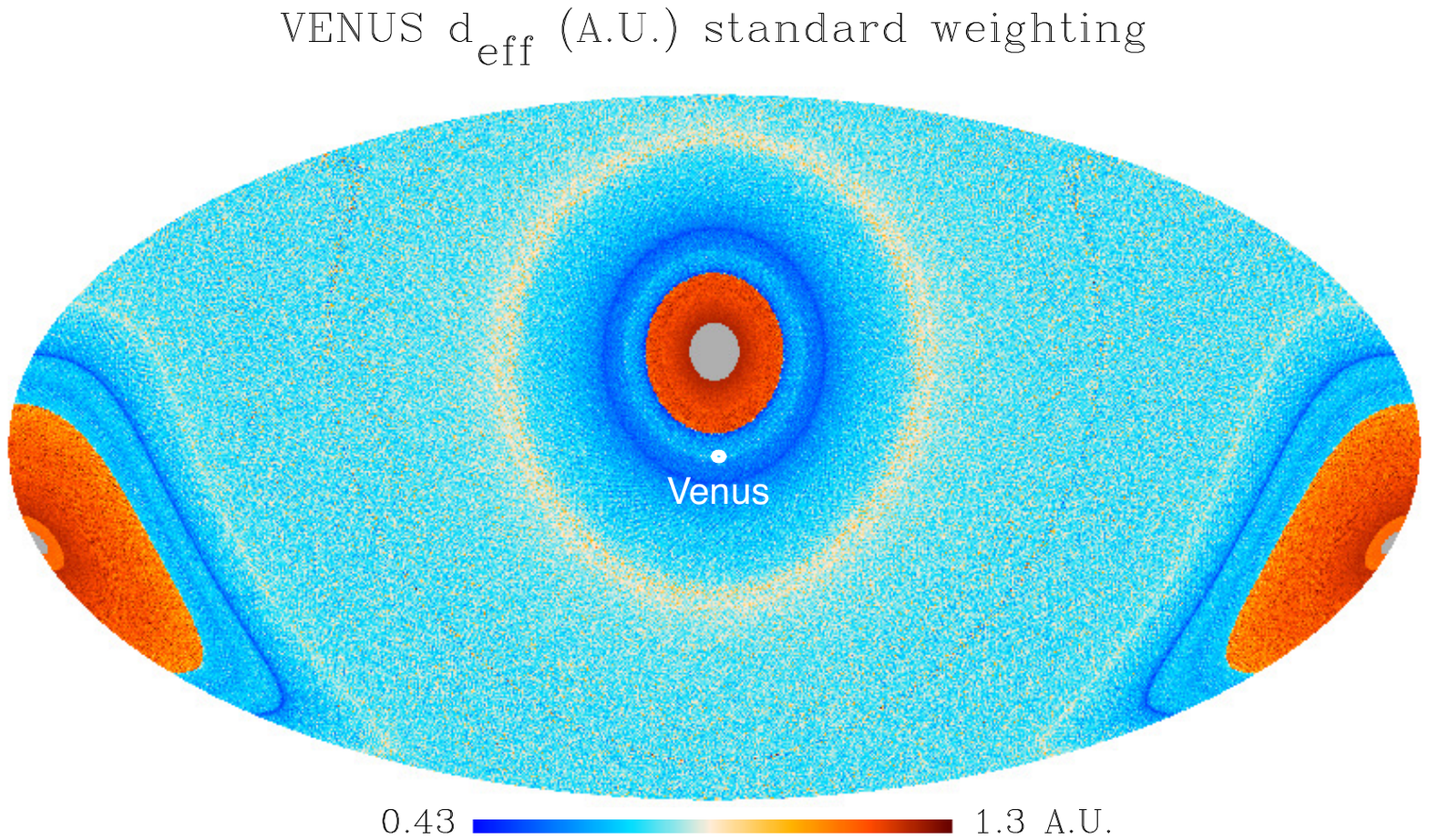}
\includegraphics[trim= 0mm 25mm 0mm 5mm, width=7.5cm]{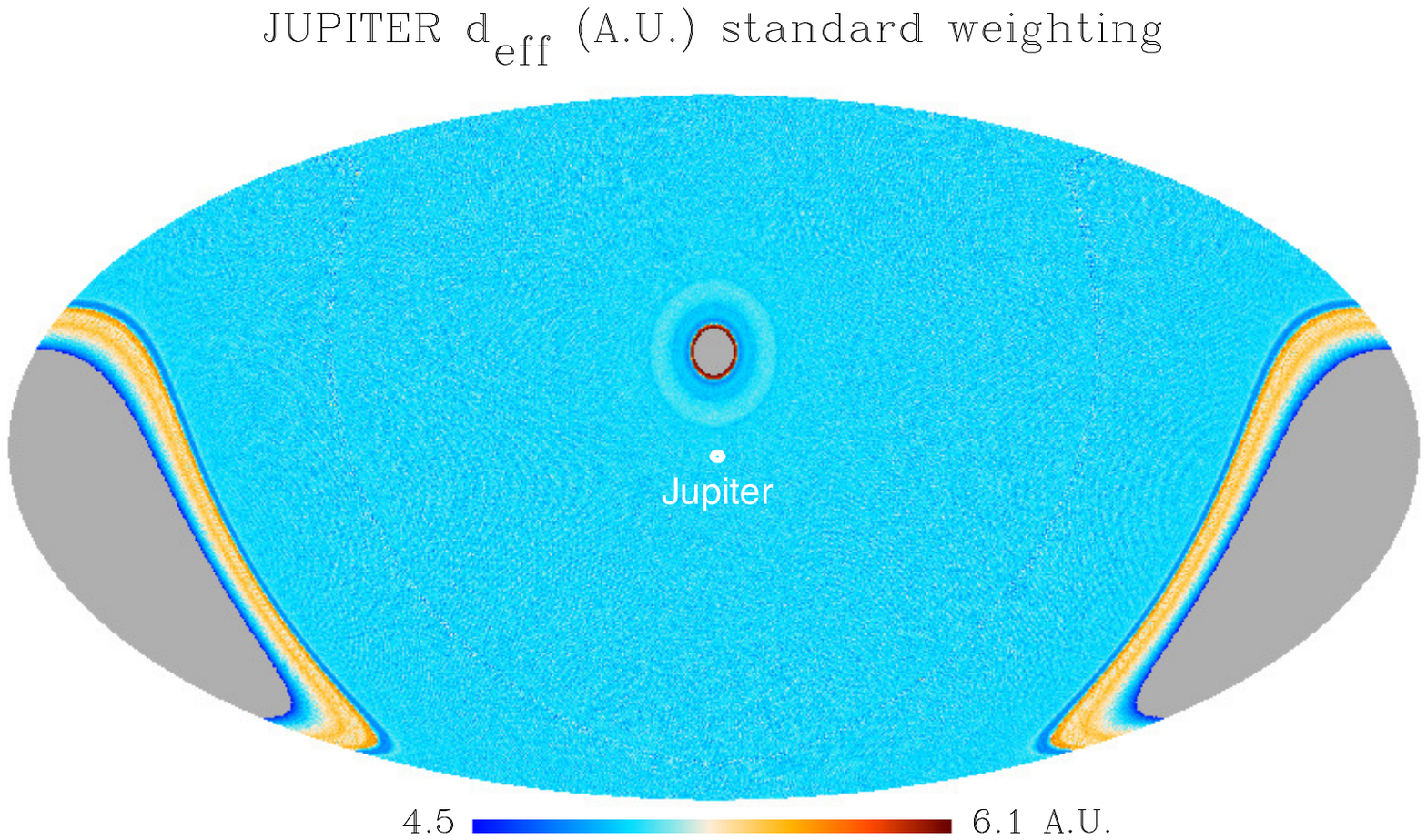}
\caption{Same as Figure~\ref{fig:dist_opt_planets} but using standard weighting.}
\label{fig:dist_std_planets}
\end{figure}

\begin{figure}
\centering
\includegraphics[trim= 0mm 25mm 0mm 5mm, width=7.5cm]{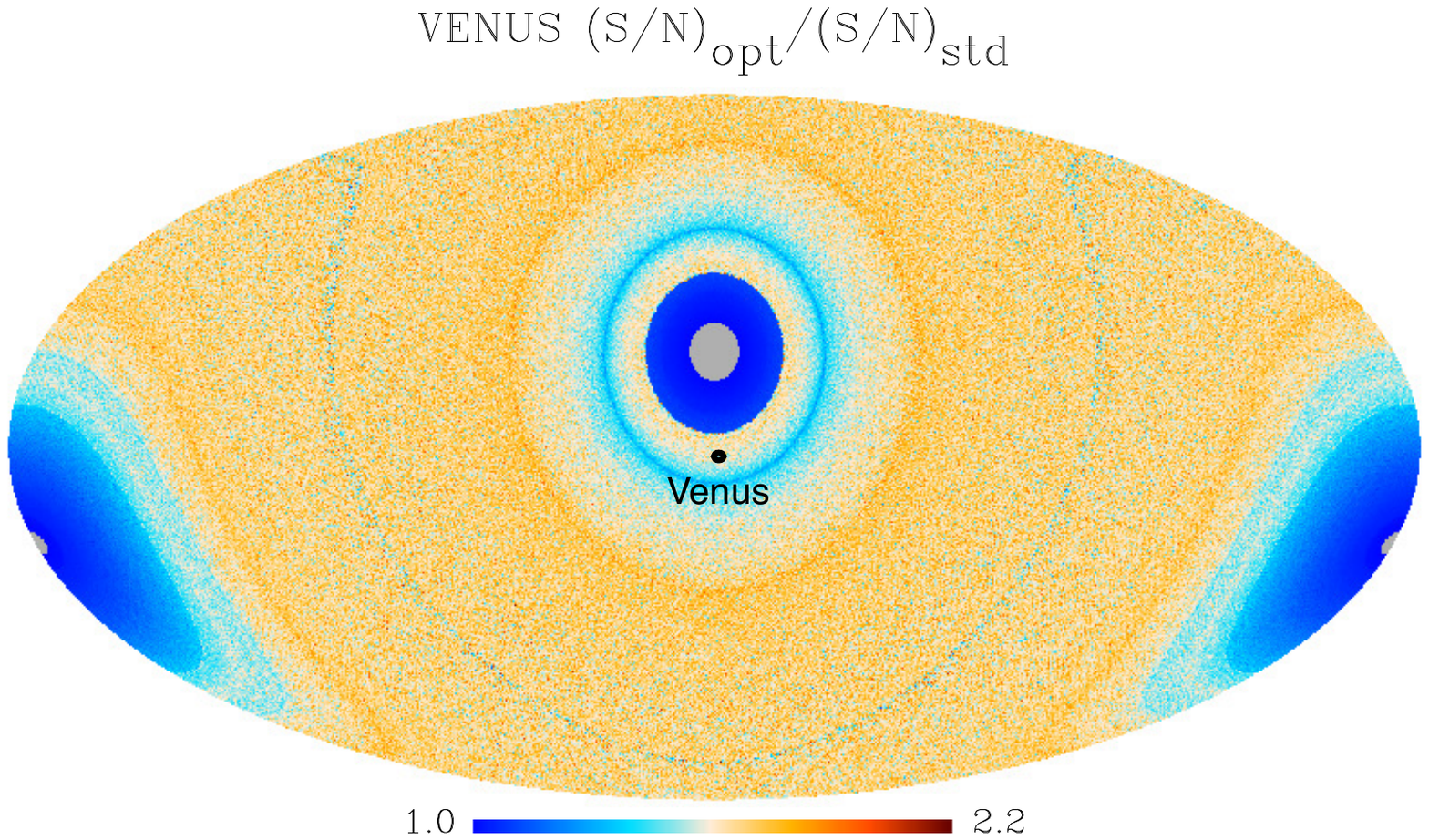}
\includegraphics[trim= 0mm 25mm 0mm 5mm, width=7.5cm]{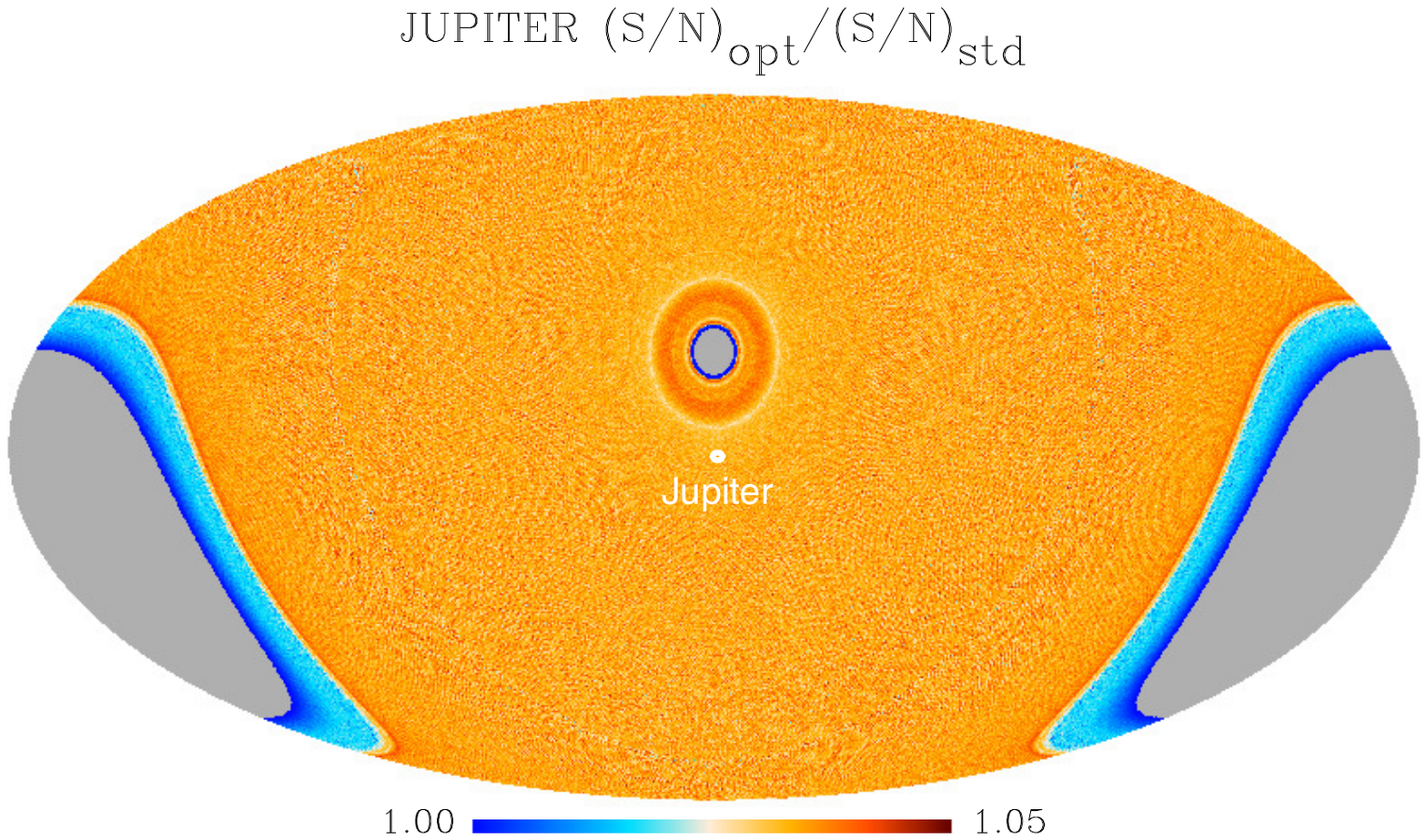}
\caption{Ratios of signal-to-noise with optimal weighting over signal-to-noise with standard weighting for Venus (left) and Jupiter (right).}
\label{fig:snr_planets}
\end{figure}

This standard weighting minimises the variance, which implies that it would maximise the signal-to-noise when the signal is constant. However, when we deal with a variable signal as that of a planet, the maximisation of the signal-to-noise requires applying weights that account for the signal-to-noise instead of for the noise only. In this case we ought use $w_i=1/(\sigma_i d_{\rm{p},i})^2$, which we call ``optimal weighting'', and then if $\sigma_i=\sigma$
\begin{equation}
m_j =  \frac{T_{\rm B}\pi r_{\rm p}^2}{\Omega_{\rm A}}\frac{\sum_i\frac{1}{d_{{\rm p},i}^4}}{\sum_i\frac{1}{d_{{\rm p},i}^2}}  =\frac{T_{\rm B}\pi}{\Omega_{\rm A}}\left(\frac{r_{\rm p}}{d_{\rm eff}}\right)^2
\end{equation}
and in this case the ``effective distance'' is defined as
\begin{equation}
d_{\rm eff}=\left(\frac{\frac{1}{N}\sum_i\frac{1}{d_{{\rm p},i}^2}}{\frac{1}{N}\sum_i\frac{1}{d_{\rm{p},i}^4}}\right)^{1/2}~~.
\label{eq:deff_opt}
\end{equation}
The signal-to-noise of pixel $j$ in this case is
\begin{equation}
{\rm SNR_j}=\frac{T_b\pi r_{\rm p}^2}{\sigma\Omega_{\rm A}}\left(\sum_i\frac{1}{d_{{\rm p},i}^4}\right)^{1/2}
\label{eq:snr_opt}
\end{equation}
Combining equations~\ref{eq:deff_std}, \ref{eq:snr_std}, \ref{eq:deff_opt} and \ref{eq:snr_opt} we can write the ratio of signal-to-noise ratios as
\begin{equation}
\frac{\rm SNR}{\rm SNR'} = \frac{d_{\rm eff}'}{d_{\rm eff}}~~.
\label{eq:snr_ratios}
\end{equation}
Throughout this paper we use this last optimal weighting, but here we present a brief comparison between the two cases. In Figure~\ref{fig:dist_opt_planets} we presented maps of the effective distance of each pixel for this case, calculated using equation~\ref{eq:deff_opt}, for Venus (left) and Jupiter (right), and for a 2-year survey starting on 1 January 2024. In Figure~\ref{fig:dist_std_planets} we show the maps of effective distance for the standard weighting, obtained using equation~\ref{eq:deff_std}. As expected, both for Venus and for Jupiter the effective distances calculated through the optimal weighting are smaller, leading to stronger signal of the planet in the final weighted maps. Also expected was that for Venus the differences between the two cases are more noticeable as this planet presents a larger variation of its distance to Earth. A more quantitative comparison is provided by the ratio of signal-to-noises in the two cases, computed through equation~\ref{eq:snr_ratios} that is shown in Figure~\ref{fig:snr_planets}. This shows that while in Jupiter differences are barely appreciable (average value $1.03$ across the map), in the case of Venus the improvement of the signal-to-noise is typically a factor $1.6$ (this is the average value across the map, and is also the value at the position of the main beam, towards the centre of the map), and reaching a maximum of $1.23$.  Note on the contrary the blue regions in the Venus map, where the improvement is modest because when those pixels are observed the planet is always very far, at around $1.3\,$A.U. (see Figures~\ref{fig:dist_opt_planets} and \ref{fig:dist_std_planets}).

\begin{thebibliography}{99}

\bibitem{smoot_1992} Smoot, G.~F., Bennett, C.~L., Kogut, A., et al.\ 1992, \apjl, 396, L1. doi:10.1086/186504
\bibitem{debernardis_00} de Bernardis, P., Ade, P.~A.~R., Bock, J.~J., et al.\ 2000, \nat, 404, 955. doi:10.1038/35010035
\bibitem[Lee et al.(2001)]{lee_01} Lee, A.~T., et al.\ 2001, \apjl, 561, L1
\bibitem{halverson_2002} Halverson, N.~W., Leitch, E.~M., Pryke, C., et al.\ 2002, \apj, 568, 38. doi:10.1086/338879
\bibitem{dickinson_2004} Dickinson, C., Battye, R.~A., Carreira, P., et al.\ 2004, \mnras, 353, 732. doi:10.1111/j.1365-2966.2004.08206.x
\bibitem{readhead_2004} Readhead, A.~C.~S., Mason, B.~S., Contaldi, C.~R., et al.\ 2004, \apj, 609, 498. doi:10.1086/421105
\bibitem{kuo_2004} Kuo, C.~L., Ade, P.~A.~R., Bock, J.~J., et al.\ 2004, \apj, 600, 32. doi:10.1086/379783
\bibitem[Tristram et al.(2005)]{tristram_2005} Tristram, M., et al.\ 2005, \aap, 436, 785
\bibitem{bennett_2013} Bennett, C.~L., Larson, D., Weiland, J.~L., et al.\ 2013, \apjs, 208, 20. doi:10.1088/0067-0049/208/2/20
\bibitem{cpp2018-6} Planck 2018 results VI. Planck Collaboration \ 2020, \aap, 641, A6. doi:10.1051/0004-6361/201833910
\bibitem{kamionkowski_1997} Kamionkowski, M., Kosowsky, A., \& Stebbins, A.\ 1997, \prd, 55, 7368. doi:10.1103/PhysRevD.55.7368
\bibitem{seljak_1997} Seljak, U. \& Zaldarriaga, M.\ 1997, \prl, 78, 2054. doi:10.1103/PhysRevLett.78.2054
\bibitem{bicep2021-13} BICEP/Keck Collaboration \ 2021, \prl, 127, 151301. doi:10.1103/PhysRevLett.127.151301
\bibitem{cpp2015-10} Planck 2015 results X. Planck Collaboration, Adam, R., Ade, P.~A.~R., et al.\ 2016, \aap, 594, A10. doi:10.1051/0004-6361/201525967
\bibitem{so_2019} The Simons Observatory Collaboration \ 2019, \jcap, 2019, 056. doi:10.1088/1475-7516/2019/02/056
\bibitem{litebird_2022} LiteBIRD Collaboration \ 2022, arXiv:2202.02773. doi:10.48550/arXiv.2202.02773
\bibitem{lspe_2021} The LSPE Collaboration\ 2021, \jcap, 2021, 008. doi:10.1088/1475-7516/2021/08/008
\bibitem{quiet_13} Bischoff, C., Brizius, A., Buder, I., et al.\ 2013, \apj, 768, 9. doi:10.1088/0004-637X/768/1/9
\bibitem{taylor_06} Taylor, A.~C.\ 2006, \nar, 50, 993. doi:10.1016/j.newar.2006.09.026
\bibitem{castro_2016} Castro-Almaz{\'a}n, J.~A., Mu{\~n}oz-Tu{\~n}{\'o}n, C., Garc{\'\i}a-Lorenzo, B., et al.\ 2016, \procspie, 9910, 99100P. doi:10.1117/12.2232646
\bibitem{vielva_2022} Vielva, P., Mart{\'\i}nez-Gonz{\'a}lez, E., Casas, F.~J., et al.\ 2022, \jcap, 2022, 029. doi:10.1088/1475-7516/2022/04/029
\bibitem{aumont_2020} Aumont, J., Mac{\'\i}as-P{\'e}rez, J.~F., Ritacco, A., et al.\ 2020, \aap, 634, A100. doi:10.1051/0004-6361/201833504
\bibitem{rubino_2023} Rubi{\~n}o-Mart{\'\i}n, J.~A., Guidi, F., G{\'e}nova-Santos, R.~T., et al.\ 2023, \mnras, 519, 3383. doi:10.1093/mnras/stac3439
\bibitem{honda_2020} Honda, S., Choi, J., G{\'e}nova-Santos, R.~T., et al.\ 2020, \procspie, 11445, 114457Q. doi:10.1117/12.2560918
\bibitem{realini_2022} Realini, S., Franceschet, C., Villa, F., et al.\ 2022, Journal of Instrumentation, 17, P01028. doi:10.1088/1748-0221/17/01/P01028
\bibitem{franceschet_2022} Franceschet, C., Del Torto, F., Villa, F., et al.\ 2022, Journal of Instrumentation, 17, P01029. doi:10.1088/1748-0221/17/01/P01029
\bibitem{peverini_2022} Peverini, O.~A., Lumia, M., Farooqui, Z., et al.\ 2022, Journal of Instrumentation, 17, P06042. doi:10.1088/1748-0221/17/06/P06042
\bibitem{jarosik_2011} Jarosik, N., Bennett, C.~L., Dunkley, J., et al.\ 2011, \apjs, 192, 14. doi:10.1088/0067-0049/192/2/14
\bibitem{cpp2013-5} Planck 2013 results V. Planck Collaboration \ 2014, \aap, 571, A5. doi:10.1051/0004-6361/201321527
\bibitem{cpp2013-8} Planck 2013 results VIII. Planck Collaboration \ 2014, \aap, 571, A8. doi:10.1051/0004-6361/201321538
\bibitem{cpp2015-26} Planck 2015 results XXVI. Planck Collaboration \ 2016, \aap, 594, A26. doi:10.1051/0004-6361/201526914
\bibitem{laing_1988} Laing, R.~A.\ 1988, \nat, 331, 149. doi:10.1038/331149a0
\bibitem{anderson_1995} Anderson, M.~C., Keohane, J.~W., \& Rudnick, L.\ 1995, \apj, 441, 300. doi:10.1086/175356
\bibitem{weiland_2011} Weiland, J.~L., Odegard, N., Hill, R.~S., et al.\ 2011, \apjs, 192, 19. doi:10.1088/0067-0049/192/2/19
\bibitem{hafez_2014} Hafez, Y.~A., Trojan, L., Albaqami, F.~H., et al.\ 2014, \mnras, 439, 2271. doi:10.1093/mnras/stt2476
\bibitem{wilson_2013} Wilson, T.~L., Rohlfs, K. \& H\"uttemeister, S.\ 2013, {\it Tools of Radio Astronomy}, Springer, ISBN 978-3-642-39950-3
\bibitem{condon_2016} Condon, J.~J. \& Ransom, S.~M.\ 2016, {\it Essential Radio Astronomy}, Princeton University Press, ISBN 978-0-691-13779-7
\bibitem{baars_1977} Baars, J.~W.~M., Genzel, R., Pauliny-Toth, I.~I.~K., et al.\ 1977, \aap, 61, 99
\bibitem{hafez_2008} Hafez, Y.~A., Davies, R.~D., Davis, R.~J., et al.\ 2008, \mnras, 388, 1775. doi:10.1111/j.1365-2966.2008.13515.x
\bibitem{ritacco_2018} Ritacco, A., Mac{\'\i}as-P{\'e}rez, J.~F., Ponthieu, N., et al.\ 2018, \aap, 616, A35. doi:10.1051/0004-6361/201731551
\bibitem{aumont_2010} Aumont, J., Conversi, L., Thum, C., et al.\ 2010, \aap, 514, A70. doi:10.1051/0004-6361/200913834
\bibitem{westerhout_1958} Westerhout, G.\ 1958, Bulletin of the Astronomical Institutes of the Netherlands, 14, 215
\bibitem{tramonte_2023} Tramonte, D., G{\'e}nova-Santos, R.~T., Rubi{\~n}o-Mart{\'\i}n, J.~A., et al.\ 2023, \mnras, 519, 3432. doi:10.1093/mnras/stac3502
\bibitem{quireza_2006} Quireza, C., Rood, R.~T., Bania, T.~M., et al.\ 2006, \apj, 653, 1226. doi:10.1086/508803
\bibitem{cpp2015-4} Planck Collaboration, Ade, P.~A.~R., Aghanim, N., et al.\ 2016, \aap, 594, A4. doi:10.1051/0004-6361/201525809
\bibitem{fahd_1992} Fahd A. K., 1992, PhD thesis, ``Study and Interpretation of the Millimeter-Wave Spectrum of Venus.'' Georgia Institute of Technology, Atlanta.
\bibitem{bellotti_2015} Bellotti A., 2015, PhD thesis, ``The millimeter-wavelength sulfur dioxide absorption spectra measured under simulated Venus conditions'', Georgia Institute of Technology, Atlanta.
\bibitem{karim_2018} Karim, R.~L., DeBoer, D., de Pater, I., et al.\ 2018, \aj, 155, 129. doi:10.3847/1538-3881/aaaab2
\bibitem{pir52} Planck Intermediate Results LII. Planck Collaboration \ 2017, \aap, 607, A122. doi:10.1051/0004-6361/201630311
\bibitem{maris21} Maris, M., Romelli, E., Tomasi, M., et al.\ 2021, \aap, 647, A104. doi:10.1051/0004-6361/202037788
\bibitem{healpix} G{\'o}rski, K.~M., Hivon, E., Banday, A.~J., et al.\ 2005, \apj, 622, 759. doi:10.1086/427976
\bibitem{realini_2020} Realini, S., 2020, PhD thesis, ``Analysis of the Strip optical system for CMB polarization measurements'', University of Milan.
\bibitem{guidi_2021} Guidi, F., Rubi{\~n}o-Mart{\'\i}n, J.~A., Pelaez-Santos, A.~E., et al.\ 2021, \mnras, 507, 3707. doi:10.1093/mnras/stab2422
\bibitem{sandri_2010} Sandri, M., Villa, F., Bersanelli, M., et al.\ 2010, \aap, 520, A7. doi:10.1051/0004-6361/200912891
\bibitem{ludwig_1973} Ludwig, A.~C.\ 1973, IEEE Transactions on Antennas and Propagation, 21, 116. doi:10.1109/TAP.1973.1140406
\bibitem{hinshaw_2003} Hinshaw, G., Barnes, C., Bennett, C.~L., et al.\ 2003, \apjs, 148, 63. doi:10.1086/37722210.48550/arXiv.astro-ph/0302222
\bibitem{pardo_2001} Pardo, J.~R., Cernicharo, J., \& Serabyn, E.\ 2001, IEEE Transactions on Antennas and Propagation, 49, 1683. doi:10.1109/8.982447
\bibitem{krachmalnicoff_2015} N. Krachmalnicoff, Challenges for Present and Future Cosmic Microwave Background Observations: Systematic Effects and Foreground Emission in Polarization, Ph.D. thesis, University of Milan, 2015.






\end{thebibliography}
\end{document}